%% file: thesis.tex
\documentclass{sdsu-thesis}

\usepackage{epsfig}
\usepackage{graphicx}
\usepackage{tikz}
\usetikzlibrary{quantikz2}
\usepackage{multirow}
\usepackage{amsmath}
\usepackage{amsfonts}
\usepackage{amssymb}
\usepackage{amsthm}
\usepackage{longtable}
\usepackage[labelsep=period,textfont=bf,labelfont=bf]{caption}
\usepackage{subcaption}
\usepackage[normalem]{ulem}
\usepackage{color, colortbl}
\definecolor{Gray}{gray}{0.9}

\newcommand{\expect}[1]{ \langle #1 \rangle} 

\newsavebox{\boxI}
\newsavebox{\boxX}
\newsavebox{\boxY}
\newsavebox{\boxZ}
\newsavebox{\boxH}
\newsavebox{\boxS}

\author{Amanda Bowman}

\title{NUCLEAR SPECTRA FROM QUANTUM LANCZOS ALGORITHM\\
 WITH REAL-TIME EVOLUTION AND MULTIPLE REFERENCE STATES}
\titleheading{Nuclear Spectra from Quantum Lanczos Algorithm\\
 with Real-Time Evolution and Multiple Reference States}

\degreeONE{Master of Science}
\degreeTWO{in}
\degreeTHREE{Computational Science}

\gradyear{2023}

\submitdate{Summer 2023}

\committeechair{Calvin Johnson}
\committeechairdept{Department of Physics}

\committeesecond{Andrew Cooksy}
\committeeseconddept{Department of Chemistry and Biochemistry}

\committeethird{Peter Blomgren}
\committeethirddept{Department of Mathematics and Statistics}

\begin{document}

% Title page 
\maketitle

% Copyright page
\begin{copyrightpage}
  Copyright~\copyright~2023 \\
  by \\
  Amanda Bowman
\end{copyrightpage}

% ======================================================================
% Abstract
% ======================================================================

\begin{abstract}
  \input{abstract}
\end{abstract}

% ======================================================================
% Table of contents
% ======================================================================

\tableofcontents

% ======================================================================
% List of tables
% ======================================================================

\listoftables

% ======================================================================
% List of figures
% ======================================================================

\listoffigures

% ======================================================================
% Acknowledgements
% ======================================================================

\begin{acknowledgments}
    I would like to express my deepest gratitude to my committee chair and research advisor, Dr. Calvin Johnson. I am grateful for the opportunity they gave me to work with them. Their unwavering support, guidance, and expertise have been invaluable throughout my journey as a master's student. I would also like to thank Dr. Andrew Cooksy and Dr. Peter Blomgren for serving on my committee.

    I would also like to thank Dr. Ionel Stetcu from Los Alamos National Lab. Their collaboration and assistance were instrumental in developing the quantum circuit simulations used in this research. Their expertise and willingness to share knowledge have contributed significantly to the success of this thesis.

    This work was supported by the U.S. Department of Energy, Office of Science, Office of High Energy Physics, under Award No.~DE-SC0019465.
\end{acknowledgments}

% ======================================================================
% Body
% ======================================================================

\input{body}

% ======================================================================
% References
% ======================================================================

\bibliographystyle{siamplain}
\bibliography{thbib}

\end{document}

%% file: abstract.tex
Models of quantum systems scale exponentially with the addition of single-particle states, which can present computationally intractable problems. Alternatively, quantum computers can store a many-body basis of $2^n$ dimensions on $n$ qubits. This motivated the quantum eigensolver algorithms developed in recent years, such as the quantum Lanczos algorithm based on the classical, iterative Lanczos algorithm. I performed numerical simulations to find the low-lying eigenstates of $^{20}$Ne, $^{22}$Na, and $^{29}$Na to compare imaginary- and real-time evolution. Though imaginary-time evolution leads to faster convergence, real-time evolution still converges within tens of iterations and satisfies the requirement for unitary operators on quantum computers. Additionally, using multiple reference states leads to faster convergences or higher accuracy for a fixed number of real-time iterations. I performed quantum circuit prototype numerical simulations on a classical computer of the QLanczos algorithm with real-time evolution and multiple reference states to find the low-lying eigenstates of $^{8}$Be. These simulations were run in both the spherical basis and Hartree-Fock basis, demonstrating that an M-scheme spherical basis leads to lower depth circuits than the Hartree-Fock basis. Finally, I present the quantum circuits for the QLanczos algorithm with real-time evolution and multiple references.

%% file: body.tex
%%%%%%%%%%%%%%%%%%%%%%%%%%%%%%%%%%%%%%%%%%%%%%%%%%%%%%%%
%%%%%%%%%%%%%%% Chapter:Introduction %%%%%%%%%%%%%%%%%%
%%%%%%%%%%%%%%%%%%%%%%%%%%%%%%%%%%%%%%%%%%%%%%%%%%%%%%%%

\chapter{Introduction}\label{chap:intro}
Solving the eigenvalue problem is a common task in physics, quantum chemistry, applied mathematics, engineering, and many other fields. The Lanczos algorithm \cite{Lanczos1950} is an iterative eigensolver that finds the extremal eigenvalues of large matrices. The Lanczos algorithm can be used in physics and quantum chemistry to find low-lying nuclear and chemical stationary states. The Hilbert space for these quantum systems scales exponentially with the number of single-particle states in the model, which presents a computational challenge for classical computers. Many-body Hamiltonian matrices can reach dimensions of billions, even when using selection rules to reduce the basis size, requiring large amounts of memory ranging from hundreds of gigabytes to petabytes \cite{Johnson2013}. Alternatively, a many-body basis of $2^n$ dimensions in Hilbert space naively requires only $n$ qubits on a quantum computer. Hence, there is interest in developing quantum eigensolver algorithms for many-body systems.

One of the first proposed quantum eigensolver algorithms is based on quantum phase estimation (QPE) \cite{Kitaev1995,Abrams1997} and used to compute the ground state energy of fermionic systems \cite{Ortiz2001} and molecules \cite{Asupuru-Guzik2005}. QPE is a multi-ancilla qubit method that estimates the phase, $\theta$, of a unitary operator, $U\ket{\psi}=e^{2\pi i \theta}\ket{\psi}$, using a quantum Fourier transformation (QFT). While it produces highly precise eigenstates, it requires deep circuits, which are impractical for current quantum computers. Circuit depth is the measure of quantum gates a quantum circuit has, corresponding to the time it takes to execute a quantum circuit on a quantum computer. Consequently, variational methods that require shallower circuits, such as the variational quantum eigensolver (VQE) \cite{Peruzzo2014,McClean2016,OMalley2016,Kandala2017,Huggins2020,Romero2022,PérezObiol2023} and quantum approximate optimization algorithm (QAOA) \cite{Farhi2022}, were developed to compute ground-state energies of molecules and nuclei and extended to compute excited states \cite{McClean2017,Colless2018,Higgot2019,Nakanishi2019,Kiss2022,Gocho2023}. Variational quantum algorithms parameterize a trial wavefunction and optimize via classical computers. Variational methods require shallower circuits, making them suitable for near-term quantum computers. However, whether these methods will be applicable to systems with large basis dimensions is unclear, and the classical optimization may prove to be a limiting factor. An emerging family of quantum eigensolver algorithms are the quantum subspace diagonalization (QSD) \cite{Seki2021,Epperly2022,Cortes2022QSD} schemes. Within this family of methods is the quantum Lanczos (QLanczos) algorithm \cite{Motta2020,Yeter2020}, which is the focus of this thesis.

In 2019, Motta et al. \cite{Motta2020} presented the QLanczos algorithm, a modified version of the Lanczos algorithm for quantum computers inspired by classical implementations of quantum Monte Carlo (QMC) methods \cite{Caffarel1991,Blunt2015}. The quantum imaginary-time evolution (QITE) operator replaces the powers of a Hamiltonian used in the classical Lanczos algorithm. In this algorithm, the non-unitary, imaginary-time evolution operator is approximated as a unitary operator, as quantum computers require unitary operators to preserve the norm of wavefunctions. A natural alternative is the unitary real-time evolution operator introduced to quantum computing by Parrish and McMahon \cite{Parrish2019} and referred to as quantum filter diagonalization (QFD). Stair et al. \cite{Stair2020} generalized this approach using the multireference selected quantum Krylov (MRSQK) algorithm, which does not increase the circuit depth and achieves higher accuracy by using more Krylov basis states. In this and other relevant literature, the basis states of the Krylov subspace are referred to as Krylov basis states. Many quantum algorithms are hybrid quantum-classical algorithms that require some post-processing on a classical computer.

In the present study, I build on the aforementioned studies by using the QLanczos algorithm with real-time evolution and multiple reference states to compute the ground and excited state energies of nuclear Hamiltonians. This combination of methods harnesses the advantages of the QLanczos algorithm, which efficiently computes off-diagonal matrix elements, the real-time evolution operator, which is unitary and, therefore, more natural on a quantum computer, and multiple reference states, which increases the accuracy without increasing the circuit depth. Additionally, I discuss the advantages of working in the spherical basis over the commonly used Hartree-Fock basis.

\begin{figure}
    \centering
    \vspace{-3mm}
    \includegraphics[width=1\textwidth]{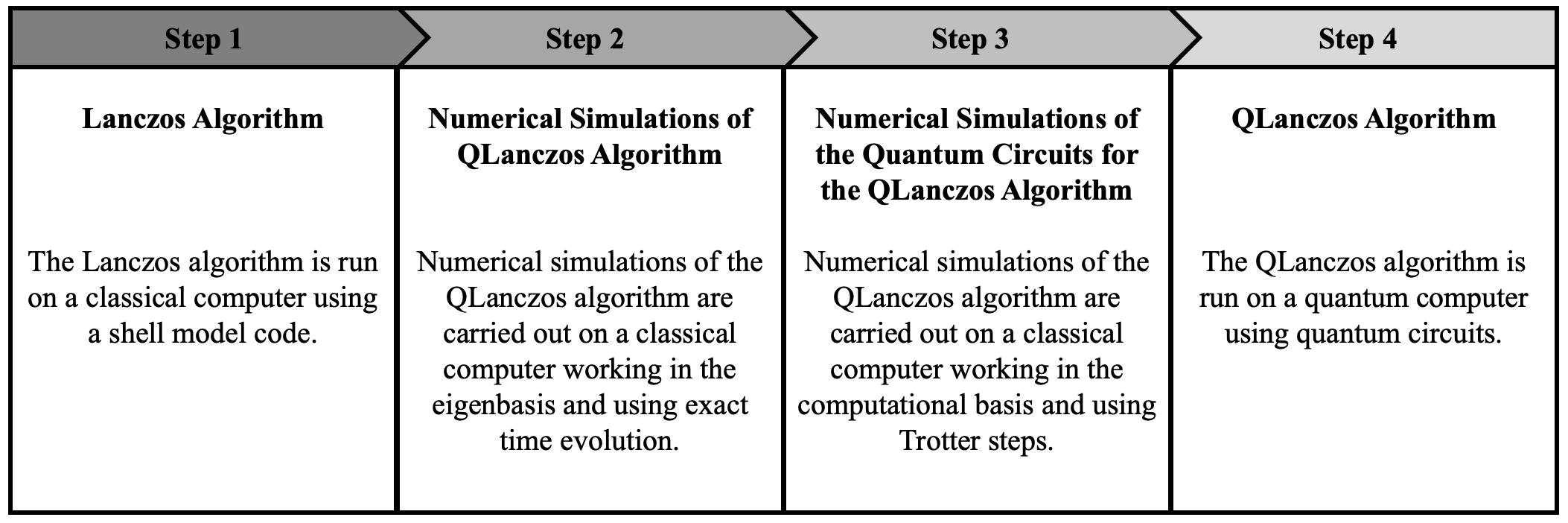}
    \caption{The four steps for developing the QLanczos algorithm. First, the Hamiltonian is fully diagonalized via the Lanczos algorithm using a shell model code (C. W. Johnson, W. E. Ormand, K. S. McElvain, and H. Shan, BIGSTICK: A flexible configuration-interaction shell-model code, arXiv:1801.08432, (2018).) Second, numerical simulations of the algorithm are run on a classical computer with exact evolution in the eigenbasis. Step three is a prototype for the quantum circuits. The Hamiltonian is mapped to qubits in the computational basis using the Jordan-Wigner mapping, and the time evolution is approximated using Trotterization. Finally, quantum circuits are designed to run on a quantum computer. I completed steps one through three in this study and sketched the quantum circuits needed to perform the QLanczos algorithm on a quantum computer (step four).}
    \vspace{-6mm}
    \label{fig:foursteps}
\end{figure}

Figure \ref{fig:foursteps} shows the process for developing the QLanczos algorithm as one goes from purely classical to mostly quantum. The first step was to fully diagonalize the nuclear Hamiltonians using the classical Lanczos algorithm using a shell model code \cite{Johnson2018}. Second, numerical simulations of the QLanczos algorithm were carried out on a classical computer using exact time evolution and working in the eigenbasis. These preliminary numerical simulations were performed as an initial evaluation of the QLanczos algorithm on a classical computer. The third step was to run numerical simulations of the quantum circuits that would be used on a quantum computer. This step was also performed on a classical computer, but using the computational basis and Trotter steps to approximate real-time evolution. Finally, the quantum circuits are designed for running the QLanczos algorithm on a quantum computer. Systems using only a few logical qubits still require extensive noise mitigation. Hence, I sketched out the quantum circuits needed to run the QLanczos algorithm on a quantum computer but did not carry out the computations.

\vspace{2mm}

In Chapter \ref{chap:nuclear-many-body-systems}, I give a brief background of nuclear many-body systems. I review the classical Lanczos algorithm and then introduce the QLanczos algorithm with real-time evolution and multiple references in Chapter \ref{chap:qlanczos-algorithm}. In Chapter \ref{chap:numerical-simulations}, I use the fully diagonalized Hamiltonians from Step 1 in Figure \ref{fig:foursteps} to construct the eigenbasis for the numerical simulations of the QLanczos algorithm (Step 2 in Figure \ref{fig:foursteps}) and present the results.

In the second half of this thesis, I cover the development of the QLanczos algorithm on a quantum computer. I begin with the fundamentals of gate-based quantum computing in Chapter \ref{chap:gate-based-quantum-computing}. In Chapter \ref{chap:preapering-qlanczos-quantum-computer}, I explain how to set up a nuclear system for a quantum computer. In Chapter \ref{chap:quantum-circuits-qlanczos}, I sketch the quantum circuits for Step 4 of Figure \ref{fig:foursteps}.

In Chapter \ref{chap:results}, I present the results from the quantum circuit prototype numerical simulations from Step 3 of Figure \ref{fig:foursteps}. In Chapter \ref{chap:conclusions}, I discuss the results.

%%%%%%%%%%%%%%%%%%%%%%%%%%%%%%%%%%%%%%%%%%%%%%%%%%%%%%%%
%%%%%%%%%%%%%%% Chapter:Background %%%%%%%%%%%%%%%%%%%%
%%%%%%%%%%%%%%%%%%%%%%%%%%%%%%%%%%%%%%%%%%%%%%%%%%%%%%%%

\chapter{Nuclear Many-Body Systems}\label{chap:nuclear-many-body-systems}
In this chapter, I will briefly overview the mathematical foundation for modeling nuclear systems. Many-body methods allow us to model quantum systems with many particles. The nuclear shell model is a framework used to describe the structure of atomic nuclei. I will first present the Hartree-Fock method and then the interacting shell model. I use these models later as a basis for nuclear Hamiltonians that will be solved by the QLanczos algorithm.

\section{Many-Body Systems}\label{sec:many-body}

The Hamiltonian for an N-particle system is

\vspace{-1mm}
\begin{equation}\label{eq:many-body-ham}
    \hat{H}=\sum_{i=1}^N-\frac{\hbar^2}{2m}\nabla_i^2+\sum_{i<j}V(\hat{r}_i-\hat{r}_j)
\end{equation}

\noindent where the first term is the kinetic energy of each particle and the second term is the two-body interaction between particles. The time-independent Schr\"{o}dinger equation is then

\vspace{-3mm}
\begin{equation}\label{eq:many-body-shro}
    \hat{H}\Psi(\vec{r}_1,\vec{r}_2,...,\vec{r}_N)=E\Psi(\vec{r}_1,\vec{r}_2,...,\vec{r}_N),
\end{equation}

\noindent where $\Psi$ is the wavefunction that describes the properties of each particle in the system. Systems with more than a few particles are not easily solved analytically except in specific cases. Instead, these complex systems can be solved numerically by choosing a basis, computing the matrix elements of the Hamiltonian, and then diagonalizing.

The wavefunction from Equation (\ref{eq:many-body-shro}) can be expanded into a basis of our choosing:

\vspace{-3mm}
\begin{equation}
    \Psi(\vec{r}_1,\vec{r}_2,...,\vec{r}_N)=\sum_\alpha c_\alpha \Phi_\alpha(\vec{r}_1,\vec{r}_2,...,\vec{r}_N).
\end{equation}

\noindent A convenient choice for the basis function, $\Phi_\alpha$, is a product wavefunction constructed of single-particle wavefunctions,

\vspace{-3mm}
\begin{equation}\label{eq:hartree-prod}
    \Phi(\vec{r}_1,\vec{r}_2,...,\vec{r}_N) = \phi_1(\vec{r}_1)\phi_2(\vec{r}_2)\dots \phi_N(\vec{r}_N),
\end{equation}

\noindent where $\phi_i$ are the orthonormal, single-particle states. However, since the goal is to model fermions, the type of particles that make up nuclei, the wavefunction should be antisymmetric under the exchange of coordinates, $\vec{r}_i\leftrightarrow\vec{r}_j$. Slater determinants are such antisymmetric product wavefunctions that describe fermion systems.

\section{Slater Determinants}

To satisfy the antisymmetry of fermions, the sign of the product wavefunction should flip under the exchange of any two fermions. A Slater determinant can represent the antisymmetric wavefunction of a system with N fermions,

\begin{equation}
    \Phi(\vec{r}_1,\vec{r}_2,...,\vec{r}_N) = \frac{1}{\sqrt{N!}}\det
    \begin{vmatrix}
        \phi_1(\vec{r}_1) & \phi_2(\vec{r}_1) & \dots & \phi_N(\vec{r}_1)\\
        \phi_1(\vec{r}_2) & \phi_2(\vec{r}_2) & \dots & \phi_N(\vec{r}_2)\\
        \vdots & \vdots & \ddots & \vdots\\
        \phi_1(\vec{r}_N) & \phi_2(\vec{r}_N) & \dots & \phi_N(\vec{r}_N)
    \end{vmatrix},
\end{equation}
\vspace{1mm}

\noindent where $\frac{1}{\sqrt{N!}}$ is the normalization factor. However, one seldom explicitly computes with determinants but instead uses second quantization.

\section{Second Quantization}\label{sec:second-quantization} 

Slater determinants can be written in occupation representation using second quantization. In occupation representation, we do not keep track of the state of each particle; instead, we keep track of how many particles are in each state. Occupation numbers, $n_i$, define the number of particles in a state, $\phi_i$. The many-body wavefunction is then an abstract vector using occupation number representation,

\vspace{-3mm}
\begin{equation}\label{eq:wfn_secquant}
    \ket{\Psi} = \ket{n_1,n_2,...,n_N},
\end{equation}

\noindent in which normalization is assumed. The anticommutation relations encode the antisymmetry of fermions,

\vspace{-4mm}
\begin{equation}
    \{\hat{a}_i,\hat{a}_j\}=\{\hat{a}^\dagger_i,\hat{a}^\dagger_j\}=0,
\end{equation}

\vspace{-3mm}
\begin{equation}
    \{\hat{a}_i,\hat{a}^\dagger_j\}=\delta_{i,j}.
\end{equation}

\noindent Two identical fermions cannot occupy the same state, so the only possible occupation number for fermions is $0$ or $1$. Applying a creation operator, $\hat{a}^\dagger$, to the vacuum state creates a particle, $\hat{a}^\dagger\ket{0}=\ket{1}$, and applying the annihilation operator, $\hat{a}$, removes a particle, $\hat{a}\ket{1}=\ket{0}$. In the many-body basis, the creation and annihilation operators can act on the single-particle states by adding or removing particles. For example, a many-body state with two particles and four single-particle states can be represented as

\vspace{-3mm}
\begin{equation}
    \hat{a}^\dagger_0\hat{a}^\dagger_3\ket{0000}=\ket{1001},
\end{equation}

\noindent where the subscript labels the single-particle state the creation operator applies to.

The Hamiltonian in second quantization is

\vspace{-2mm}
\begin{equation}\label{eq:Ham_secondquantization}
    \hat{H} = \sum_{ij}\bra{i}\hat{T}\ket{j}\hat{a}^\dagger_i \hat{a}_j + \frac{1}{4}\sum_{ijkl} \bra{ij}\hat{V}\ket{kl} \hat{a}^\dagger_i \hat{a}^\dagger_j \hat{a}_l \hat{a}_k,
\end{equation}

\noindent where the first term is the one-body operator, which defines the kinetic energy, and the second term is the two-body operator, which defines the interactions between particles. 

\section{Nuclear Shell Model}\label{sec:shell-model}
The nuclear shell model is a framework for choosing the single-particle states, $\phi_i$. The interaction between particles, the second term of Equation (\ref{eq:Ham_secondquantization}), can be approximated as a mean field where each nucleon interacts with the average potential of the other nucleons. The single-particle states are taken as time-independent solutions to the potentials. A common approximation to the mean-field potential is the 3-dimensional harmonic oscillator.

The potential is chosen to be rotationally invariant, resulting in four good quantum numbers: the radial quantum number ($n$), the orbital quantum number ($l$), the total angular momentum ($j$), and the z-component of total angular momentum ($m$). The radial quantum number is an integer value that describes the number of nodes in the radial wavefunction. Some authors start at $n=1$, although it does not coincide with the physical meaning. Here I use the convention of starting the radial quantum number at $n=0$. In spectroscopic notation, letters label the levels of orbital quantum number:  $s$ is $l=0$, $p$ is $l=1$, $d$ is $l=2$, $f$ is $l=3$, and so on. Combinations of the quantum numbers label unique orbitals written as $n(l)_j$, where $(l)$ is the letter that labels the orbital quantum number. There are $2j+1$ degeneracies for each orbital with $m=-j, -j+1, ..., j-1, j$. The first orbital is $0s_{1/2}$ which means $n=0$, $l=0$ and $j=1/2$ and there are two possible states, $m=+1/2$ and $m=-1/2$. Table \ref{tab:orbitals} shows the first few orbitals.

\begin{table}
\begin{center}
\begin{tabular}{c | c | c | c | c } 
 orbital & $n$ & $l$ & $j$ & $m$ \\ [0.5ex] 
 \hline
 $0s_{1/2}$ & 0 & 0 & 1/2 & +1/2, -1/2 \\
 $0p_{3/2}$ & 0 & 1 & 3/2 & +3/2, +1/2, -1/2, -3/2 \\
 $0p_{1/2}$ & 0 & 1 & 1/2 & +1/2, -1/2 \\
 $0d_{5/2}$ & 0 & 2 & 5/2 & +5/2, +3/2, +1/2, -1/2, -3/2, -5/2 \\
 $1s_{1/2}$ & 1 & 0 & 1/2 & +1/2, -1/2 \\
 $0d_{3/2}$ & 0 & 2 & 3/2 & +3/2, +1/2, -1/2, -3/2 \\
\end{tabular}
\bigskip
\caption{The first six orbitals of the nuclear shell model. Three quantum numbers, $n(l)_j$, describe an orbital. Each orbital can contain up to $2j+1$ nucleons with a unique $m$ for each species. \label{tab:orbitals}}

\end{center} 
\vspace{-13mm}
\end{table}

The harmonic oscillator has an infinite number of single-particle states. Since we cannot model an infinite number of states, we ignore states with a low probability of being occupied. The states with a very high probability of being occupied are considered to be an inert core that adds an effective energy to the Hamiltonian. The remaining particles are the valence particles considered to orbit around the inert core. The model space is constructed from the states that the valence particles are likely to fill. For example, $^{14}$N has seven protons and seven neutrons. An inert core of six protons and six neutrons could be assumed, $^{12}$C, which fills the $0s_{1/2}$ and $0p_{3/2}$ shells. Then the valence particles are one proton and one neutron, which would likely exist in the $0p_{1/2}$ shell. The model space has two possible states the valence particle could fill for each species.

\subsection{Hartree-Fock}\label{sec:HF}
The Hartree-Fock method is an independent particle model that approximates the interacting term of Equation (\ref{eq:Ham_secondquantization}) as a mean field where each particle interacts with the average potential of all other particles. We begin with a trial wavefunction that is assumed to be a single Slater determinant,

\begin{equation}
    \ket{\Psi_T}=\prod_\alpha\hat{c}^\dagger_\alpha\ket{0}.
\end{equation}

\noindent From the variational theorem, we know that the expectation energy of the Hamiltonian,

\vspace{-3mm}
\begin{equation}\label{eq:expectation_value}
    \frac{\bra{\Psi_T}\hat{H}\ket{\Psi_T}}{\braket{\Psi_T}{\Psi_T}},
\end{equation}

\noindent is an upper bound to the ground state energy. The minimum is found by varying the expectation energy with respect to the single-particle states. The single-particle states can be linear combinations of states with different $j$, $l$, and $m$. The state with the lowest single-particle energy is the Hartree-Fock state, an approximation of the ground state. This first-order method is not highly accurate and can be used as a starting point for other algorithms, such as the Lanczos algorithm. In this thesis, I will use the Hartree-Fock state from a code by Stetcu and Johnson \cite{Stetcu2002}.

\subsection{Interacting Shell Model}\label{sec:interacting}
The interacting shell model basis is constructed using the shell model space described in Section \ref{sec:shell-model}. I will refer to this basis as the spherical basis. An exact solution would require an infinite number of Slater determinants. However, the model space is restricted to states that valence particles are most likely to occupy, as previously described, and further reduce the basis using a selection rule which I will discuss in this section. I will use the nucleus of $^{14}$N as an example to illustrate this model.

Assume an inert core of six protons and six neutrons and two valence particles, one proton and one neutron, the nucleus of $^{14}$N. I restrict the valence particles to the $0p_{1/2}$ shell. Each nucleon has two possible states for a total of four single-particle states in this model space, as shown in Table \ref{tab:orbitals-N14}.

\vspace{-1mm}
\begin{table}
\begin{center}
\begin{tabular}{ c | c | c | c } 
state & species & j & m \\ [0.5ex] 
\hline
 0 & p & 1/2 & +1/2 \\
 1 & p & 1/2 & -1/2 \\
 2 & n & 1/2 & +1/2 \\
 3 & n & 1/2 & -1/2 
\end{tabular}
\bigskip
\caption{\label{tab:orbitals-N14}The single-particle states that make up the shell model space for the valence particles of $^{14}$N, one proton and one neutron in the $0p_{1/2}$ shell.}
\end{center}
\vspace{-11mm}
\end{table}

\noindent The wavefunction of a nucleus is constructed of both proton and neutron single-particle states $\ket{\Psi}=\ket{j_p,m_p}\otimes\ket{j_n,m_n}$. For example, for one proton and one neutron in the $0p_{1/2}$ shell, there are four Slater determinants,

\vspace{-3mm}
\begin{equation}
    \begin{aligned}
        \ket{0}\otimes\ket{2} &= \ket{1/2,1/2}\otimes\ket{1/2,1/2} \\
        \ket{0}\otimes\ket{3} &= \ket{1/2,1/2}\otimes\ket{1/2,-1/2} \\
        \ket{1}\otimes\ket{2} &= \ket{1/2,-1/2}\otimes\ket{1/2,1/2} \\
        \ket{1}\otimes\ket{3} &= \ket{1/2,-1/2}\otimes\ket{1/2,-1/2}.
    \end{aligned}
\end{equation}

\noindent In shorthand, the tensor product is omitted. For example, $\ket{0}\otimes\ket{2}$ can be written as $\ket{0,2}$. These states can be represented in second quantization as

\vspace{-3mm}
\begin{equation}
    \begin{aligned}
        \ket{0,2} &= \ket{1010} \\
        \ket{0,3} &= \ket{1001} \\
        \ket{1,2} &= \ket{0110} \\
        \ket{1,3} &= \ket{0101}.
    \end{aligned}
\end{equation}

The many-body basis grows exponentially with the addition of single-particle states. The basis can be reduced using a selection rule called the M-scheme. States with a specified total $M$ are selected, where $M$ is the sum of the z-component of total angular momentum for protons and neutrons, $M = m_p+m_n$. For the case of $^{14}$N described above, if $M=0$, the only Slater determinants allowed are $\ket{0,3}$ and $\ket{1,2}$. The matrix elements of the Hamiltonian, $\hat{H}$, in an M-scheme spherical basis are computed by

\vspace{-3mm}
\begin{equation}
    H_{m,n}=\bra{\Psi_m}\hat{H}\ket{\Psi_n},
\end{equation} 

\noindent where \ket{\Psi_i} are the M-scheme Slater determinants. In the next Chapter, I discuss how the Lanczos algorithm diagonalizes the matrix representation of the Hamiltonian.

%%%%%%%%%%%%%%%%%%%%%%%%%%%%%%%%%%%%%%%%%%%%%%%%%%%%%%%%%%%%%%%%
%%%%%%%%%%%%%%% Chapter:QLanczos Algorithm %%%%%%%%%%%%%%%%%%%%
%%%%%%%%%%%%%%%%%%%%%%%%%%%%%%%%%%%%%%%%%%%%%%%%%%%%%%%%%%%%%%%%

\chapter{QLanczos Algorithm}\label{chap:qlanczos-algorithm}

The QLanczos algorithm is based on the well-known classical Lanczos algorithm \cite{Lanczos1950}. The Lanczos algorithm is an iterative eigensolver for Hermitian matrices. It is useful when one does not need the entire eigenspectrum or when large dimensions make traditional methods of fully solving the eigenvalue problem insurmountable. The Lanczos algorithm was introduced to nuclear physics by Whitehead \cite{Whitehead1977} in 1977 and has since been a common method to solve for low-lying eigenstates of nuclear Hamiltonians. 

For nuclear systems, we are concerned with solving the many-body eigenvalue problem in Equation (\ref{eq:many-body-shro}). The Lanczos algorithm repeatedly applies the Hamiltonian, $\hat{H}$, to a pivot vector, $\ket{v_1}$, $S$ times. Each iteration creates a new vector

\vspace{-3mm}
\begin{equation}
\ket{v_k}=\hat{H}^k\ket{v_1},
\end{equation}

\noindent where $k=0,1,2,...S$. Each new vector is orthogonalized against all previous vectors such that $\braket{v_k}{v_l}=\delta_{kl}$. This set of Lanczos vectors creates a $(S+1)$-dimensional Krylov subspace in which the Hamiltonian is tridiagonal,

\vspace{-3mm}
\begin{equation}
    \hat{H}_\mathrm{Krylov}=
    \begin{pmatrix}
        \alpha_1 & \beta_1  &          &             & 0\\
        \beta_1  & \alpha_2 & \beta_2  &             & \\
                 & \beta_2  & \alpha_3 & \ddots      & \\
                 &          & \ddots   & \ddots      & \beta_{k-1} \\
        0        &          &          & \beta_{k-1} & \alpha_k
    \end{pmatrix},
\end{equation}

\noindent where $\alpha_k=\bra{v_k}\hat{H}\ket{v_k}$ and $\beta_k=\bra{v_{k+1}}\hat{H}\ket{v_k}=\bra{v_k}\hat{H}\ket{v_{k+1}}$. Then a traditional eigensolver can be used to solve for the eigenvalues of $\hat{H}_\mathrm{Krylov}$. The extremal eigenvalues of the Hamiltonian in the Krylov subspace converge to the extremal eigenvalues of the Hamiltonian in the full space. Hence, the full transformation does not need to be carried out to get the low-lying eigenvalues. Although there is no general rule, and it depends upon the Hamiltonian and chosen reference state, the ground state energy can usually be found within 50 iterations.

In 2019, Motta et al. \cite{Motta2020} developed a modified version of the Lanczos algorithm called the quantum Lanczos (QLanczos) algorithm. The QLanczos algorithm is based on classical applications of the Lanczos algorithm and diffusion Monte Carlo methods in which the imaginary-time propagator, $e^{-\tau \hat{H}}$, replaces $\hat{H}$ \cite{Caffarel1991,Blunt2015}. Imaginary-time evolution is used to find low-lying eigenstates because, for large $\tau$, the ground-state is projected out,

\begin{equation}
    \ket{\Psi_{\mathrm{gs}}}=\underset{\tau \rightarrow \infty}{\mathrm{lim}}e^{-\tau \hat{H}}\ket{\Psi_0},
\end{equation}

\noindent where $\ket{\Psi_0}$ is a reference state and $\ket{\Psi_{\mathrm{gs}}}$ is the ground state. Time is broken up into small, fixed steps, $\Delta \tau$, and used to generate basis states for the Krylov subspace,

\vspace{-3mm}
\begin{equation}\label{eq:qite_Qlanczos}
        \ket{\Psi_k}=\left(e^{-\Delta \tau \hat{H}}\right)^k\ket{\Psi_0}.
\end{equation}

\noindent While imaginary-time evolution is useful on classical computers, it is not easily translated onto a quantum computer. Quantum computers require unitary operators, and the imaginary-time evolution operator is not unitary. To overcome this, Motta et al. replaces the imaginary-time evolution steps with unitary updates.

\section{Real-Time Evolution}\label{sec:RTE}

Alternatively, the real-time evolution operator, $e^{-it\hat{H}}$, is naturally unitary. Parrish and McMahon \cite{Parrish2019} were the first to introduce a real-time evolution eigensolver to quantum computing. Although real-time evolution has limited use on classical computers because it oscillates rather than decays like imaginary-time evolution, it is natural for quantum computers because it is unitary. Quantum computers require quantum circuits to be unitary; therefore, real-time evolution can be directly realized as a quantum circuit. Like imaginary-time evolution, real-time evolution can be used to generate Krylov basis states. The QLanczos algorithm with real-time evolution replaces the imaginary-time evolution operator in Equation (\ref{eq:qite_Qlanczos}) with the real-time evolution operator. Likewise, the real-time evolution operator broken up into fixed time steps is repeatedly applied to a reference state,

\vspace{-3mm}
\begin{equation}\label{eq:single-ref}
    \ket{\Psi_k} = \left(e^{-i\Delta t\hat{H}}\right)^k \ket{\Psi_0},
\end{equation}

\noindent where $\ket{\Psi_0}$ is a reference vector and $k=0,1,...S$. If the real-time evolution operator is applied $S$ times, then the Krylov subspace has a dimension of $D=S+1$. Since the basis states cannot be orthogonalized against each other easily on a quantum computer as in the classical Lanczos algorithm, an overlap between each state is computed. The matrix elements of the overlap matrix, $N_{k,l}$ and Hamiltonian, $H_{k,l}$, in the Krylov subspace are

\vspace{-3mm}
\begin{equation}\label{eq:norm}
    N_{k,l} = \braket{\Psi_k}{\Psi_l}
\end{equation}

\noindent and

\vspace{-3mm}
\begin{equation}\label{eq:Ham}
    H_{k,l} = \bra{\Psi_k}\hat{H}\ket{\Psi_l}.
\end{equation}

\noindent Then the generalized eigenvalue problem is solved on a classical computer to find the eigenvalues of the Hamiltonian in the Krylov subspace,

\begin{equation}\label{eq:gen-eig}
    \sum_lH_{k,l}\nu_l=E\sum_lN_{k,l}\nu_l. 
\end{equation}

\section{Multiple Reference States}

The multiple reference state generalization was introduced to the QLanczos algorithm by Stair et al. \cite{Stair2020}. This version uses a set of reference states to construct the Krylov basis states, which increases the potential overlap of the initial wavefunction with the ground state. The multiple reference generalization is a linear combination of $R$ reference states that replaces the single reference state from Equation (\ref{eq:single-ref}). The basis states are then,

\vspace{-3mm}
\begin{equation}
    \ket{\Psi_{a,k}} = \left(e^{-i\Delta t\hat{H}}\right)^k\ket{\Psi_a},
\end{equation}

\noindent where $a=1,2,...,R$. The dimension of the Krylov subspace is now $D=R(S+1)$. The basis states are naturally normalized because the real-time evolution operator is unitary. The overlap matrix and Hamiltonian in the Krylov subspace are then,

\vspace{-3mm}
\begin{equation}\label{eq:norm-multi}
    N_{a,k;b,l} = \braket{\Psi_{a,k}}{\Psi_{b,l}}
\end{equation}

\noindent and

\vspace{-3mm}
\begin{equation}\label{eq:Ham-multi}
    H_{a,k;b,l} = \bra{\Psi_{a,k}}\hat{H}\ket{\Psi_{b,l}}. 
\end{equation}

\noindent Then Equations (\ref{eq:norm-multi}) and (\ref{eq:Ham-multi}) can be substituted into the generalized eigenvalue problem,

\vspace{-3mm}
\begin{equation}\label{eq:gen-eig-multi}
    \sum_{b,l}H_{a,k;b,l}\nu_{b,l}=E\sum_{b,l}N_{a,k;b,l}\nu_{b,l},
\end{equation}

\noindent and solved on a classical computer. 

%%%%%%%%%%%%%%%%%%%%%%%%%%%%%%%%%%%%%%%%%%%%%%%%%%%%%%%%%%%%%%%%%%%
%%%%%%%%%%%%%%% Chapter:Numerical Simulations %%%%%%%%%%%%%%%%%%%%
%%%%%%%%%%%%%%%%%%%%%%%%%%%%%%%%%%%%%%%%%%%%%%%%%%%%%%%%%%%%%%%%%%%

\chapter{Preliminary Numerical Simulations}\label{chap:numerical-simulations}

In this chapter, I discuss the details and results of Step 2 of developing the QLanczos algorithm as described in Figure \ref{fig:foursteps}. This preliminary step was done to evaluate if the nonstandard Krylov basis, described in Chapter \ref{chap:qlanczos-algorithm}, is efficient for computing extremal eigenvalues. I carried out numerical simulations of the QLanczos algorithm on a classical computer using Fortran and Python to compare using exact real-time evolution to exact imaginary-time evolution. Additionally, I ran simulations of the QLanczos algorithm with exact real-time evolution and multiple reference states. I also introduced noise to the simulations to simulate running the algorithm on real quantum hardware. I present and discuss the details of the numerical simulations and how the results informed the next step.

All of the numerical simulations in this chapter were carried out in the eigenbasis. Three different nuclei that could be fully diagonalized were chosen: $^{20}$Ne, $^{22}$Na, and $^{29}$Na. For each nucleus, an inert core of $^{16}$O was assumed, and the valence particles were restricted to the $sd$-shell ($0d_{5/2}$, $1s_{1/2}$, and $0d_{3/2}$) which can have up to 12 particles for each species. For $^{20}$Ne there are two valence protons and two valence neutrons. For $^{22}$Na there are three valence protons and three valence neutrons. For $^{29}$Na there are three valence protons and eight valence neutrons. Brown and Richter's Universal $sd$-shell interaction version B (USDB) \cite{Brown2006} was used to compute the two-body term in Equation (\ref{eq:Ham_secondquantization}). USDB is a set of single-particle energies and two-body matrix elements empirically fitted to reproduce experimental nuclei data between $^{18}$O and $^{38}$Ca. The nuclei were diagonalized using a shell model code \cite{Johnson2018} producing the full spectrum of energies, \{$E_j$\}. In this basis, the M-scheme dimensions of $^{20}$Ne, $^{22}$Na, and $^{29}$Na are 640, 6,116, and 1,935, respectively. This was Step 1 in developing the QLanczos algorithm as described in Figure \ref{fig:foursteps}.

For the computation of the ground state energy, the simulations ran until the computed energies converged to a specified percentage of the correlation energy,

\vspace{-3mm}
\begin{equation}
    E_c = E_{exact}-E_{HF},
\end{equation}

\noindent which is the difference between the exact ground state energy, $E_{exact}$, and the Hartree-Fock energy, $E_{HF}$, found using a random phase approximation method \cite{Stetcu2002}. I use the simulated Hartree-Fock state in these preliminary numerical simulations. In later chapters, I investigate if this choice of reference state yields an advantage.

\section{Real- vs Imaginary-Time Evolution}\label{sec:real-time-sims}
Synthetic data was generated using the entire spectrum of energies, \{$E_j$\}, from the full diagonalization of the Hamiltonians. For the QLanczos algorithm using exact time evolution, the reference state, $\ket{\Psi_0}$, was expanded in the eigenbasis,

\begin{equation}
    \ket{\Psi_0}= \sum_j\psi_j\ket{j},
\end{equation}

\noindent where $\hat{H}\ket{j}=E_j\ket{j}$. The reference state was generated by evolving a random vector using imaginary-time evolution until the expectation energy of the Hamiltonian was close to the Hartree-Fock energy. Using the Hartree-Fock state as a reference state allows us to start at a lower energy and therefore requires fewer iterations to converge.

Two variables,

\vspace{-3mm}
\begin{equation}\label{eq:nu-single}
    \nu_k = \bra{\Psi_0}\left(e^{-i\Delta t\hat{H}}\right)^k\ket{\Psi_0} = \sum_j\psi_j^*\psi_j\left(e^{-i\Delta tE_j}\right)^k
\end{equation}

\noindent and

\vspace{-3mm}
\begin{equation}\label{eq:epsilom-single}
    \epsilon_k = \bra{\Psi_0}\left(e^{-i\Delta t\hat{H}}\right)^k\hat{H}\ket{\Psi_0} = \sum_j\psi_j^*\psi_j\left(e^{-i\Delta tE_j}\right)^kE_j
\end{equation}

\noindent are defined to save the computational cost of repeatedly calculating the time evolved states. For the imaginary-time version, $\Delta \tau$ replaces $i\Delta t$ in Equations (\ref{eq:nu-single}) and (\ref{eq:epsilom-single}). Now, the matrix elements for the overlap matrix are

\vspace{-3mm}
\begin{equation}\label{eq:Norm_real}
    N_{k,l}=\nu_{l-k}=\nu^*_{k-l}=N^*_{l,k},
\end{equation}

\noindent and for the Hamiltonian, they are

\vspace{-3mm}
\begin{equation}\label{eq:Ham_real}
    H_{k,l}=\epsilon_{l-k}=\epsilon^*_{k-l}=H^*_{l,k}.
\end{equation}

\noindent For the imaginary-time version, the overlap matrix and Hamiltonian must be normalized since the evolution is non-unitary:

\vspace{-3mm}
\begin{equation}\label{eq:Norm_im}
    N_{k,l}=\frac{\nu_{k+l}}{\sqrt{\nu_{2k}\nu_{2l}}},
\end{equation}

\begin{equation}\label{eq:Ham_im}
    H_{k,l}=\frac{\epsilon_{k+l}}{\sqrt{\nu_{2k}\nu_{2l}}}.
\end{equation}

\noindent Solving the generalized eigenvalue problem takes some finesse due to singular eigenvalues in the overlap matrix. First, near-singular eigenvalues are thrown out.

\begin{equation}\label{eq:sqrt-overlap}
    \left(N^{-\frac{1}{2}}\right)_{ij}=\sum_{r,v(r)>\delta}u^{(r)}_j\frac{1}{\sqrt{v(r)}}u^{(r)}_i,
\end{equation}

\noindent where $N\vec{u}=v\vec{u}$, $r$ labels the eigenpairs, and $\delta$ is a cutoff for near singular eigenvalues. Then the effective Hamiltonian is calculated,

\vspace{-3mm}
\begin{equation}\label{eq:effective-Ham}
    H_{eff}=N^{-\frac{1}{2}}HN^{-\frac{1}{2}}.
\end{equation}

\begin{figure}
    \centering
    \includegraphics[width=0.6\textwidth]{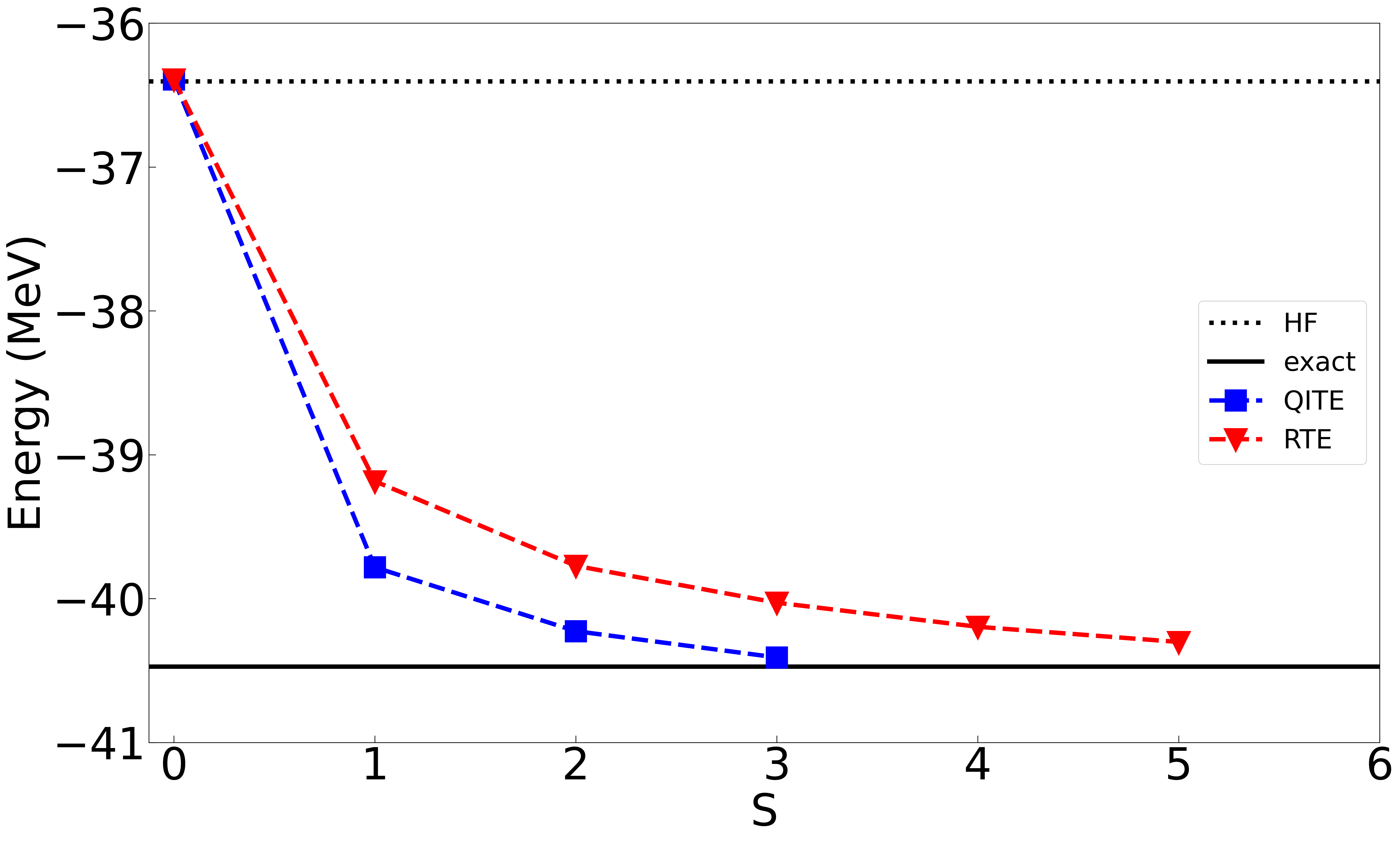}
    \caption{Numerical simulation of the QLanczos algorithm with imaginary-time evolution (QITE) and real-time evolution (RTE) in the eigenbasis to solve for the ground state energy of $^{20}$Ne. The dotted black line is the simulated Hartree-Fock energy, $E_{HF}$. The solid black line is the exact energy, $E_{exact}$. $S$ is the total number of time iterations with time step sizes of $\Delta \tau = \Delta t = 0.1 \textrm{MeV\textsuperscript{-1}}$. The simulations ran until the energies converged to within 5\% of the correlation energy, $E_c$.}
    \label{fig:cmpr-evo-ground-ne20}
\end{figure}

\begin{figure}[htbp!]
    \centering
    \includegraphics[width=0.6\textwidth]{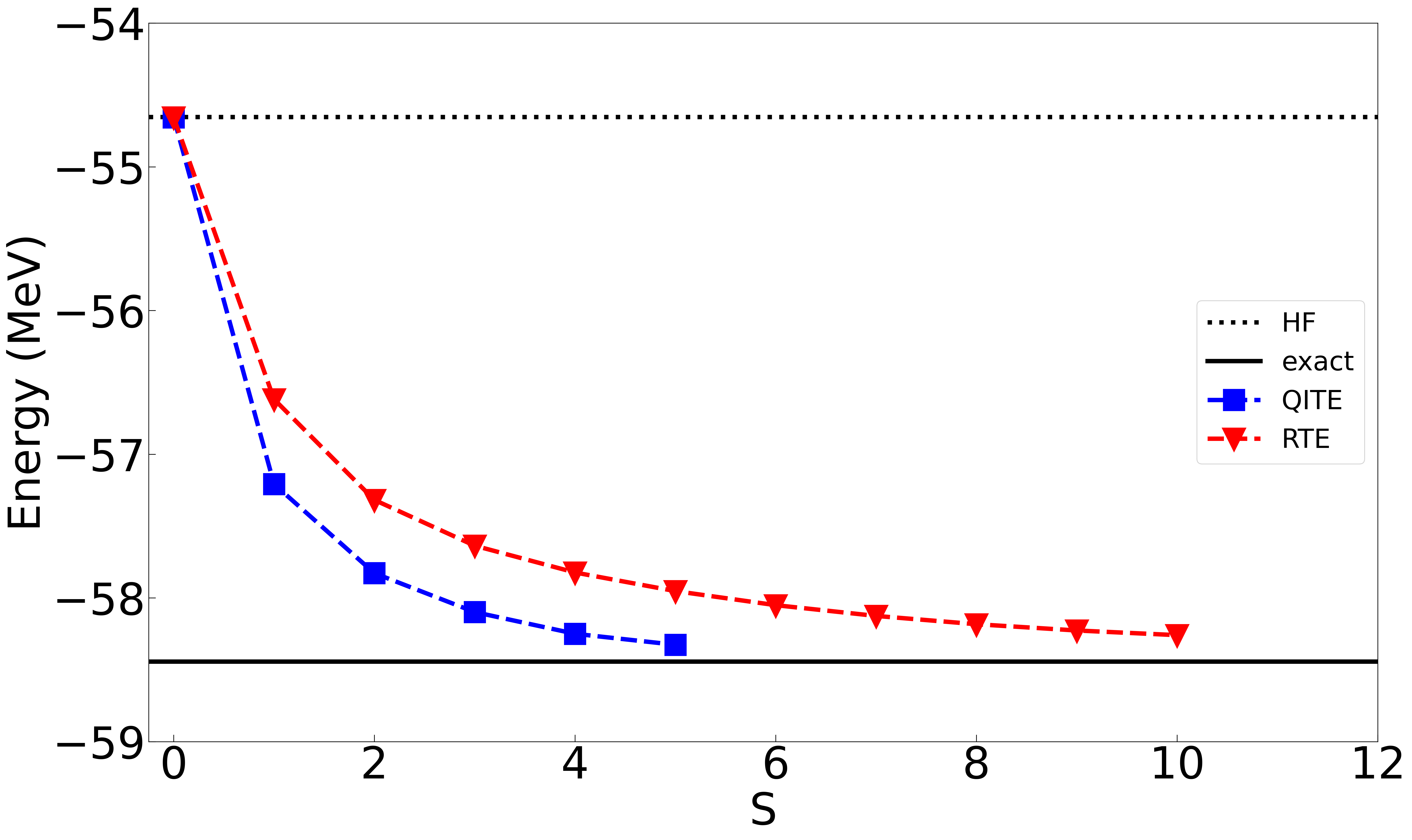}
    \caption{Numerical simulation of the QLanczos algorithm with imaginary-time evolution (QITE) and real-time evolution (RTE) in the eigenbasis to solve for the ground state energy of $^{22}$Na. The dotted black line is the simulated Hartree-Fock energy, $E_{HF}$. The solid black line is the exact energy, $E_{exact}$. $S$ is the total number of time iterations with time step sizes of $\Delta \tau = \Delta t = 0.1 \textrm{MeV\textsuperscript{-1}}$. The simulations ran until the energies converged to within 5\% of the correlation energy, $E_c$.}
    \label{fig:cmpr-evo-ground-na22}

    \bigskip
    
    \includegraphics[width=0.6\textwidth]{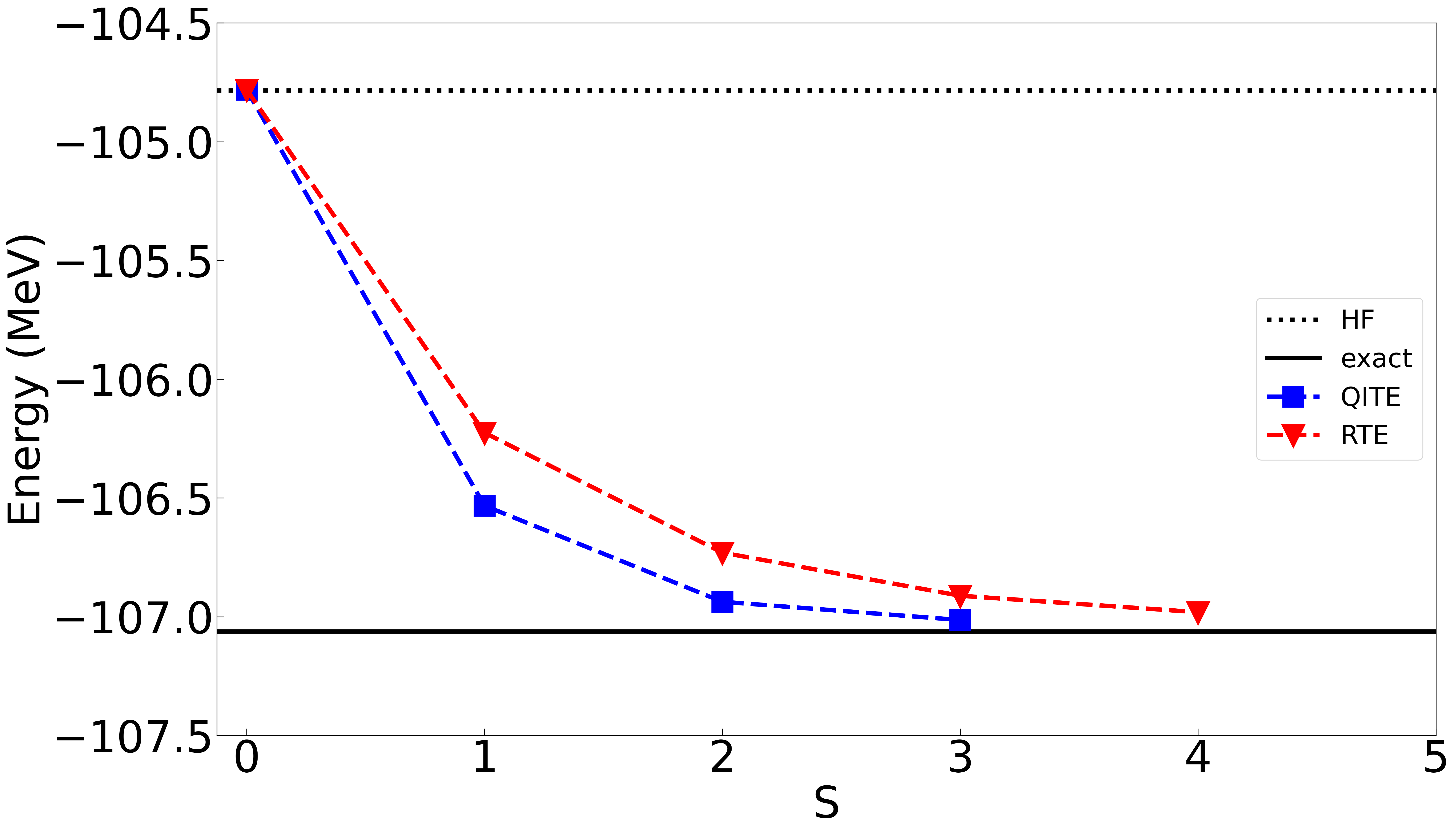}
    \caption{Numerical simulation of the QLanczos algorithm with imaginary-time evolution (QITE) and real-time evolution (RTE) in the eigenbasis to solve for the ground state energy of $^{29}$Na. The dotted black line is the simulated Hartree-Fock energy, $E_{HF}$. The solid black line is the exact energy, $E_{exact}$. $S$ is the total number of time iterations with time step sizes of $\Delta \tau = \Delta t = 0.1 \textrm{MeV\textsuperscript{-1}}$. The simulations ran until the energies converged to within 5\% of the correlation energy, $E_c$.}
    \label{fig:cmpr-evo-ground-na29}
\end{figure}

Figures \ref{fig:cmpr-evo-ground-ne20}-\ref{fig:cmpr-evo-ground-na29} shows the results of numerical simulations comparing QLanczos with real- and imaginary-time evolution in the eigenbasis. The simulations were carried out until the computed energies were within $5\%$ of the correlation energy, $E_c$. The time step size for imaginary-time evolution was $\Delta \tau = 0.1 \textrm{MeV\textsuperscript{-1}}$, and for real-time evolution was $\Delta t = 0.1 \textrm{MeV\textsuperscript{-1}}$. The QLanczos algorithm using imaginary-time evolution converges in fewer iterations than real-time evolution for all three cases. However, real-time evolution still converges to the ground state within ten iterations, a fraction of the full space. This is because the Lanczos algorithm alone works very well at finding eigenvalues. The advantage of real-time evolution is that it is unitary and, therefore, more straightforward to implement on a quantum computer.

\pagebreak
\section{Multiple Reference States}\label{sec:multiple-refs-sims}
The basis states for the multiple reference generalization with real-time evolution are

\vspace{-5mm}
\begin{equation}
    \ket{\Psi_{a,k}} = \left(e^{-i\Delta t\hat{H}}\right)^k \ket{\Psi_a},
\end{equation}

\noindent where $\ket{\Psi_a}$ are the reference states. The reference states were generated by evolving a linear combination of random vectors using imaginary-time evolution until the expectation energy of the Hamiltonian was close to the Hartree-Fock energy.

The overlap matrix and Hamiltonian are then

\vspace{-3mm}
\begin{equation}
    N_{a,k;b,l} = \nu_{l-k}(a,b)=\nu^*_{k-l}(b,a)=N^*_{b,l;a,k}
\end{equation}

\noindent and

\vspace{-3mm}
\begin{equation}
    H_{a,k;b,l} = \epsilon_{l-k}(a,b)=\epsilon^*_{k-l}(b,a)=H^*_{b,l;a,k},
\end{equation}

\noindent where

\vspace{-3mm}
\begin{equation}
    \nu_k(a,b)=\bra{\Psi_a}\left(e^{-i\Delta t\hat{H}}\right)^k\ket{\Psi_b}=\sum_j\psi_j^*(a)\psi_j(b)\left(e^{-i\Delta tE_j}\right)^k
\end{equation}

\noindent and

\vspace{-3mm}
\begin{equation}
    \epsilon_k(a,b)=\bra{\Psi_a}\left(e^{-i\Delta t\hat{H}}\right)^k\hat{H}\ket{\Psi_b}=\sum_j\psi_j^*(a)\psi_j(b)\left(e^{-i\Delta tE_j}\right)^kE_j.
\end{equation}

\noindent Again, for the imaginary-time version, $\Delta \tau$ replaces $i\Delta t$ and the overlap matrix and Hamiltonian are normalized:

\vspace{-3mm}
\begin{equation}
    N_{k,l}=\frac{\nu_{k+l}(a,b)}{\sqrt{\nu_{2k}(a,a)\nu_{2l}(b,b)}}
\end{equation}

\begin{equation}
    H_{k,l}=\frac{\epsilon_{k+l}(a,b)}{\sqrt{\nu_{2k}(a,a)\nu_{2l}(b,b)}}.
\end{equation}

\noindent Then the same process for throwing out singular eigenvalues (\ref{eq:sqrt-overlap}-\ref{eq:effective-Ham}) is followed, and the generalized eigenvalue problem (\ref{eq:gen-eig-multi}) is solved.

\begin{figure}[b!]
    \centering
    \includegraphics[width=0.6\textwidth]{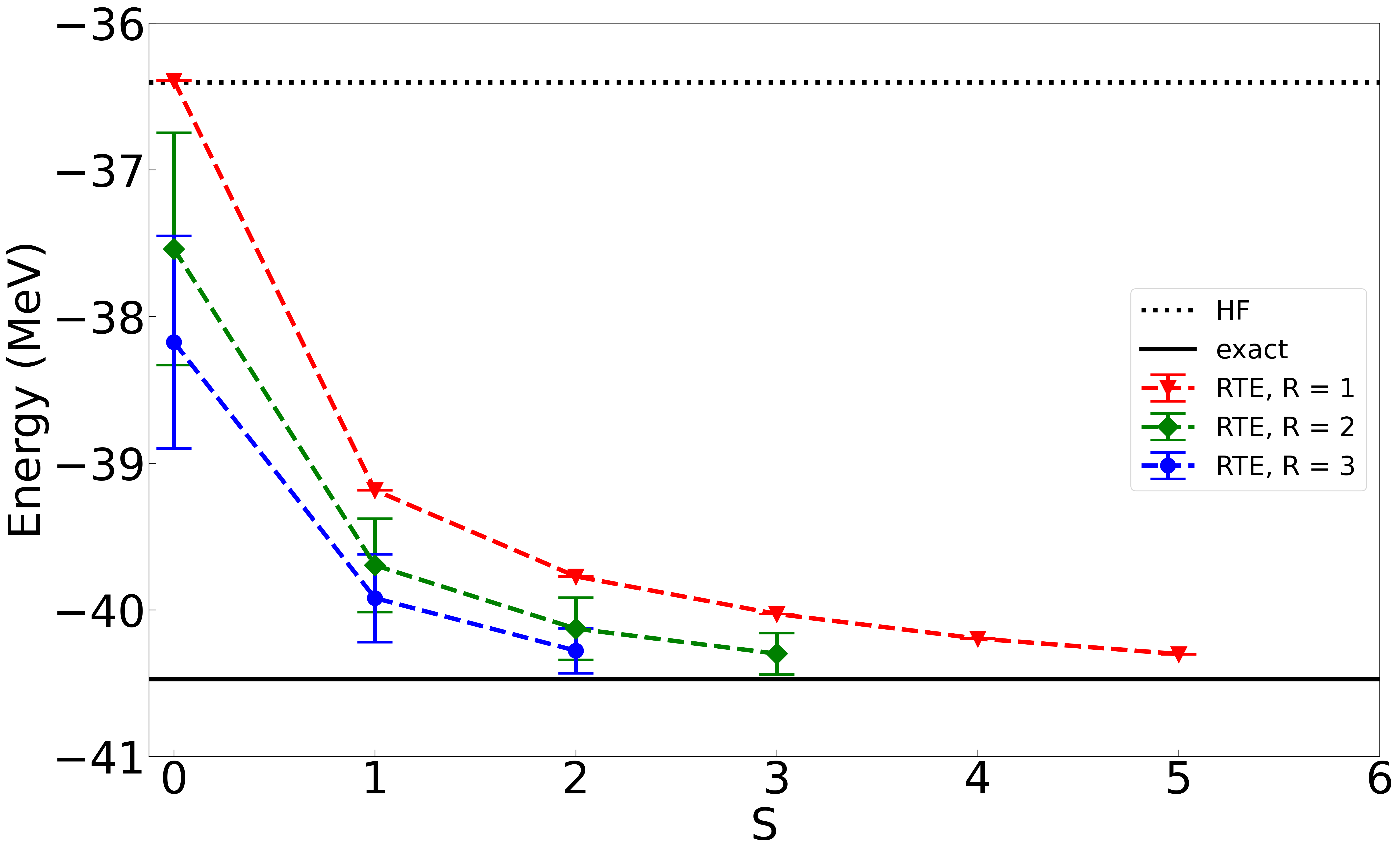}
    \caption{Numerical simulations of the QLanczos algorithm with real-time evolution (RTE) in the eigenbasis to solve for the ground state energy of $^{20}$Ne using different numbers of reference states, R. The dotted black line is the simulated Hartree-Fock energy, $E_{HF}$. The solid black line is the exact energy, $E_{exact}$. $S$ is the total number of real-time iterations with a time step size of $\Delta t = 0.1 \textrm{MeV\textsuperscript{-1}}$. The simulations ran until the energies converged within 5\% of the correlation energy, $E_c$. Error bars represent the standard deviation of 100 runs.}
    \label{fig:multi-refs-ground-ne20}

    \bigskip
    
    \includegraphics[width=0.6\textwidth]{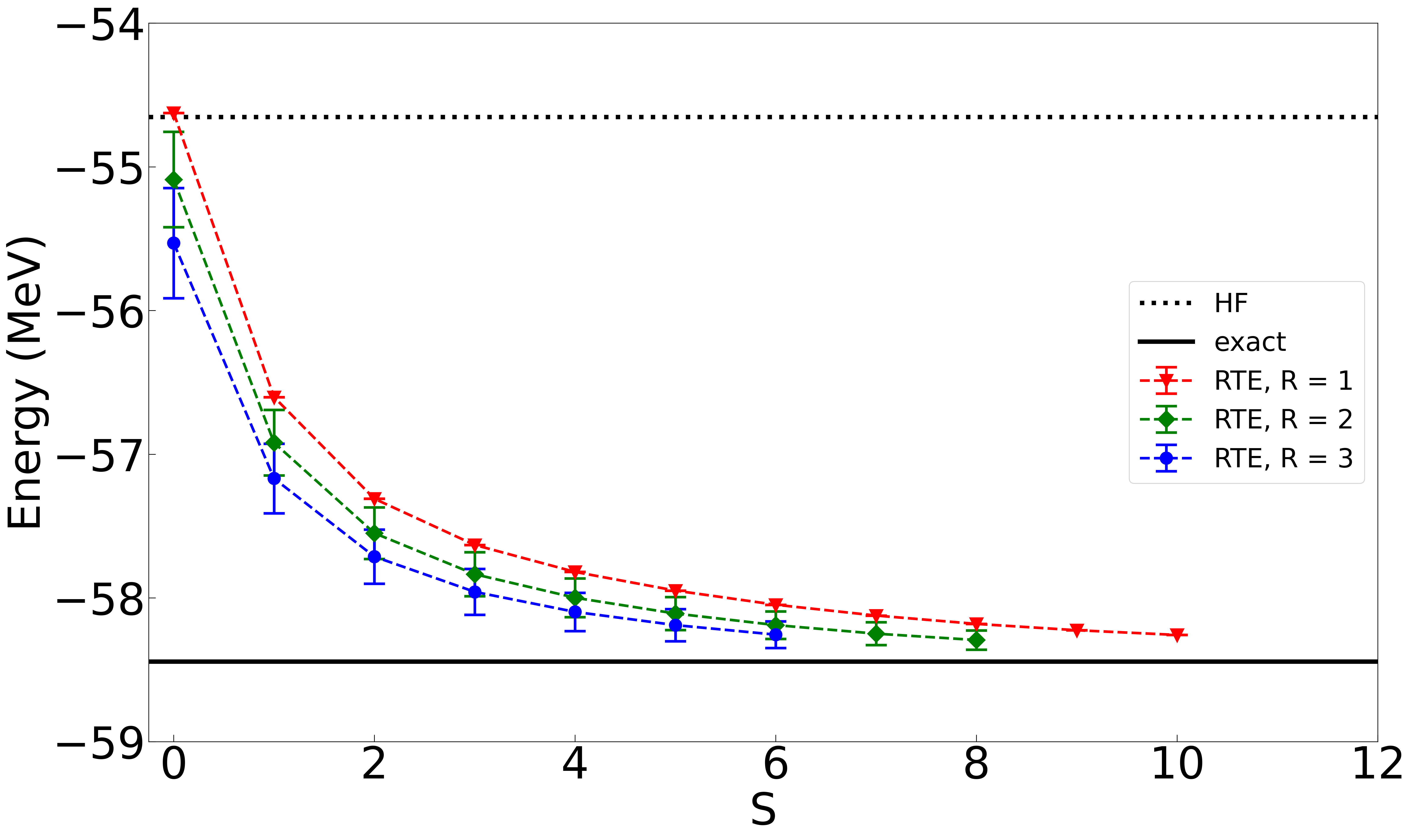}
    \caption{Numerical simulations of the QLanczos algorithm with real-time evolution (RTE) in the eigenbasis to solve for the ground state energy of $^{22}$Na using different numbers of reference states, R. The dotted black line is the simulated Hartree-Fock energy, $E_{HF}$. The solid black line is the exact energy, $E_{exact}$. $S$ is the total number of real-time iterations with a time step size of $\Delta t = 0.1 \textrm{MeV\textsuperscript{-1}}$. The simulations ran until the energies converged within 5\% of the correlation energy, $E_c$. Error bars represent the standard deviation of 100 runs.}
    \label{fig:multi-refs-ground-na22}
\end{figure}

\begin{figure}[ht!]
    \centering
    \includegraphics[width=0.6\textwidth]{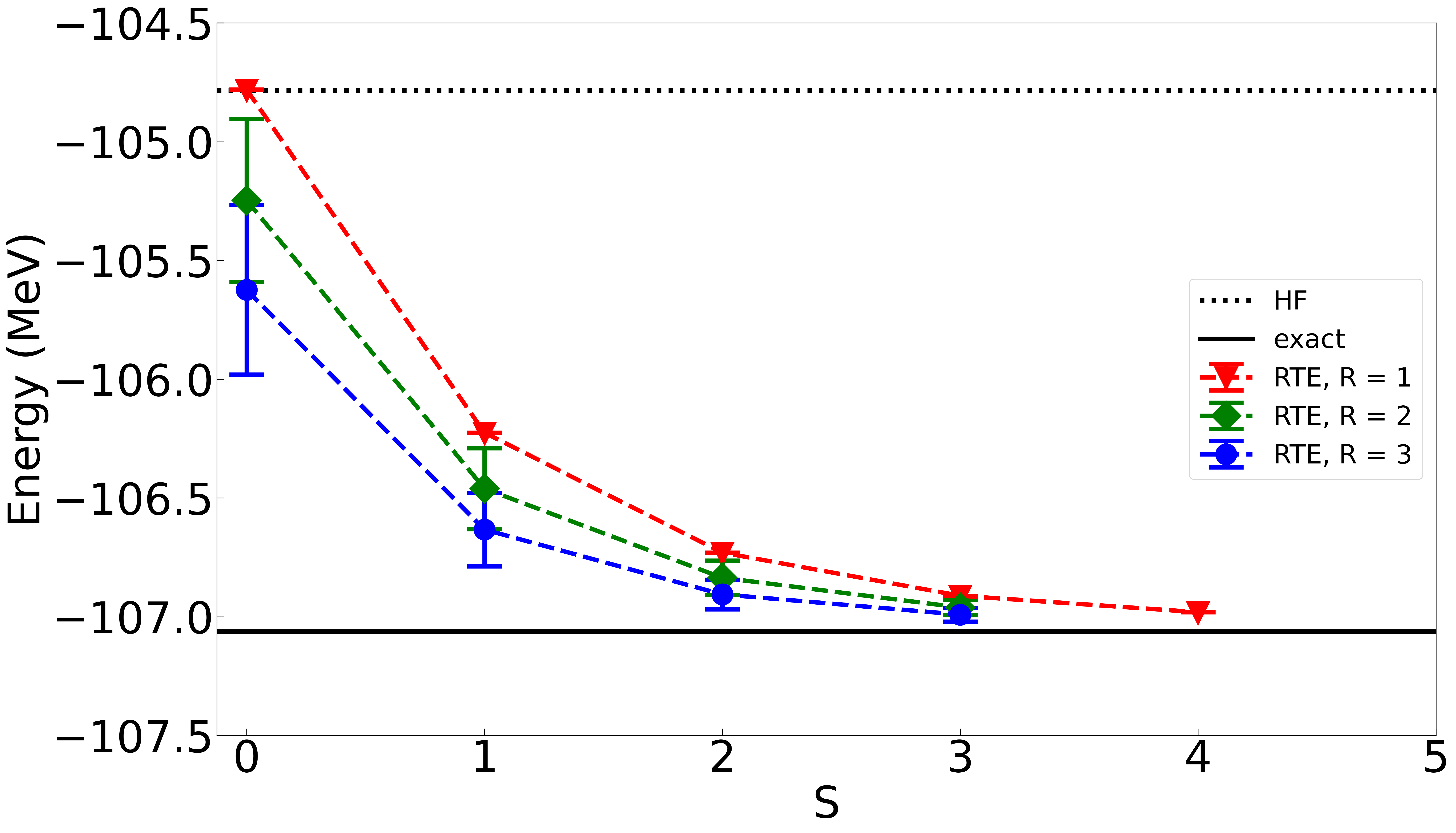}
    \caption{Numerical simulations of the QLanczos algorithm with real-time evolution (RTE) in the eigenbasis to solve for the ground state energy of $^{29}$Na using different numbers of reference states, R. The dotted black line is the simulated Hartree-Fock energy, $E_{HF}$. The solid black line is the exact energy, $E_{exact}$. $S$ is the total number of real-time iterations with a time step size of $\Delta t = 0.1 \textrm{MeV\textsuperscript{-1}}$. The simulations ran until the energies converged within 5\% of the correlation energy, $E_c$. Error bars represent the standard deviation of 100 runs.}
    \label{fig:multi-refs-ground-na29}
\end{figure}

I carried out numerical simulations of the QLanczos with exact real-time evolution with multiple references to compute the ground state energies of $^{20}$Ne, $^{22}$Na, and $^{29}$Na shown in Figures \ref{fig:multi-refs-ground-ne20}-\ref{fig:multi-refs-ground-na29}. The numerical simulations ran until the ground state energy converged to 5\% of the correlation energy with a time step size of $\Delta t = 0.1 \textrm{ MeV\textsuperscript{-1}}$. Using multiple reference states decreased the number of real-time iterations needed to converge to the ground state energy for all three cases. For $^{20}$Ne and $^{22}$Na, there is even more of an improvement when three reference states are used. However, for $^{29}$Na, adding a third reference state did not further reduce the number of real-time iterations needed for convergence.

\begin{figure}[htbp!]
    \centering
    \includegraphics[width=0.6\textwidth]{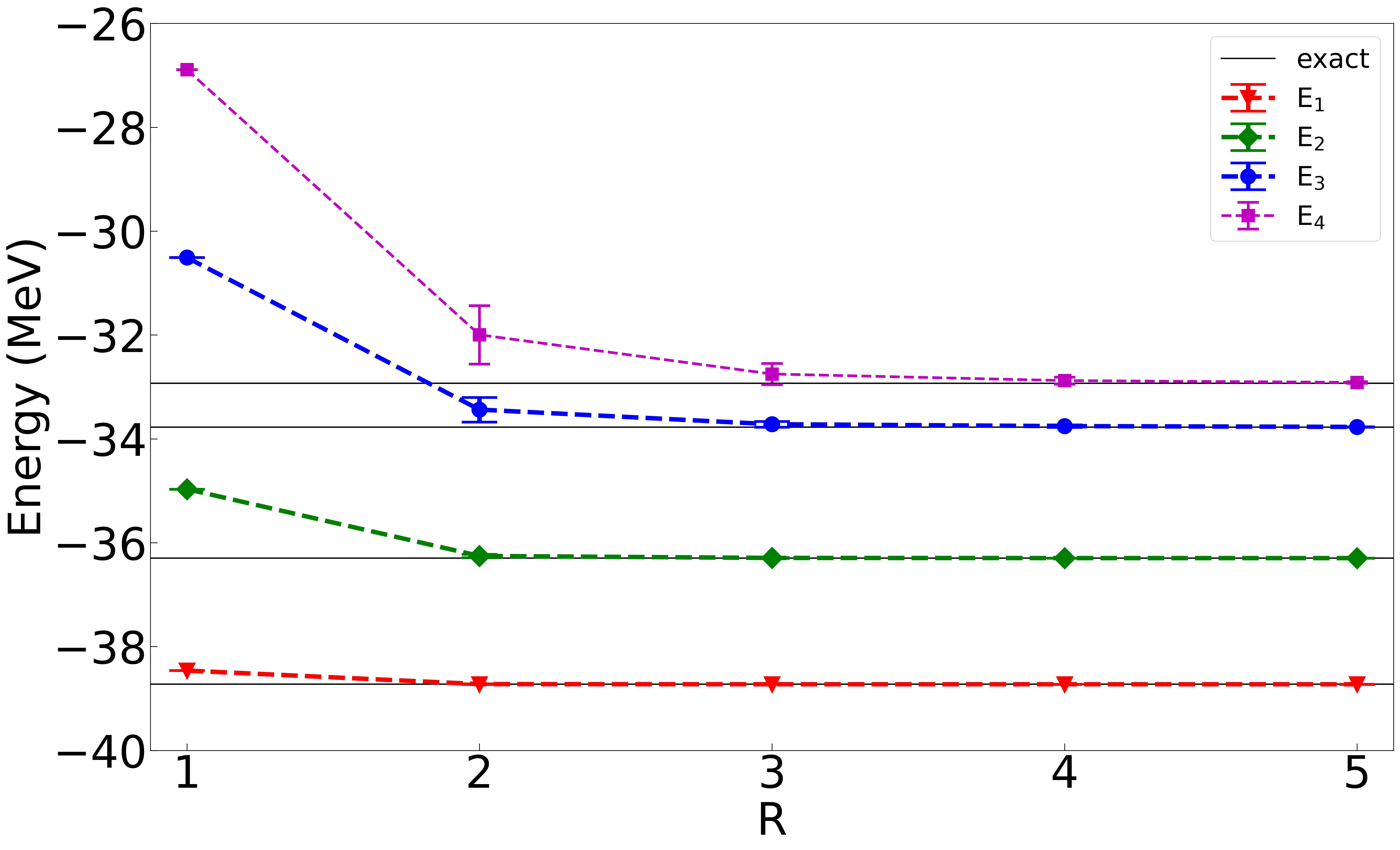}
    \caption{Numerical simulations of the QLanczos algorithm with real-time evolution in the eigenbasis to solve for the first four excited state energies of $^{20}$Ne using different numbers of reference states, R. The solid black lines are the exact energies. $E_n$ labels the $n$th excited state. The simulations ran for $S=9$ iterations with a time step size of $\Delta t = 0.1 \textrm{MeV\textsuperscript{-1}}$. Error bars represent the standard deviation of 100 runs.}
    \label{fig:multi-refs-excited-ne20}

    \bigskip

    \includegraphics[width=0.6\textwidth]{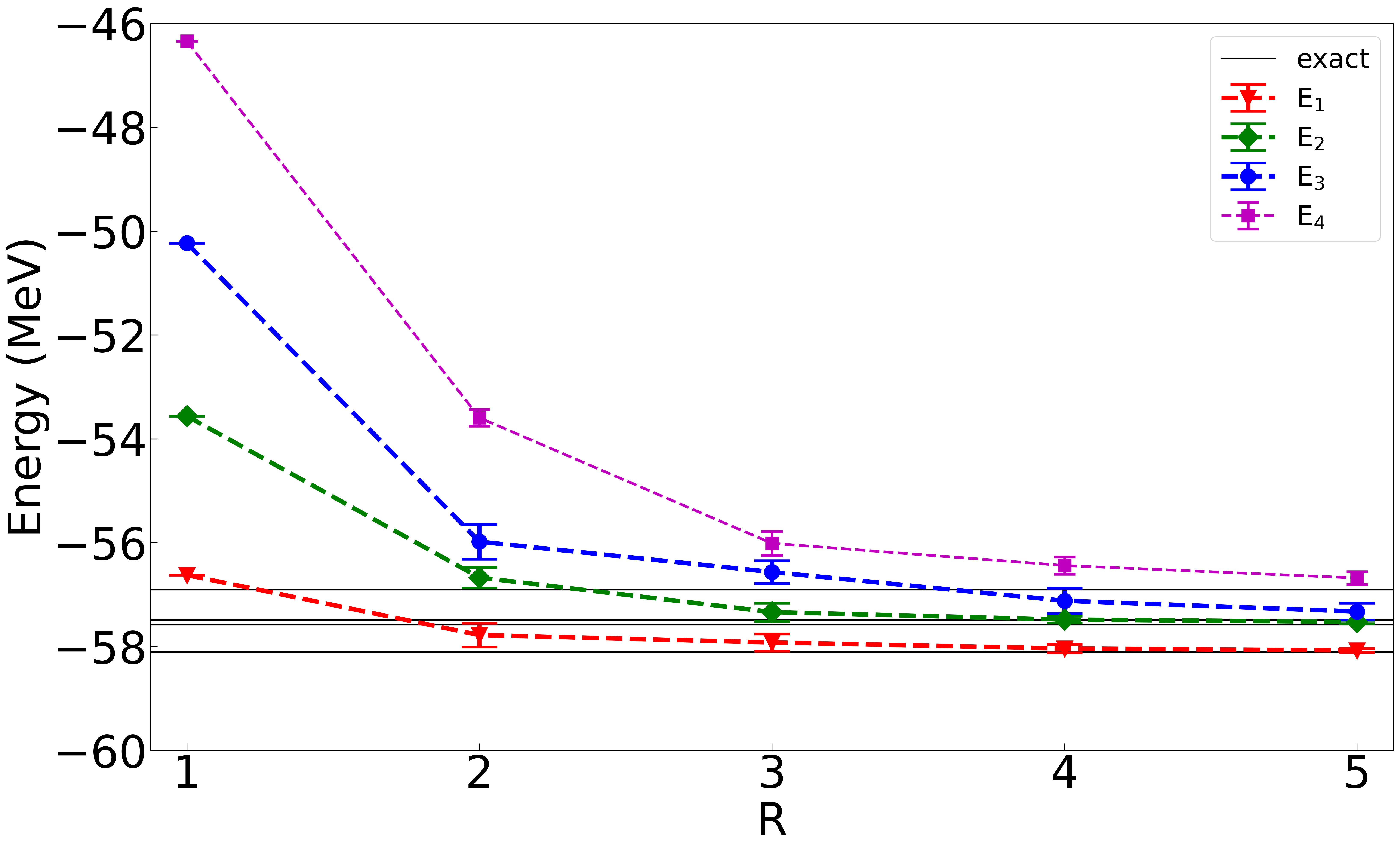}
    \caption{Numerical simulations of the QLanczos algorithm with real-time evolution in the eigenbasis to solve for the first four excited state energies of $^{22}$Na using different numbers of reference states, R. The solid black lines are the exact energies. $E_n$ labels the $n$th excited state. The simulations ran for $S=9$ iterations with a time step size of $\Delta t = 0.1 \textrm{MeV\textsuperscript{-1}}$. Error bars represent the standard deviation of 100 runs.}
    \label{fig:multi-refs-excited-na22}
\end{figure}

\begin{figure}[ht!]
    \centering
    \includegraphics[width=0.6\textwidth]{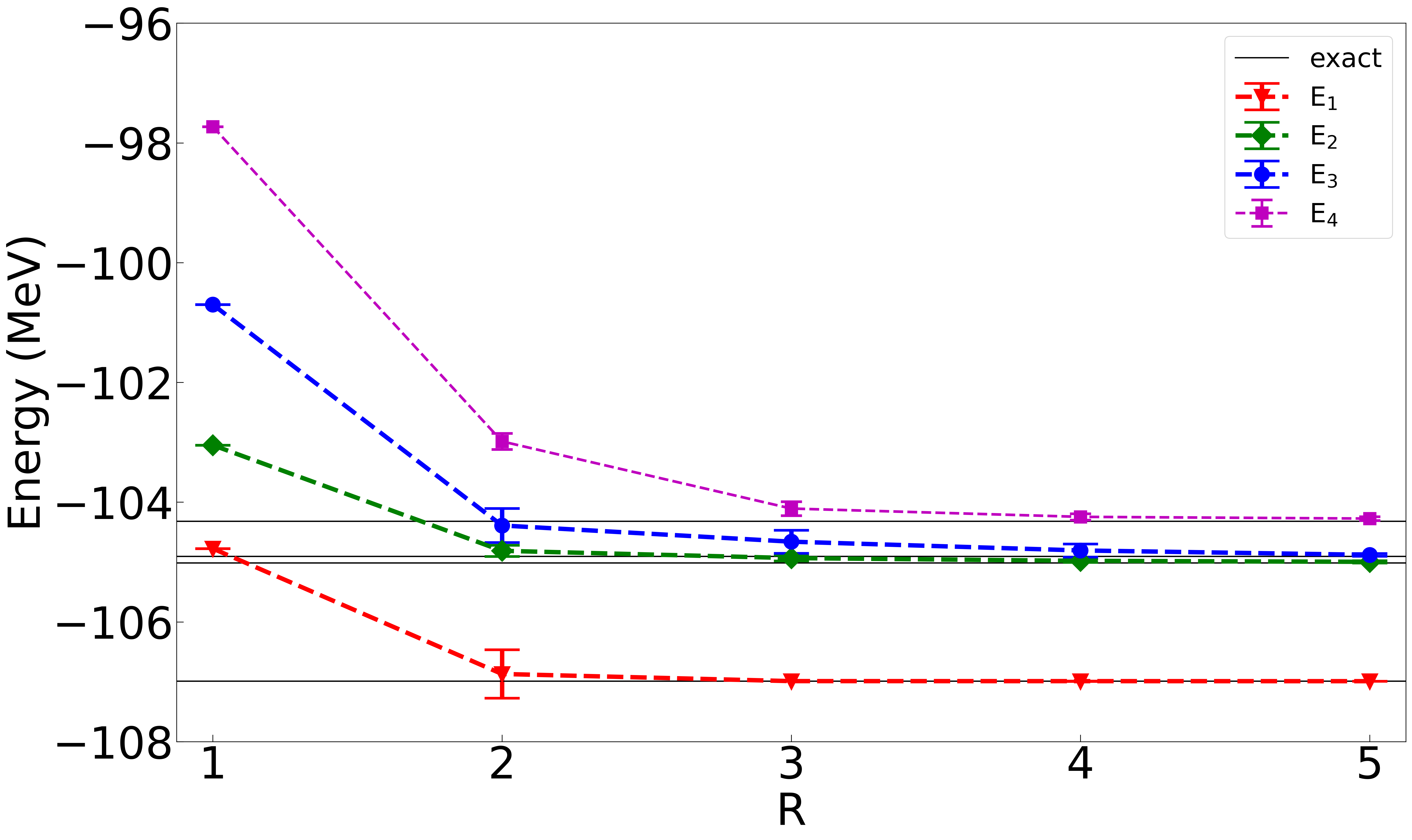}
    \caption{Numerical simulations of the QLanczos algorithm with real-time evolution in the eigenbasis to solve for the first four excited state energies of $^{29}$Na using different numbers of reference states, R. The solid black lines are the exact energies. $E_n$ labels the $n$th excited state. The simulations ran for $S=9$ iterations with a time step size of $\Delta t = 0.1 \textrm{MeV\textsuperscript{-1}}$. Error bars represent the standard deviation of 100 runs.}
    \label{fig:multi-refs-excited-na29}
\end{figure}

Multiple reference states also found excited state energies using fewer real-time iterations. I carried out numerical simulations of the QLanczos algorithm with exact real-time evolution and multiple references to solve for the first four excited state energies of $^{20}$Ne, $^{22}$Na, and $^{29}$Na shown in Figures \ref{fig:multi-refs-excited-ne20}-\ref{fig:multi-refs-excited-na29}. The number of real-time iterations is fixed (S=8), and the time step size used was $\Delta t= 0.1$. In these simulations, the computed excited energies converge to the exact energies as the number of reference states increases for all three cases.

\pagebreak
\section{Adding Noise}\label{sec:adding-noise}
Current quantum computers are in the Noisy-Intermediate Scale Quantum (NISQ) era, which means they are susceptible to noise from the environment and cannot implement continuous quantum error correction. I added noise to the numerical simulations to simulate how the QLanczos algorithm with real-time and multiple references would perform on a NISQ computer. Uniform noise was added or subtracted by a random percentage, $\eta$, of the $\nu_k$ and $\epsilon_k$ values,

\begin{align}
    \nu_k \rightarrow \nu_k \pm \eta\nu_k \label{eq:new-nu}\\ 
    \epsilon_k \rightarrow \epsilon_k \pm \eta\epsilon_k \label{eq:new-epsilon},
\end{align}

\noindent which are the values that would be computed on a quantum computer via a quantum circuit.

\begin{figure}[b!]
    \centering
    \includegraphics[width=0.6\textwidth]{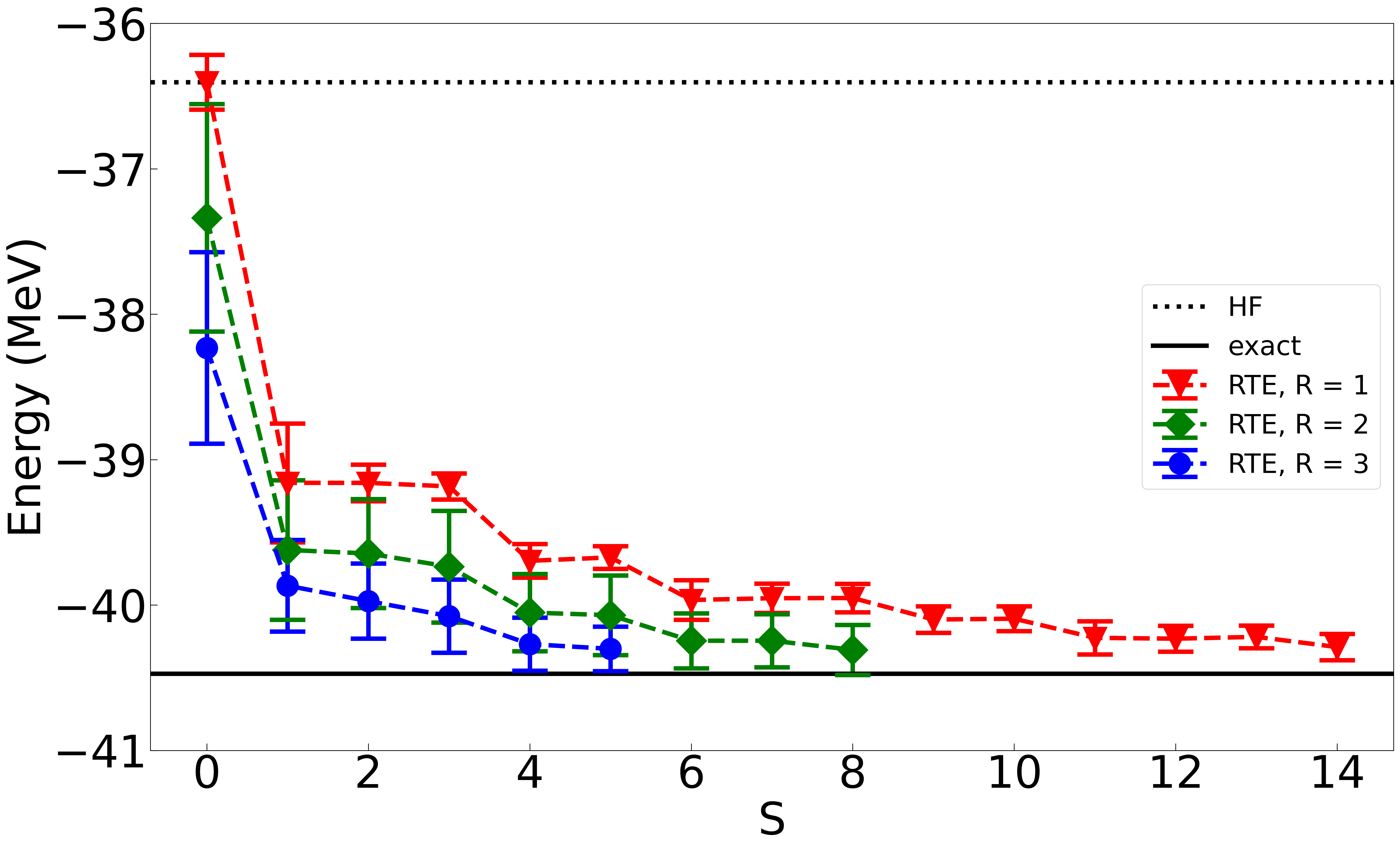}
    \caption{Numerical simulations of the QLanczos algorithm with real-time evolution (RTE) in the eigenbasis to solve for the ground state energy of $^{20}$Ne using different numbers of reference states, R, and less than 1\% of noise. The dotted black line is the simulated Hartree-Fock energy, $E_{HF}$. The solid black line is the exact energy, $E_{exact}$. $S$ is the total number of real-time iterations with a time step size of $\Delta t = 0.1 \textrm{MeV\textsuperscript{-1}}$. The simulations ran until the energies converged within 5\% of the correlation energy, $E_c$. Error bars represent the standard deviation of 100 runs.}
    \label{fig:multi-refs-ground-noise-ne20}

    \bigskip

    \includegraphics[width=0.6\textwidth]{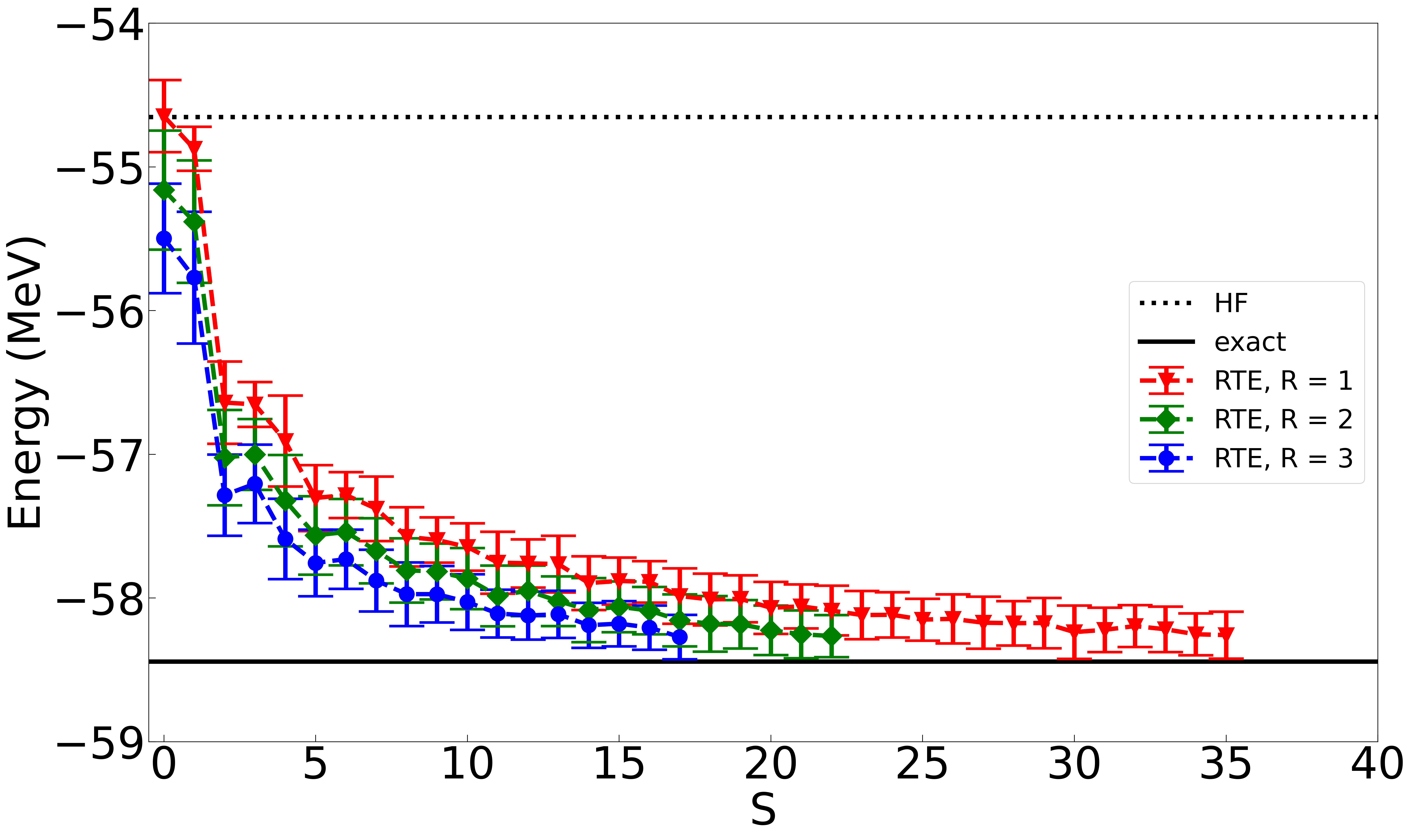}
    \caption{Numerical simulations of the QLanczos algorithm with real-time evolution (RTE) in the eigenbasis to solve for the ground state energy of $^{22}$Na using different numbers of reference states, R, and less than 1\% of noise. The dotted black line is the simulated Hartree-Fock energy, $E_{HF}$. The solid black line is the exact energy, $E_{exact}$. $S$ is the total number of real-time iterations with a time step size of $\Delta t = 0.1 \textrm{MeV\textsuperscript{-1}}$. The simulations ran until the energies converged within 5\% of the correlation energy, $E_c$. Error bars represent the standard deviation of 100 runs.}
    \label{fig:multi-refs-ground-noise-na22}
\end{figure}

\begin{figure}
    \centering
    \includegraphics[width=0.6\textwidth]{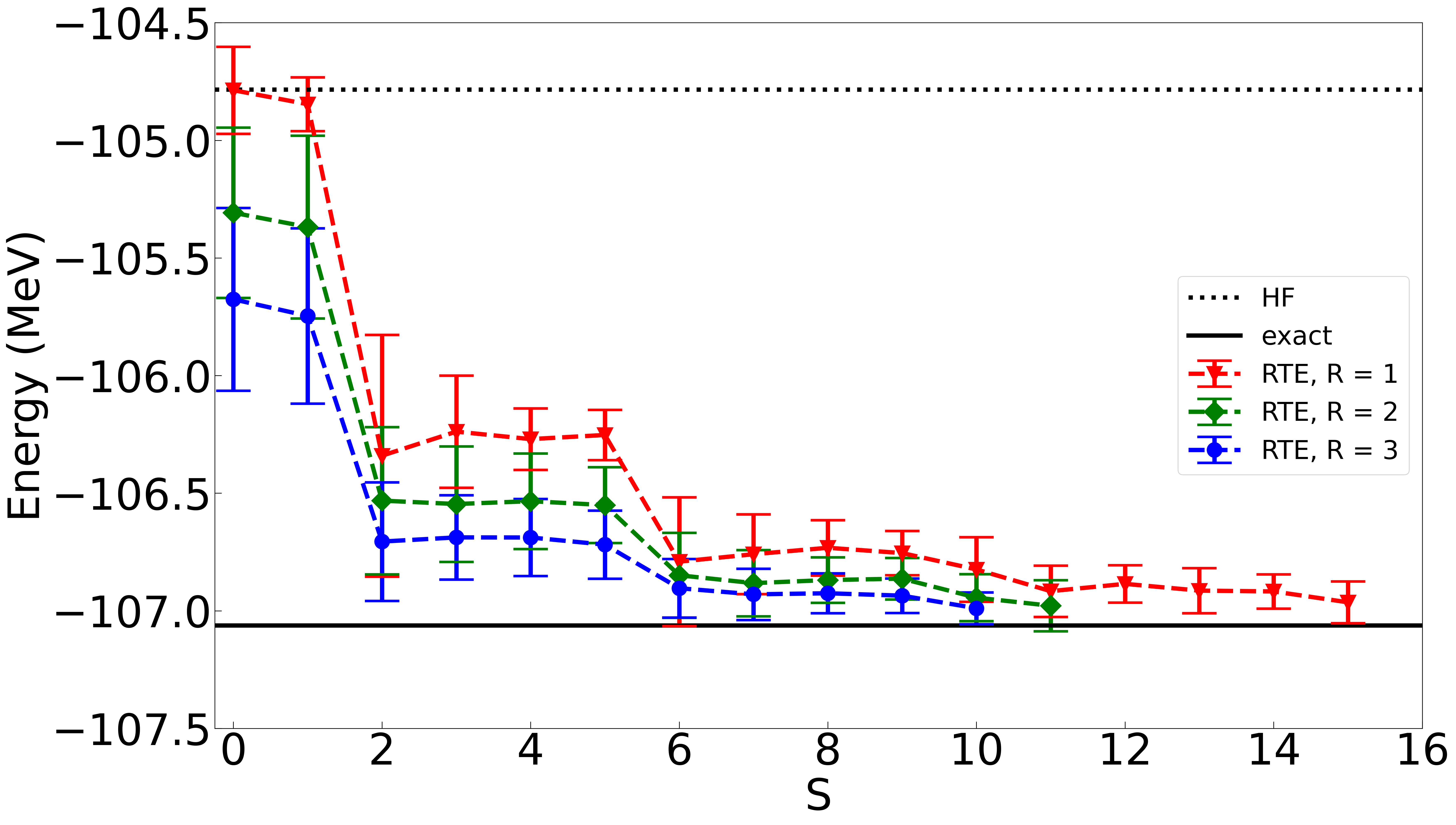}
    \caption{Numerical simulations of the QLanczos algorithm with real-time evolution (RTE) in the eigenbasis to solve for the ground state energy of $^{29}$Na using different numbers of reference states, R, and less than 1\% of noise. The dotted black line is the simulated Hartree-Fock energy, $E_{HF}$. The solid black line is the exact energy, $E_{exact}$. $S$ is the total number of real-time iterations with a time step size of $\Delta t = 0.1 \textrm{MeV\textsuperscript{-1}}$. The simulations ran until the energies converged within 5\% of the correlation energy, $E_c$. Error bars represent the standard deviation of 100 runs.}
    \label{fig:multi-refs-ground-noise-na29}
\end{figure}

I carried out numerical simulations of the QLanczos algorithm using the same method as in Section \ref{sec:multiple-refs-sims} but replacing $\nu_k$ and $\epsilon_k$ with Equations (\ref{eq:new-nu}) and (\ref{eq:new-epsilon}). Figures \ref{fig:multi-refs-ground-noise-ne20}-\ref{fig:multi-refs-ground-noise-na29} shows the ground state energies of $^{20}$Ne, $^{22}$Na, and $^{29}$Na with added noise of $\eta \le 1\%$ and a time step size of $\Delta t = 0.1 \textrm{ MeV\textsuperscript{-1}}$. The numerical simulations ran until they converged to 5\% of the correlation energy. The number of real-time iterations increased nearly threefold compared to Figures \ref{fig:multi-refs-ground-ne20}-\ref{fig:multi-refs-ground-na29} for all three cases when using a single reference state. Using three reference states required nearly the same number of iterations needed to converge when using a single reference state with no added noise, shown in Figures \ref{fig:multi-refs-ground-ne20}-\ref{fig:multi-refs-ground-na29}. This suggests that using multiple reference states can help mitigate errors due to noise.

The numerical simulations in this chapter demonstrate that the QLanczos algorithm with real-time evolution is a viable method for solving for the ground state and excited states energies of nuclei, demonstrating that this nonstandard Krylov basis leads to relatively quick convergences. Although it is not apparent why real-time evolution is so effective, a recent study provided a justification for real-time evolution \cite{Shen2023realtime}. Additionally, using multiple reference states reduced the number of real-time iterations needed to converge to the exact energies. Using multiple reference states also reduced the number of real-time iterations needed when noise was added to the system, representing a more realistic implementation of the QLanczos algorithm on a quantum computer. These preliminary results motivated the next step to map the system to qubits and develop the algorithm to be implemented on a quantum computer.

%%%%%%%%%%%%%%%%%%%%%%%%%%%%%%%%%%%%%%%%%%%%%%%%%%%%%%%%%%%%%%%%
%%%%%%%%%%%%%%%% Chapter:Quantum Computing %%%%%%%%%%%%%%%%%%%%
%%%%%%%%%%%%%%%%%%%%%%%%%%%%%%%%%%%%%%%%%%%%%%%%%%%%%%%%%%%%%%%%

\chapter{Gate-Based Quantum Computing}\label{chap:gate-based-quantum-computing}
The advantages of quantum computing lie in its ability to employ non-classical properties of quantum mechanics: superposition and entanglement. Quantum algorithms leverage these properties to represent and solve complex systems using less memory and time. The superposition of states on quantum computers allows us to store $2^n$ states on only $n$ qubits, exponentially reducing the memory required, which is advantageous for many-body systems whose Hilbert space grows exponentially. Entanglement, in which measuring one qubit can reveal information about one or more other qubits, has promising applications in quantum cryptography, quantum teleportation, and superdense coding. However, current quantum processors are not fault-tolerant and the scaling of these processors is a significant obstacle that must be overcome for quantum computers to claim quantum supremacy \cite{Preskill2018}. Quantum supremacy is when a quantum computer solves a problem that is intractable on a classical computer.

Currently, gate-based quantum computing is one of the leading methods, used by Google, IBM, and others. Gate-based quantum computations are represented as quantum circuits comprised of qubits, quantum gates, and measurements \cite{Nielsen2010}. Qubits describe the state of a wavefunction, quantum gates are unitary transformations that act on qubits, and measurements are Hermitian operators that measure the state of a qubit. Quantum circuits are run and measured many times resulting in the probabilities of each possible state. In this chapter, I will briefly cover each of these components and present a quantum circuit that demonstrates both superposition and entanglement.

\section{Qubits}
A quantum bit (qubit) is the smallest unit of information on a quantum computer, analogous to the bit in classical computing. On a quantum computer, we work in the so-called computational basis,

\begin{equation}
    \ket{0} =
    \begin{pmatrix}
        1 \\
        0
    \end{pmatrix}
\end{equation}

\noindent and

\begin{equation}
    \ket{1} =
    \begin{pmatrix}
        0 \\
        1
    \end{pmatrix},
\end{equation}
\vspace{1mm}

\noindent which are the eigenvectors of the Pauli matrix, $\sigma_z$. In quantum computing, $\sigma_z$ is written as $Z$, given in Table \ref{tab:gates}. A qubit is a linear combination of the basis vectors,

\vspace{-4mm}
\begin{equation}\label{eq:qubit}
    \ket{q}=a\ket{0}+b\ket{1},
\end{equation}

\noindent where $a$ is the probability of measuring $\ket{0}$ and $b$ is the probability of measuring $\ket{1}$. On a gate-based quantum computer, qubits are represented as lines:

\[
    \centering
    \begin{quantikz}
    \lstick{\ket{q}} && \\
    \end{quantikz}.
\]

\noindent Qubits are usually initialized in the $\ket{0}$ state. A multi-qubit state is the tensor production of multiple qubits,

\vspace{-4mm}
\begin{equation}
    \ket{\Psi}=\ket{q_1}\otimes\ket{q_2}\otimes...\otimes\ket{q_N}.
\end{equation}

\section{Quantum Gates}\label{sec:quant-gates}
Quantum gates evolve qubits. Quantum gates are restricted to unitary operators because quantum systems evolve by unitary transformation, preserving the norm so all probabilities sum to one. Some common gates are the identity gate, the Pauli gates, the Hadamard gate, and the $S$ gate. The matrix and circuit representations of these gates are listed in Table \ref{tab:gates}.

\savebox{\boxI}{\begin{quantikz} & \gate{I} & \end{quantikz}}
\savebox{\boxX}{\begin{quantikz} & \gate{X} & \end{quantikz}}
\savebox{\boxY}{\begin{quantikz} & \gate{Y} & \end{quantikz}}
\savebox{\boxZ}{\begin{quantikz} & \gate{Z} & \end{quantikz}}
\savebox{\boxH}{\begin{quantikz} & \gate{H} & \end{quantikz}}
\savebox{\boxS}{\begin{quantikz} & \gate{S} & \end{quantikz}}

\vspace{3mm}
\begin{table}
\centering
\begin{tabular}{ c | c c c c c c} 
    Gate & $I$ & $X$ & $Y$ & $Z$ & $H$ & $S$ \\[1ex]
    Matrix & $\begin{pmatrix} 1 & 0 \\ 0 & 1 \end{pmatrix}$ & $\begin{pmatrix} 0 & 1 \\ 1 & 0 \end{pmatrix}$ & $\begin{pmatrix} 0 & -i \\ +i & 0 \end{pmatrix}$ & $\begin{pmatrix} 1 & 0 \\ 0 & -1 \end{pmatrix}$ & $\frac{1}{\sqrt{2}}\begin{pmatrix} 1 & 1 \\ 1 & -1 \end{pmatrix}$ & $\begin{pmatrix} 1 & 0 \\ 0 & i \end{pmatrix}$ \\[2ex]
    Circuit & \usebox\boxI & \usebox\boxX & \usebox\boxY & \usebox\boxZ & \usebox\boxH & \usebox\boxS
\end{tabular}
\bigskip
\caption{\label{tab:gates} The matrix and circuit representation of some common quantum gates. The first row is the identity gate which returns the input unchanged. The second to fourth rows are the Pauli gates. The fifth row is the Hadamard ($H$) gate that puts a qubit in a superposition and transforms the basis from the $Z$ basis to the $X$ basis. The sixth row is the $S$ gate, when used in combination with the $H$ gate, transforms the basis from the $Z$ basis to the $Y$ basis.}
\vspace{-13mm}
\end{table}

\pagebreak
\noindent It is possible to take measurements in a different basis such as the $X$ basis or $Y$ basis. The $H$ and $S$ gates take a qubit from the computational ($Z$) basis to the $X$ basis or $Y$ basis,

\begin{equation}\label{eq:ZtoX_basis}
    \bra{\Psi}HZH\ket{\Psi}=\bra{\Psi}X\ket{\Psi}
\end{equation}

\noindent and

\begin{equation}\label{eq:ZtoY_basis}
    \bra{\Psi}HSZ S^\dagger H\ket{\Psi}=\bra{\Psi}Y\ket{\Psi}.
\end{equation}

\noindent Rotation gates are quantum gates that apply a parameterized rotation to a qubit. The $R_Z(\theta)$ gate rotates a qubit $\theta$ radians about the $z$-axis:

\[
    \begin{quantikz}
         & \gate{R_Z(\theta)} & \\
    \end{quantikz}
    \equiv
    \begin{pmatrix}
        \cos\frac{\theta}{2} & -\sin\frac{\theta}{2} \\
        \sin\frac{\theta}{2} & \cos\frac{\theta}{2}
    \end{pmatrix}.
\]

\noindent Similarly, the $R_X(\theta)$ and $R_Y(\theta)$ gates rotate about the $x$-axis and $y$-axis, respectively.

There are also multi-qubit gates, such as the controlled-not ($CNOT$) gate,

\[
    \begin{quantikz}
    \lstick{control $\rightarrow$ $\ket{q_1}$} & \ctrl{1} & \\
    \lstick{target $\rightarrow$ $\ket{q_2}$} & \targ{} &
    \end{quantikz},
\]

\noindent that applies a specified Pauli gate to the target qubit, $\ket{q_2}$, if the controlled qubit, $\ket{q_1}$, is $\ket{1}$. For example, the controlled-$X$ ($C_X$) flips the target qubit by applying the $X$ gate if the control qubit is $\ket{1}$. Similar to classical logic gates, truth tables can be constructed for quantum gates. Table \ref{tab:CNOT} is the truth table for the $C_X$ gate.

\begin{table}
\begin{center}
\begin{tabular}{ c | c } 
    input & output \\
    $\ket{q_1}$ $\ket{q_2}$ & $\ket{q_1}$ $\ket{q_2}$ \\ [0.5ex] 
    \hline
    $\ket{0}$ $\ket{0}$ & $\ket{0}$ $\ket{0}$ \\
    $\ket{0}$ $\ket{1}$ & $\ket{0}$ $\ket{1}$ \\
    $\ket{1}$ $\ket{0}$ & $\ket{1}$ $\ket{1}$ \\
    $\ket{1}$ $\ket{1}$ & $\ket{1}$ $\ket{0}$ \\
\end{tabular}
\vspace{2mm}
\caption{\label{tab:CNOT}The truth table of the controlled-$X$ ($C_X$) quantum gate. The target qubit, $\ket{q_2}$, flips if the control qubit, $\ket{q_1}$, is $\ket{1}$.}
\end{center}
\vspace{-13mm}
\end{table}

\section{Quantum Circuits}
The last component of quantum circuits is measurements. Measurements are taken using Hermitian operators. Pauli matrices are both unitary and Hermitian and can be used as quantum gates or measurements. Typically, measurements are taken using the $Z$ operator. The circuit below is the representation of a measurement in a quantum circuit:

\[
    \begin{quantikz}
     & \meter{} \\
    \end{quantikz}.
\]

\noindent Quantum circuits are run many times and measured. The number of runs resulting in a measurement of $\ket{0}$ or $\ket{1}$ is counted and divided by the number of runs resulting in the probability that the qubit is in state $\ket{0}$ or $\ket{1}$.

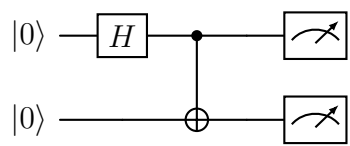
\begin{figure}
    \centering
    \begin{quantikz}
    \lstick{\ket{0}} & \gate{H} & \ctrl{1} && \meter{} \\
    \lstick{\ket{0}} && \targ{} && \meter{} \\
    \end{quantikz}
    \caption{The quantum circuit to produce a Bell state, the maximum entanglement of two qubits.}
    \label{cir:bell-state}
\end{figure}

A Bell state is a maximally entangled state between two qubits, created using the $H$ and $C_X$ gates as shown in Figure \ref{cir:bell-state}.

\noindent The resulting wavefunction from the circuit in Figure \ref{cir:bell-state} is $\ket{\Psi}=\frac{1}{\sqrt{2}}\ket{00}+\frac{1}{\sqrt{2}}\ket{11}$. If we were to only measure $\ket{q_1}$ and the results was $\ket{q_1}=\ket{0}$ then we would know $\ket{q_2}$ would also be $\ket{0}$ without having to measure it. Likewise if we measure $\ket{q_1}=\ket{1}$ then we would know $\ket{q_2}=\ket{1}$. There is a 50\% probability of measuring either state, $P_{00}=\left|\frac{1}{\sqrt{2}}\right|^2=0.5$ and $P_{11}=\left|\frac{1}{\sqrt{2}}\right|^2=0.5$.

This chapter covered the basic building blocks of gate-based quantum computing: qubits, quantum gates, and measurements. Quantum computers can compute the overlap matrix (\ref{eq:norm}) and Hamiltonian (\ref{eq:Ham}) in the Krylov subspace. Then we can use a classical computer to solve the generalized eigenvalue problem (\ref{eq:gen-eig}). In the next chapter, I will discuss how to prepare nuclear systems and real-time evolution to run the QLanczos algorithm on a quantum computer.

%%%%%%%%%%%%%%%%%%%%%%%%%%%%%%%%%%%%%%%%%%%%%%%%%%%%%%%%%%%%%%%%%%%%%%%%%%%%%%%%%%%%%%%
%%%%%%%%%%%%%%% Chapter:Preparing QLanczos for a Quantum Computer %%%%%%%%%%%%%%%%%%%%
%%%%%%%%%%%%%%%%%%%%%%%%%%%%%%%%%%%%%%%%%%%%%%%%%%%%%%%%%%%%%%%%%%%%%%%%%%%%%%%%%%%%%%%

\chapter{Preparing QLanczos for a Quantum Computer}\label{chap:preapering-qlanczos-quantum-computer}

There are three steps to prepare the QLanczos for a quantum computer. First, prepare the nuclear state. Second, map the system to qubits. Third, approximate the time evolution as a sequence of elementary operations. To demonstrate these methods, I continue with the example case, $^{14}$N, introduced in Section \ref{sec:shell-model}.

\section{State Preparation}\label{sec:state-prep}

The choice of the reference state in the QLanczos algorithm influences the success of finding the ground state energy. As discussed in Section \ref{sec:HF}, the Hartree-Fock state can be used as a reference state for the QLanczos algorithm. I used the simulated Hartree-Fock state as a reference state for the numerical simulations in Chapter \ref{chap:numerical-simulations}. In later chapters, I will also use the lowest energy configuration in the spherical basis as the reference state. This section will describe how these reference states are prepared on a quantum computer.

As shown in Chapter \ref{chap:nuclear-many-body-systems}, the many-body states in second quantization can be represented as products of $\ket{0}$'s and $\ket{1}$'s. On a quantum computer, these states are mapped to qubits. Recall the Slater determinants written in second quantization for $^{14}$N in Sections \ref{sec:interacting},

\vspace{-2mm}
\begin{equation}\label{eq:n14_slaters}
    \begin{aligned}
        \ket{0,2} &= \ket{1010} \\
        \ket{0,3} &= \ket{1001} \\
        \ket{1,2} &= \ket{0110} \\
        \ket{1,3} &= \ket{0101} .
    \end{aligned}
\end{equation}
\vspace{1mm}

\noindent Qubits are normally initialized in the $\ket{0}$ state, while an application of the $X$ gate puts a qubit in the $\ket{1}$ state. For example, the following quantum circuit creates the Slater determinant $\ket{0,3}$.

\[
    \centering
    \begin{quantikz}
    \lstick{\ket{0}} & \gate{X} && \rstick{$\rightarrow$ $\ket{1}$} \\
    \lstick{\ket{0}} &&& \rstick{$\rightarrow$ $\ket{0}$} \\
    \lstick{\ket{0}} &&& \rstick{$\rightarrow$ $\ket{0}$} \\
    \lstick{\ket{0}} & \gate{X} && \rstick{$\rightarrow$ $\ket{1}$} \\
    \end{quantikz}
\]
\vspace{-2mm}

\noindent The circuits above can also be written as $X_0X_3\ket{0000}$, where the subscript labels the qubit the quantum gate applies to.

\section{Mapping Hamiltonians to Qubits}\label{sec:JW}
There are two common methods for mapping the creation, $\hat{a}^\dagger$, and annihilation, $\hat{a}$, operators to qubits, the Jordan-Wigner transformation \cite{Jordan1928} and the Bravyi-Kitaev transformation \cite{Bravyi2002}. The Hamiltonian in second quantization (\ref{eq:Ham_secondquantization}) can be mapped to qubits using these methods. The Bravyi-Kitaev transformation is more efficient but also more complex to implement. However, a study comparing the two methods \cite{Tranter2018} showed that using an optimized Trotter number diminishes the advantage of the Bravyi-Kitaev method over the Jordan-Wigner method. I use the Jordan-Wigner transformation because it is straightforward and more commonly used.

The Jordan-Wigner transformation works by transforming Hamiltonians in second quantization to a tensor product of Pauli operators, referred to as Pauli strings. The Jordan-Wigner transformation defines the creation operator as

\begin{equation}
    \hat{a}^\dagger_n \xrightarrow[]{} \frac{1}{2}\left(\prod^{n-1}_{j=0}-Z_j\right)\left(X_n-i Y_n\right),
\end{equation}

\noindent and the annihilation operator as

\vspace{-3mm}
\begin{equation}
    \hat{a}_n \xrightarrow[]{} \frac{1}{2}\left(\prod^{n-1}_{j=0}-Z_j\right)\left(X_n+i Y_n\right),
\end{equation}

\noindent where $n$ labels the qubit the operators apply to \cite{Dumitrescu2018}. The string of $Z$ operators is necessary to preserve the anticommuting of fermions.

Using the $^{14}$N example, I will demonstrate the mapping. Recall, the Slater determinants are $\ket{0,2}$, $\ket{0,3}$, $\ket{1,2}$, and $\ket{1,3}$ and the bit strings are written out in Equation (\ref{eq:n14_slaters}). The single-particle energies, which are all the same, are omitted, leaving us with only the two-body part of the Hamiltonian in Equation (\ref{eq:Ham_secondquantization}). Using an M-scheme of $M=-1,0,1$, there are only six terms to be mapped to qubits using the Jordan-Wigner transformation,

\vspace{-5mm}
\begin{equation}\label{eq:N14-ham}
\begin{aligned}
    \hat{H} = 
    &\frac{1}{4} ( \bra{0,2} \hat{V} \ket{0,2} \hat{a}^\dagger_0 \hat{a}^\dagger_2 \hat{a}_2 \hat{a}_0 + \bra{0,3} \hat{V} \ket{0,3} \hat{a}^\dagger_0 \hat{a}^\dagger_3 \hat{a}_3 \hat{a}_0 + \bra{1,2} \hat{V} \ket{1,2} \hat{a}^\dagger_1 \hat{a}^\dagger_2 \hat{a}_2 \hat{a}_1 \\
    &+ \bra{0,3} \hat{V} \ket{1,2} \hat{a}^\dagger_0 \hat{a}^\dagger_3 \hat{a}_2 \hat{a}_1 + \bra{1,2} \hat{V} \ket{0,3} \hat{a}^\dagger_1 \hat{a}^\dagger_2 \hat{a}_3 \hat{a}_0 + \bra{1,3}\hat{V}\ket{1,3}\hat{a}^\dagger_1 \hat{a}^\dagger_3 \hat{a}_3 \hat{a}_1 ).
\end{aligned}
\end{equation}

\noindent From the Jordan-Wigner transformation the creation operators are

\vspace{-3mm}
\begin{equation}\label{eq:JW-creation}
\begin{aligned}
    \hat{a}^\dagger_0 &= \frac{1}{2}(X_0-iY_0) \\
    \hat{a}^\dagger_1 &= \frac{1}{2}(Z_0X_1-iZ_0Y_1) \\
    \hat{a}^\dagger_2 &= \frac{1}{2}(Z_0Z_1X_2-iZ_0Z_1Y_2) \\
    \hat{a}^\dagger_3 &= \frac{1}{2}(Z_0Z_1Z_2X_3-iZ_0Z_1Z_2Y_3),
\end{aligned}
\end{equation}

\noindent and the annihilation operators are

\vspace{-3mm}
\begin{equation} \label{eq:JW-annihilation}
\begin{aligned}
    \hat{a}_0 &= \frac{1}{2}(X_0+iY_0) \\
    \hat{a}_1 &= \frac{1}{2}(Z_0X_1+iZ_0Y_1) \\
    \hat{a}_2 &= \frac{1}{2}(Z_0Z_1X_2+iZ_0Z_1Y_2) \\
    \hat{a}_3 &= \frac{1}{2}(Z_0Z_1Z_2X_3+iZ_0Z_1Z_2Y_3).
\end{aligned}   
\end{equation}

\noindent When the creation (\ref{eq:JW-creation}) and annihilation (\ref{eq:JW-annihilation}) operators are substituted into Equation (\ref{eq:N14-ham}), the resulting Hamiltonian is a summation of Pauli strings,

\begin{equation}\label{eq:N14-ham-jw}
\begin{aligned}
    \hat{H} &= c_0I+c_1Z_0+c_2Z_1+c_3Z_2+c_4Z_3+c_5Z_0Z_3+c_6Z_1Z_2 \\
              &+c_7Z_0Z_2+c_8Z_1Z_3+c_9Y_0Y_1Y_2Y_3+c_{10}X_0X_1X_2X_3 \\ 
              &+c_{11}Y_0Y_1X_2X_3+c_{1,2}X_0X_1Y_2Y_3+c_{13}Y_0X_1Y_2X_3 \\ 
              &+c_{14}X_0Y_1X_2Y_3+c_{15}X_0Y_1Y_2X_3+c_{16}Y_0X_1X _2Y_3,
\end{aligned}    
\end{equation}
\vspace{1mm}

\noindent where the subscript of the Pauli matrices labels the qubit that the Pauli or identity operator applies to and $c_m$ are the coefficients.

I chose this problem to work through because it is small enough to write by hand. For a slightly larger case, such as two protons and two neutrons in the full $0p$-shell, which is $^8$Be, the Hamiltonian contains 975 Pauli string terms in the spherical basis. This process is tedious when done by hand, but software development kits for quantum computing, such as Qiskit \cite{Qiskit} by IBM, have packages that can compute these automatically.

\section{Simulating Hamiltonian Dynamics}\label{sec:trotterization}
Trotterization (also called Suzuki-Trotter splitting) \cite{Trotter1959} is used in simulations of time evolution for non-commuting operators \cite{whitfield2011}. When the exponential Hamiltonian is decomposed, each term may not commute. The Hamiltonian is defined as a sum of local terms, $\hat{H}=\sum_mc_mP_m$, where $P_m$ are the Pauli strings and $c_m$ are the coefficients. The first-order Trotter decomposition of real-time evolution is,

\vspace{-3mm}
\begin{equation}
    e^{-i\hat{H}\Delta tk} \approx \left(\prod_m e^{-ic_mP_m\Delta tk/N}\right)^N = U_k,
\end{equation}
\vspace{1mm}

\noindent where $k = 0, 1, ...S$ and $N$ is the Trotter number \cite{whitfield2011}. The Trotter decomposition of $^{14}$N with $N = 1$ is

\vspace{-2mm}
\begin{equation}\label{eq:trotter-decomp}
\begin{aligned}
    U_k
    =&\exp(-ic_0I\Delta tk)\exp(-ic_1Z_0\Delta tk)\exp(-ic_2Z_1\Delta tk) \\ 
    &\exp(-ic_3Z_2\Delta tk)\exp(-ic_4Z_3\Delta tk)\exp(-ic_5Z_0Z_3\Delta tk) \\
    &\exp(-ic_6Z_1Z_2\Delta tk)\exp(-ic_7Z_0Z_2\Delta tk)\exp(-ic_8Z_1Z_3\Delta tk) \\ 
    &\exp(-ic_9Y_0Y_1Y_2Y_3\Delta tk)\exp(-ic_{10}X_0X_1X_2X_3\Delta tk) \\ 
    &\exp(-ic_{11}Y_0Y_1X_2X_3\Delta tk)\exp(-ic_{12}X_0X_1Y_2Y_3\Delta tk) \\ 
    &\exp(-ic_{13}Y_0X_1Y_2X_3\Delta tk)\exp(-ic_{14}X_0Y_1X_2Y_3\Delta tk) \\ 
    &\exp(-ic_{15}X_0Y_1Y_2X_3\Delta tk)\exp(-ic_{16}Y_0X_1X _2Y_3\Delta tk) 
\end{aligned}
\end{equation}
\vspace{1mm}

\noindent The next step is to design the quantum circuits to compute these equations on a quantum computer.

%%%%%%%%%%%%%%%%%%%%%%%%%%%%%%%%%%%%%%%%%%%%%%%%%%%%%%%%%%%%%%%%%%%%%
%%%%%%%%%%%%%%% Chapter:QLanczos Quantum Circuits %%%%%%%%%%%%%%%%%%
%%%%%%%%%%%%%%%%%%%%%%%%%%%%%%%%%%%%%%%%%%%%%%%%%%%%%%%%%%%%%%%%%%%%%

\chapter{Quantum Circuits for QLanczos}\label{chap:quantum-circuits-qlanczos}

In this chapter, I will introduce the quantum circuits needed to perform the QLanczos algorithm on a gate-based quantum computer which prepares the algorithm for Step 4 in Figure \ref{fig:foursteps}. I will discuss how to find the inner product of two vectors on a quantum computer using an ancillary (ancilla) qubit. Using this method, I will present the quantum circuits to compute the overlap matrix (\ref{eq:norm}) and Hamiltonian (\ref{eq:Ham}). I will sketch the real-time unitary gates necessary for computing the overlap matrix and Hamiltonian using the $^{14}$N example discussed in Chapter \ref{chap:preapering-qlanczos-quantum-computer}.

\section{Computing the Overlap Matrix}
The diagonal matrix elements of the overlap matrix, $\braket{\Psi_{a,k}}{\Psi_{a,k}}$, are simply one since the quantum gates, which are unitary, preserve the norm. To compute the off-diagonal matrix elements of the overlap matrix, $\braket{\Psi_{a,k}}{\Psi_{b,l}}$, a quantum circuit that can compute the inner product between two vectors is needed. This can be done using an ancilla qubit which is simply an additional qubit that is not part of the qubits representing the nuclear system. The quantum circuit in Figure \ref{cir:dot-prod} computes the inner product of two vectors \cite{Zhao2021}.

\vspace{-3mm}
\begin{figure}
    \centering
    \begin{quantikz}[wire types={q,b}]
    \lstick{\ket{q_\alpha}} & \slice{1} &&& \slice{2} && \ctrl{1} & \slice{3} && \gate{H} & \slice{4} && \meter{} \\
    \lstick{$\ket{0}^{\otimes n}$} &&& \gate{\hat{U}_i} &&& \gate{\hat{U}_j} &&&&&& 
    \end{quantikz}
    \caption{The quantum circuit to compute the overlap between two vectors, $\braket{\Psi_{a,k}}{\Psi_{b,l}}$, using an ancilla qubit, $\ket{q_\alpha}$. The first unitary operator, $\hat{U}_i$, creates the first vector, $\hat{U}_i\ket{0}^{\otimes n}=\ket{\Psi_{a,k}}$, from the multi-qubit zeroth state. The second unitary operator, $\hat{U}_j$, creates the second vector, $\hat{U}_j\ket{\Psi_{a,k}}=\ket{\Psi_{b,l}}$ and entangles the system qubits with the ancilla qubit. To compute the real part of the inner product, the ancilla qubit is $\ket{q_\alpha}=H\ket{0}=\frac{1}{\sqrt{2}}\ket{0}+\frac{1}{\sqrt{2}}\ket{1}$ and to compute the imaginary part of the inner product, the ancilla qubit is $\ket{q_\alpha}=S^\dagger H\ket{0}=\frac{1}{\sqrt{2}}\ket{0}-\frac{i}{\sqrt{2}}\ket{1}$. The quantum circuit is run many times, and the ancilla qubit is measured. The red dashed lines mark each step in the quantum circuit.}
    \label{cir:dot-prod}
\end{figure}

\noindent The multi-qubit zeroth state, $\ket{0}^{\otimes n}$, represents $n$ qubits for each single particle state in the system initialized in the $\ket{0}$ state. To compute the real part of the overlap, the ancilla qubit is $\ket{q_\alpha}=\frac{1}{\sqrt{2}}\ket{0}+\frac{1}{\sqrt{2}}\ket{1}$. The unitary operators, $U_i$ and $U_j$, create the Krylov basis states from the multi-qubit zero state $\ket{0}^{\otimes n}$,

\begin{equation}
    \hat{U}_i\ket{0}^{\otimes n}=\ket{\Psi_{a,k}}
\end{equation}

and

\begin{equation}
    \hat{U}_j\ket{\Psi_{a,k}}=\ket{\Psi_{b,l}}.
\end{equation}

\noindent Using the red dashed lines, I broke up the quantum circuit in Figure \ref{cir:dot-prod} into four steps to explain how the wavefunction changes at each step.

\begin{enumerate}
    \item The initial wavefunction is $\ket{\Psi_1}=\frac{1}{\sqrt{2}}\ket{0}\ket{0}^{\otimes n}+\frac{1}{\sqrt{2}}\ket{1}\ket{0}^{\otimes n}$.
    \item The first unitary operator is $U_i = U_aU_k$ where $U_a$ creates a reference state $\ket{\Psi_a}$ as described in Section \ref{sec:state-prep}. Then $U_k$ evolves the reference state $k$ time steps. The overall wavefunction is then $\ket{\Psi_2}=\frac{1}{\sqrt{2}}\ket{0}\ket{\Psi_{a,k}}+\frac{1}{\sqrt{2}}\ket{1}\ket{\Psi_{a,k}}$.
    \item The system qubits are entangled with the ancilla qubit using a $C_X$ gate. The next unitary operator creates another reference state at $l$ time steps, $\hat{U}_j\ket{\Psi_{a,k}}=\ket{\Psi_{b,l}}$ and is entangled with the ancilla qubit. So the overall wavefunction at step three is $\ket{\Psi_3}=\frac{1}{\sqrt{2}}\ket{0}\ket{\Psi_{a,k}}+\frac{1}{\sqrt{2}}\ket{1}\ket{\Psi_{b,l}}$.
    \item Then, the $H$ gate creates a superposition of the two reference states. The overall wavefunction is then $\ket{\Psi_4}=\frac{1}{2}\ket{0}(\ket{\Psi_{a,k}}+\ket{\Psi_{b,l}})+\frac{1}{2}\ket{1}(\ket{\Psi_{a,k}}-\ket{\Psi_{b,l}})$.
\end{enumerate}

\noindent Only the ancilla qubit is measured. Analytically, the probability of measuring the ancilla qubit to be \ket{0} is

\begin{equation}\label{eq:prob_0}
    \begin{aligned}
    P_\alpha(\ket{0}) &= \frac{1}{4}(\braket{\Psi_{a,k}}{\Psi_{a,k}}+\braket{\Psi_{b,l}}{\Psi_{b,l}}+\braket{\Psi_{a,k}}{\Psi_{b,l}}+\braket{\Psi_{b,l}}{\Psi_{a,k}}) \\
    &= \frac{1}{2}(1+\mathrm{Re}(\braket{\Psi_{a,k}}{\Psi_{b,l}})),
    \end{aligned}
\end{equation}

\noindent and the probability of it being \ket{1} is

\begin{equation}\label{eq:prob_1}
    \begin{aligned}
    P_\alpha(\ket{1}) &= \frac{1}{4}(\braket{\Psi_{a,k}}{\Psi_{a,k}}+\braket{\Psi_{b,l}}{\Psi_{b,l}}-\braket{\Psi_{a,k}}{\Psi_{b,l}}-\braket{\Psi_{b,l}}{\Psi_{a,k}}) \\
    &= \frac{1}{2}(1-\mathrm{Re}(\braket{\Psi_{a,k}}{\Psi_{b,l}})).
    \end{aligned}
\end{equation}

\noindent Running and measuring the quantum circuit many times results in the empirical probability of $P_\alpha(\ket{0})$ and $P_\alpha(\ket{1})$. Using Equations (\ref{eq:prob_0}) and (\ref{eq:prob_1}), the real part of the overlap between two vectors is

\vspace{-3mm}
\begin{equation}
    \mathrm{Re}(\braket{\Psi_{a,k}}{\Psi_{b,l}})=P_\alpha(\ket{0})-P_\alpha(\ket{1}).
\end{equation}

The process of computing the imaginary part of the overlap is similar. The same quantum circuit in Figure \ref{cir:dot-prod} is used except the ancilla qubit is $\ket{q_a}=\frac{1}{\sqrt{2}}(\ket{0}-i\ket{1})$ making the overall wavefunction

\vspace{-5mm}
\begin{equation}
     \ket{\Psi_4}=\frac{1}{2}\ket{0}(\ket{\Psi_{a,k}}-i\ket{\Psi_{b,l}})+\frac{1}{2}\ket{1}(\ket{\Psi_{a,k}}+i\ket{\Psi_{b,l}}).
 \end{equation}

\noindent Analytically, the probability of measuring the ancilla qubit to be \ket{0} is $P_\alpha(\ket{0}) = \frac{1}{2}(1-\mathrm{Im}\braket{\Psi_{a,k}}{\Psi_{b,l}})$ and the probability of it being \ket{1} is $P_\alpha(\ket{1}) = \frac{1}{2}(1+\mathrm{Im}\braket{\Psi_{a,k}}{\Psi_{b,l}})$. Running and measuring the quantum circuit many times results in the empirical probability of $P_\alpha(\ket{0})$ and $P_\alpha(\ket{1})$. So the imaginary part of the overlap is

\vspace{-5mm}
\begin{equation}
    \mathrm{Im}(\braket{\Psi_{a,k}}{\Psi_{b,l}})=P_\alpha(\ket{1})-P_\alpha(\ket{0}).
\end{equation}

\section{Computing Matrix Elements of the Hamiltonian}
To compute the diagonal matrix elements of the Hamiltonian, $\bra{\Psi_{a,k}}\hat{H}\ket{\Psi_{a,k}}$, the Hamiltonian needs to be decomposed into Pauli operators. For example, a Hermitian operator, 

\vspace{-3mm}
\begin{equation}
    \hat{A} = 
    \begin{pmatrix}
        a & b^* \\
        b & c
    \end{pmatrix},
\end{equation}

\noindent can be decomposed using the Pauli and identity operators,

\vspace{-3mm}
\begin{equation}\label{eq:pauli_decomp}
    \hat{A}=\frac{a+c}{2}I+\mathrm{Re}(b)X+\mathrm{Im}(b)Y+\frac{a-c}{2}Z.
\end{equation}

\noindent To measure the expectation value of $\hat{A}$, $\bra{v}\hat{A}\ket{v}=\expect{\hat{A}}$, the expectation value for each term in Equation (\ref{eq:pauli_decomp}) needs to be computed,

\vspace{-3mm}
\begin{equation}\label{eq:pauli_decomp_exp}
    \expect{\hat{A}}=\frac{a+c}{2}\expect{I}+\mathrm{Re}(b)\expect{X}+\mathrm{Im}(b)\expect{Y}+\frac{a-c}{2}\expect{Z}.
\end{equation}

\noindent Each expectation value is computed using individual quantum circuits. For example, the quantum circuit to compute the expectation value of the second term in Equation (\ref{eq:pauli_decomp_exp}), $\expect{X}$, is,

\[
    \begin{quantikz}
    \lstick{\ket{v}} & \gate{H} & \meter{}
    \end{quantikz}.
\]

\noindent The $H$ gate takes the qubit from the computational basis ($Z$ basis) to the $X$ basis (see Equation (\ref{eq:ZtoX_basis})). After running and measuring the quantum circuit many times, the expectation value can be computed, $\expect{X} = P_\alpha(\ket{0})-P_\alpha(\ket{1})$.

To compute the off-diagonal matrix elements of the Hamiltonian, $\bra{\Psi_{a,k}}\hat{H}\ket{\Psi_{b,l}}$, the Hamiltonian is decomposed using Equation (\ref{eq:pauli_decomp}):

\begin{equation}\label{eq:pauli-decomp-off-diag}
    \begin{aligned}
    \bra{v}\hat{A}\ket{w} &=\frac{a+c}{2}\bra{v}I\ket{w}+\mathrm{Re}(b)\bra{v}X\ket{w} \\
    &+\mathrm{Im}(b)\bra{v}Y\ket{w}+\frac{a-c}{2}\bra{v}Z\ket{w}.
    \end{aligned}
\end{equation}

\noindent Figure \ref{cir:ham} shows a quantum circuit that computes the off-diagonal matrix elements of a Hamiltonian.

\vspace{-3mm}
\begin{figure}
    \centering
    \begin{quantikz}[wire types={q,b}]
    \lstick{\ket{q_\alpha}} && \ctrl{1} & \gate{H} & \meter{} \\
    \lstick{$\ket{0}^{\otimes n}$} & \gate{\hat{U}_i} & \gate{\hat{U}_j} && \meter{}
    \end{quantikz}
    \caption{The quantum circuit to compute the off-diagonal matrix elements of $Z$, $\bra{\Psi_{a,k}}Z\ket{\Psi_{b,l}}$, using an ancilla qubit, $\ket{q_\alpha}$. The first unitary operator, $\hat{U}_i$, creates the first vector, $\hat{U}_i\ket{0}^{\otimes n}=\ket{\Psi_{a,k}}$, from the multi-qubit zeroth state. The second unitary operator, $\hat{U}_j$, creates the second vector, $\hat{U}_j\ket{\Psi_{a,k}}=\ket{\Psi_{b,l}}$ and entangles the system qubits with the ancilla qubit. To compute the real part of the inner product, the ancilla qubit is $\ket{q_\alpha}=H\ket{0}=\frac{1}{\sqrt{2}}\ket{0}+\frac{1}{\sqrt{2}}\ket{1}$ and to compute the imaginary part of the inner product, the ancilla qubit is $\ket{q_\alpha}=S^\dagger H \ket{0}=\frac{1}{\sqrt{2}}\ket{0}-\frac{i}{\sqrt{2}}\ket{1}$. The quantum circuit is run many times, and the ancilla qubit and system qubits are measured.}
    \label{cir:ham}
\end{figure}

\noindent This is almost the same circuit for computing the inner product (Figure \ref{cir:dot-prod}), except the system qubits are also measured.

Let us look at how to compute the fourth term of Equation (\ref{eq:pauli-decomp-off-diag}) using the example case, $^{14}$N, \bra{\Psi_{a,k}}Z\ket{\Psi_{b,l}}. The analytical expectation value of $Z$ when $\ket{q_\alpha}=\ket{0}$ is

\begin{equation}\label{eq:exp_0}
\begin{aligned}
    \expect{Z} &= \frac{1}{4}(\bra{\Psi_{a,k}}Z\ket{\Psi_{a,k}}+\bra{\Psi_{b,l}}Z\ket{\Psi_{b,l}}+\bra{\Psi_{a,k}}Z\ket{\Psi_{b,l}}+\bra{\Psi_{b,l}}Z\ket{\Psi_{a,k}}) \\
    &= \frac{1}{4}\bra{\Psi_{a,k}}Z\ket{\Psi_{a,k}}+\frac{1}{4}\bra{\Psi_{b,l}}Z\ket{\Psi_{b,l}}+\frac{1}{2}\mathrm{Re}(\bra{\Psi_{a,k}}Z\ket{\Psi_{b,l}}).
\end{aligned}
\end{equation}

\noindent The analytical expectation value of $Z$ when $\ket{q_\alpha}=\ket{1}$ is 

\begin{equation}\label{eq:exp_1}
\begin{aligned}
    \expect{Z} &= \frac{1}{4}(\bra{\Psi_{a,k}}Z\ket{\Psi_{a,k}}+\bra{\Psi_{b,l}}Z\ket{\Psi_{b,l}}-\bra{\Psi_{a,k}}Z\ket{\Psi_{b,l}}-\bra{\Psi_{b,l}}Z\ket{\Psi_{a,k}}) \\
    &= \frac{1}{4}\bra{\Psi_{a,k}}Z\ket{\Psi_{a,k}}+\frac{1}{4}\bra{\Psi_{b,l}}Z\ket{\Psi_{b,l}}-\frac{1}{2}\mathrm{Re}(\bra{\Psi_{a,k}}Z\ket{\Psi_{b,l}}.
\end{aligned}
\end{equation}

\noindent When the quantum circuit in Figure \ref{cir:ham} is run and measured many times, the empirical expectation values of $Z$ can be computed using the results. We can compute the expectation value of an operator using its eigenvalues and the probabilities found by the quantum computation,

\begin{equation}\label{eq:exp}
    \expect{A}=\sum_n\left|c_n\right|^2\lambda_n=\sum_nP(n)\lambda_n,
\end{equation}

\noindent where $c_n$ are the coefficients, $\lambda_n$ are the eigenvalues, and $P(n)$ are the probabilities. Using Equation (\ref{eq:exp}), the expectation value of $Z$ when $\ket{q_\alpha}=\ket{0}$ is

\begin{equation}\label{eq:exp_prob_0}
\begin{aligned}
    \expect{Z} &= P_{\alpha0123}(\ket{00000})-P_{\alpha0123}(\ket{01000})-P_{\alpha0123}(\ket{00100})\\
    &-P_{\alpha0123}(\ket{00010})-P_{\alpha0123}(\ket{00001})+P_{\alpha0123}(\ket{01100})\\
    &+P_{\alpha0123}(\ket{00011})+P_{\alpha0123}(\ket{01001})+P_{\alpha0123}(\ket{00110})\\
    &+P_{\alpha0123}(\ket{00101})+P_{\alpha0123}(\ket{01010})-P_{\alpha0123}(\ket{00111})\\
    &-P_{\alpha0123}(\ket{01011})-P_{\alpha0123}(\ket{01101})-P_{\alpha0123}(\ket{01110})\\
    &+P_{\alpha0123}(\ket{01111}).
\end{aligned}
\end{equation}

\noindent The expectation value of $Z$ when $\ket{q_\alpha}=\ket{1}$ is

\begin{equation}\label{eq:exp_prob_1}
\begin{aligned}
    \expect{Z} &= P_{\alpha0123}(\ket{10000})-P_{\alpha0123}(\ket{11000})-P_{\alpha0123}(\ket{10100})\\
    &-P_{\alpha0123}(\ket{10010})-P_{\alpha0123}(\ket{10001})+P_{\alpha0123}(\ket{11100})\\
    &+P_{\alpha0123}(\ket{10011})+P_{\alpha0123}(\ket{11001})+P_{\alpha0123}(\ket{10110})\\
    &+P_{\alpha0123}(\ket{10101})+P_{\alpha0123}(\ket{11010})-P_{\alpha0123}(\ket{10111})\\
    &-P_{\alpha0123}(\ket{11011})-P_{\alpha0123}(\ket{11101})-P_{\alpha0123}(\ket{11110})\\
    &+P_{\alpha0123}(\ket{11111}).
\end{aligned}
\end{equation}

\noindent The real part of the inner product is found by substituting Equation (\ref{eq:exp_prob_0}) into Equation (\ref{eq:exp_0}) and Equation (\ref{eq:exp_prob_1}) into Equation (\ref{eq:exp_1}) and taking the difference resulting in Equation (\ref{eq:re_mat_el}).

\begin{equation}\label{eq:re_mat_el}
\begin{aligned}
    \mathrm{Re}(\bra{\Psi_{a,k}}Z\ket{\Psi_{b,l}}) &= P_{\alpha0123}(\ket{00000})-P_{\alpha0123}(\ket{01000})-P_{\alpha0123}(\ket{00100})\\
    &-P_{\alpha0123}(\ket{00010})-P_{\alpha0123}(\ket{00001})+P_{\alpha0123}(\ket{01100})\\
    &+P_{\alpha0123}(\ket{00011})+P_{\alpha0123}(\ket{01001})+P_{\alpha0123}(\ket{00110})\\
    &+P_{\alpha0123}(\ket{00101})+P_{\alpha0123}(\ket{01010})-P_{\alpha0123}(\ket{00111})\\
    &-P_{\alpha0123}(\ket{01011})-P_{\alpha0123}(\ket{01101})-P_{\alpha0123}(\ket{01110})\\
    &+P_{\alpha0123}(\ket{01111})-P_{\alpha0123}(\ket{10000})+P_{\alpha0123}(\ket{11000})\\
    &+P_{\alpha0123}(\ket{10100})+P_{\alpha0123}(\ket{10010})+P_{\alpha0123}(\ket{10001})\\
    &-P_{\alpha0123}(\ket{11100})-P_{\alpha0123}(\ket{10011})-P_{\alpha0123}(\ket{11001})\\
    &-P_{\alpha0123}(\ket{10110})-P_{\alpha0123}(\ket{10101})-P_{\alpha0123}(\ket{11010})\\
    &+P_{\alpha0123}(\ket{10111})+P_{\alpha0123}(\ket{11011})+P_{\alpha0123}(\ket{11101})\\
    &+P_{\alpha0123}(\ket{11110})-P_{\alpha0123}(\ket{11111}).
\end{aligned}
\end{equation}

\noindent To compute the imaginary part of the off-diagonal matrix elements of the Hamiltonian, the same quantum circuit in Figure \ref{cir:ham} is run except the ancilla qubit is $\ket{q_\alpha}=\frac{1}{\sqrt{2}}\ket{0}-\frac{i}{\sqrt{2}}\ket{1}$.

\section{Constructing Real-Time Unitary Gates}
Continuing with the $^{14}$N example, I will sketch the quantum circuits for the Trotterized real-time evolution operator, $\hat{U}_k$, where $k$ is the number of real-time evolution iterations. For one Trotter step, there are 17 terms to compute in Equation (\ref{eq:trotter-decomp}). Each exponential term is computed using a series of $C_X$, $H$, $S$, and $R_Z(\theta)$ gates. For example, the circuit to compute the eighth term from Equation (\ref{eq:trotter-decomp}), $\exp(-ic_7Z_0Z_2\Delta tk)$, is

\vspace{-3mm}
\[
    \begin{quantikz}
    \lstick{\ket{q_0}} & \ctrl{2} && \ctrl{2} & \\
    \lstick{\ket{q_1}} &&&& \\
    \lstick{\ket{q_2}} & \targ{}  & \gate{R_Z(\theta_7)} & \targ{}  & \\
    \lstick{\ket{q_3}} &&&&
    \end{quantikz},
\]

\noindent where $\theta_7 = c_7\Delta tk$. If the Pauli string includes $X$ or $Y$ Pauli operators, then $H$ and $S$ gates are used to change the basis from the computational basis (see Equations (\ref{eq:ZtoX_basis}) and (\ref{eq:ZtoY_basis})). For example, the circuit for the 12\textsuperscript{th} term from Equation (\ref{eq:trotter-decomp}), $\exp(-ic_{11}Y_0Y_1X_2X_3\Delta tk)$, is

\vspace{-5mm}
\[
    \begin{quantikz}
    \lstick{\ket{q_0}} & \gate{H} & \gate{S} & \ctrl{1} &&&&&& \ctrl{1} & \gate{S^\dagger} & \gate{H} & \\
    \lstick{\ket{q_1}} & \gate{H} & \gate{S} & \targ{} & \ctrl{1} &&&& \ctrl{1} & \targ{}  & \gate{S^\dagger} & \gate{H} & \\
    \lstick{\ket{q_2}} & \gate{H} &&& \targ{} & \ctrl{1} && \ctrl{1} & \targ{} &&& \gate{H} & \\
    \lstick{\ket{q_3}} & \gate{H} &&&& \targ{} & \gate{R_Z(\theta_{11})} & \targ{} &&&& \gate{H} & 
    \end{quantikz},
\]

\noindent where $\theta_{11} = c_{11}\Delta tk$. The $H$ and $S$ gates puts $\ket{q_0}$ and $\ket{q_1}$ in the $Y$ basis and the $H$ gate puts $\ket{q_2}$ and $\ket{q_3}$ in the $X$ basis. The sequence of the 17 terms in Equation (\ref{eq:trotter-decomp}) would generate the state $\ket{\Psi_k}$. In the next chapter, I give the results for running the quantum operations of these quantum circuits on a classical computer.

%%%%%%%%%%%%%%%%%%%%%%%%%%%%%%%%%%%%%%%%%%%%%%%%%%%%%%%%
%%%%%%%%%%%%%%%%%% Chapter:Results %%%%%%%%%%%%%%%%%%%%
%%%%%%%%%%%%%%%%%%%%%%%%%%%%%%%%%%%%%%%%%%%%%%%%%%%%%%%%

\chapter{Results}\label{chap:results} %show quantum simulation results for Be8

In this chapter, I present the results of numerical simulations of the quantum operations described in Chapter \ref{chap:quantum-circuits-qlanczos} for the QLanczos algorithm run on a classical computer as described in Step 3 of Figure \ref{fig:foursteps}. The results in this chapter are from the numerical simulations of the QLanczos algorithm with real-time evolution and multiple references to solve for the low-lying energies of the valence particles of $^8$Be in the spherical and Hartree-Fock basis. A helium core is assumed for this system, and the valence particles are two protons and two neutrons in the full $0p$-shell. The model space is six possible single-particle states for each species in the full $0p$-shell ($0p_{1/2}$ and $0p_{3/2}$). The interactions for the $0p$-shell were from Cohen and Kurath \cite{Cohen1965effective}. The Hamiltonian was decomposed into Pauli strings using the Jordan-Wigner transformation, and real-time evolution was approximated using Trotter-Suzuki decomposition, methods developed in Chapter \ref{chap:preapering-qlanczos-quantum-computer}.

The matrix elements of the Hamiltonian in the spherical basis and Hartree-Fock basis were computed using a shell model code \cite{Johnson2018}. An M-scheme spherical basis of $M=0$ was used. The Jordan-Wigner transformation leads to a Hamiltonian containing 975 Pauli strings in the spherical basis, requiring approximately 22,151 quantum gates to compute a real-time evolution step, $U_k$, for a Trotter number of $N=1$. The Hartree-Fock basis was unrestricted. In the Hartree-Fock basis, the Jordan-Wigner transformation leads to a Hamiltonian containing 2,431 Pauli strings requiring approximately 56,583 quantum gates to compute $U_k$ for $N=1$. The gate approximations were based on the number of Pauli stings and whether the Pauli stings contained $I$, $X$, $Y$, or $Z$ operators. For example, a Pauli sting containing $XXYZ$ would require two $H$ gates for each $X$ operator and two $S$ (or $S^\dagger$) and two $H$ gates for the $Y$ operator. Additionally, each operator that is not an identity gate, $I$, is entangled, requiring a pair of $CNOT$ gates. Then, one rotation gate. For this example, the single Pauli string, $XXYZ$, would require 17 quantum gates.

\begin{figure}
\centering
\begin{subfigure}{.5\textwidth}
    \centering
    \includegraphics[width=1\textwidth]{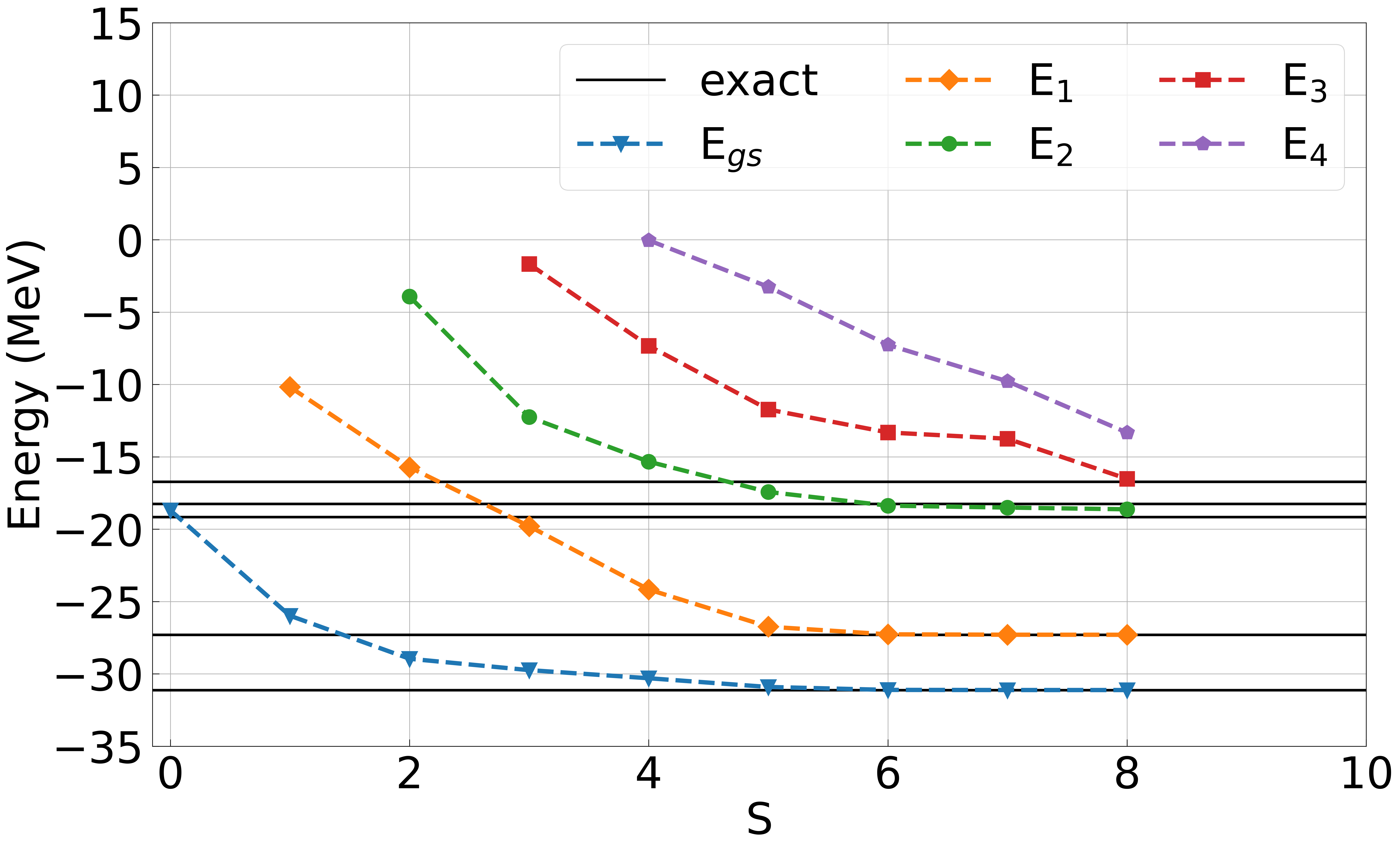}
    \subcaption[]{}
\end{subfigure}%
\begin{subfigure}{.5\textwidth}
    \centering
    \includegraphics[width=1\textwidth]{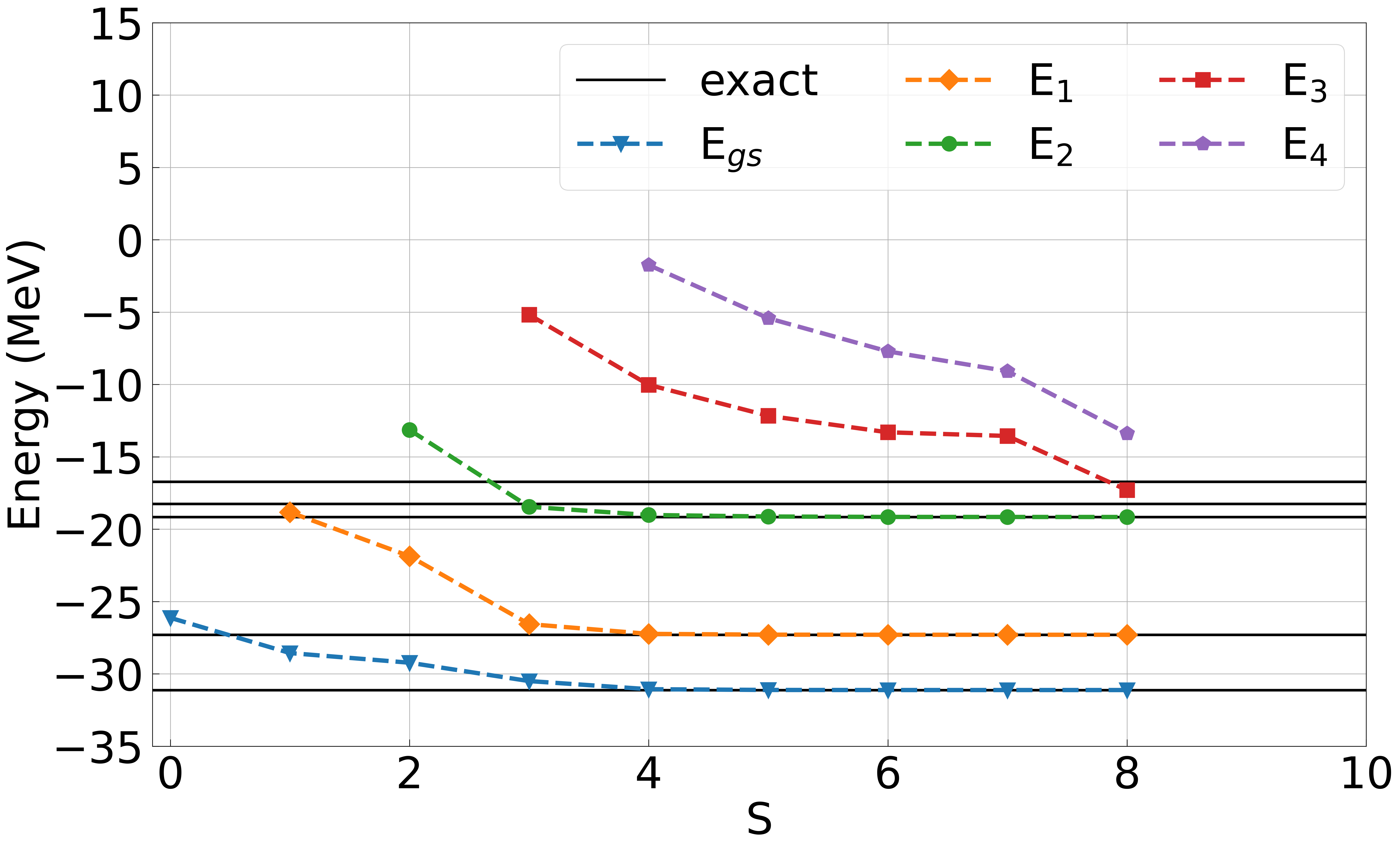}
    \subcaption[]{}
\end{subfigure}
\caption{Numerical simulations of the QLanczos algorithm with exact real-time evolution to solve for the lowest five energy states of the valence particles of $^8$Be (two protons and two neutrons in the full $0p$-shell). The simulation was run using a single reference state; (a) the lowest energy configuration in the spherical basis and (b) the Hartree-Fock state. The solid black lines are the exact energies. $E_n$ labels the $n$th excited state. A fixed number of real-time evolution iterations was used ($S=8$) with a time step size of $\Delta t = 0.1 \textrm{MeV\textsuperscript{-1}}$.}
\label{fig:spectrum_be8_hf_sph}
\end{figure}

Figure \ref{fig:spectrum_be8_hf_sph} show the results of numerical simulations of the QLanczos algorithm quantum circuits to find the lowest five energy eigenstates using exact real-time evolution and a single reference state. The quantum operations were applied in a basis of all possible configurations in the $0p$-shell written as bit strings described in Section \ref{sec:second-quantization}. The reference state for the Hartree-Fock basis was the Hartree-Fock state, $\ket{\Psi_1}=\ket{110000110000}$. For the spherical basis, the reference state was the lowest energy configuration, $\ket{\Psi_1}=\ket{000110000110}$. Table \ref{tab:spectrum_be8_errors} compares the exact energies to the energies computed in Figure \ref{fig:spectrum_be8_hf_sph} with a fixed number of time steps, $S=8$. The Hartree-Fock basis resulted in energies closer to the exact solution than the simulations using the spherical basis for all five energies computed. This is because the Hartree-Fock state starts at a lower energy.

\begin{table}
\begin{center}
\begin{tabular}{ c || c | c | c | c | c} 
    & $E$\textsubscript{exact} & $E$\textsubscript{sph} & $\Delta E$\textsubscript{sph} & $E$\textsubscript{HF} & $\Delta E$\textsubscript{HF} \\[0.5ex]
    \hline
    \hline
    $E$\textsubscript{gs} (MeV) & -31.119 & -31.119 & 1.64e-06 & -31.119 & 5.86e-09 \\[0.5ex]
    \hline
    $E_1$ (MeV) & -27.300 & -27.299 & 1.33e-05 & -27.300 & 3.91e-07 \\[0.5ex]
    \hline
    $E_2$ (MeV) & -19.162 & -18.628 & 2.79e-02 & -19.160 & 9.75e-05 \\[0.5ex]
    \hline
    $E_3$ (MeV) & -18.249 & -16.522 & 9.46e-02 & -17.305 & 5.17e-02 \\[0.5ex]
    \hline
    $E_4$ (MeV) & -16.722 & -13.332 & 2.03e-01 & -13.384 & 2.00e-01 \\[0.5ex]
\end{tabular}
\bigskip
\caption{\label{tab:spectrum_be8_errors} The lowest five energy states of two protons and two neutrons in the full $0p$-shell (nucleus of $^8$Be). $E$\textsubscript{exact} are the exact energies computed in the Hartree-Fock basis. $\Delta E$\textsubscript{sph} are the energies computed in the spherical basis from the simulations in Figure \ref{fig:spectrum_be8_hf_sph} at $S=8$ and $\Delta E$\textsubscript{sph} are their relative errors. $E$\textsubscript{HF} are the energies computed in the Hartree-Fock basis from the simulations in Figure \ref{fig:spectrum_be8_hf_sph} at $S=8$ and $\Delta E$\textsubscript{HF} are their relative errors.}
\end{center}
\vspace{-13mm}
\end{table}

I carried out numerical simulations to find the five lowest energies in the Hartree-Fock and spherical basis using different numbers of reference states, $R$. In the spherical basis, the additional reference states used were 

\begin{equation}
    \begin{aligned}
     \ket{\Psi_2}&=\ket{001001001001} \\
     \ket{\Psi_3}&=\ket{000110100010} \\
     \ket{\Psi_4}&=\ket{000110010100} \\
     \ket{\Psi_5}&=\ket{100010000110},
    \end{aligned}
\end{equation}

\noindent which are states with next the lowest energy configurations that weren't degenerate. For the Hartree-Fock basis, the additional reference state used were the particle-hole excitations of the Hartree-Fock state, 

\begin{equation}
    \begin{aligned}
     \ket{\Psi_2}&=\frac{1}{\sqrt{2}}\ket{110000101000}+\frac{1}{\sqrt{2}}\ket{101000110000} \\
     \ket{\Psi_3}&=\frac{1}{\sqrt{2}}\ket{110000100100}+\frac{1}{\sqrt{2}}\ket{100100110000} \\
     \ket{\Psi_4}&=\frac{1}{\sqrt{2}}\ket{110000100010}+\frac{1}{\sqrt{2}}\ket{100010110000} \\
     \ket{\Psi_5}&=\frac{1}{\sqrt{2}}\ket{110000100001}+\frac{1}{\sqrt{2}}\ket{100001110000}.       
    \end{aligned}
\end{equation}

\begin{figure}
\centering
\begin{subfigure}{.5\textwidth}
    \centering
    \includegraphics[width=1\textwidth]{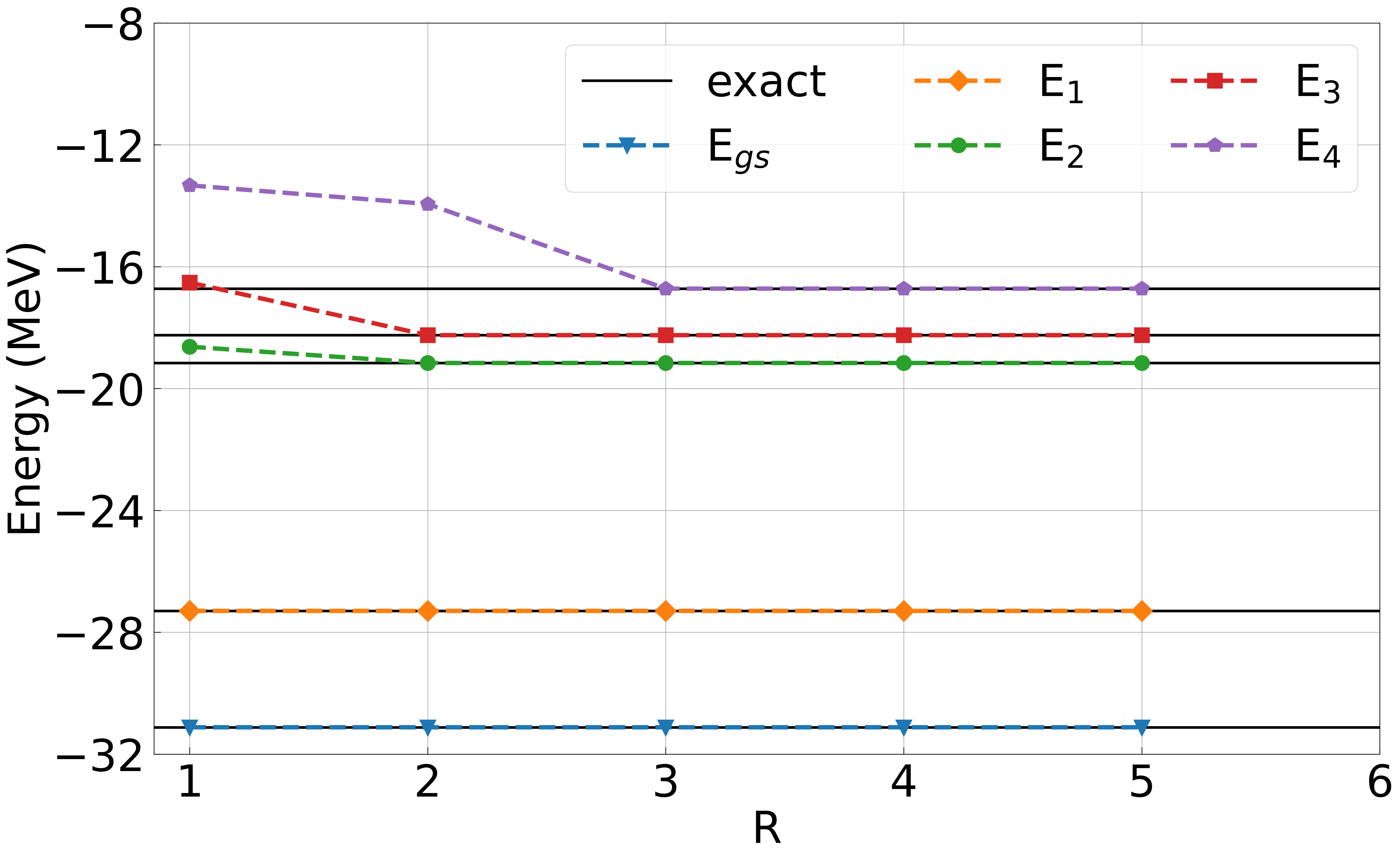}
    \subcaption[]{}
\end{subfigure}%
\begin{subfigure}{.5\textwidth}
    \centering
    \includegraphics[width=1\textwidth]{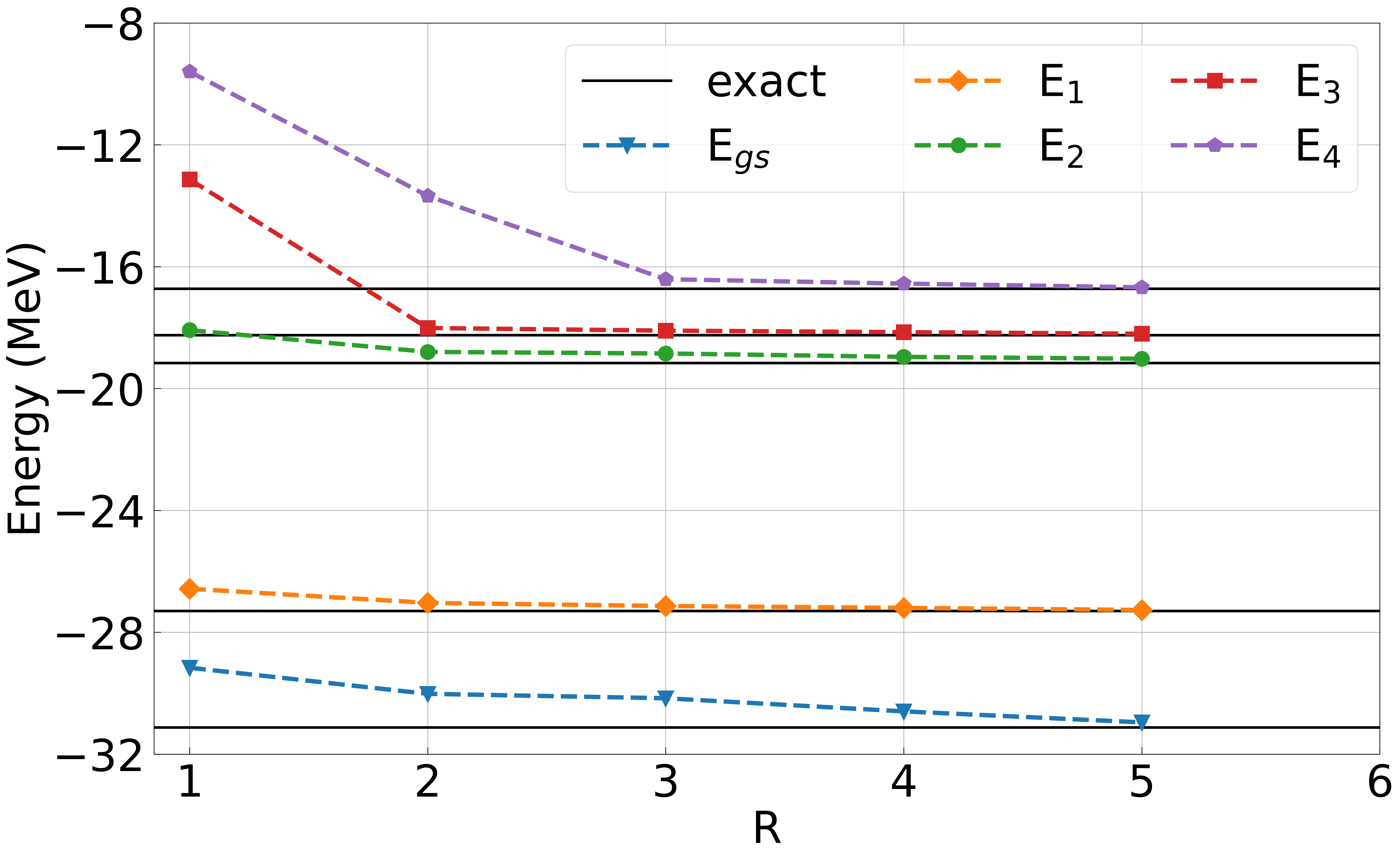}
    \subcaption[]{}
\end{subfigure}
\begin{subfigure}{.5\textwidth}
    \centering
    \includegraphics[width=1\textwidth]{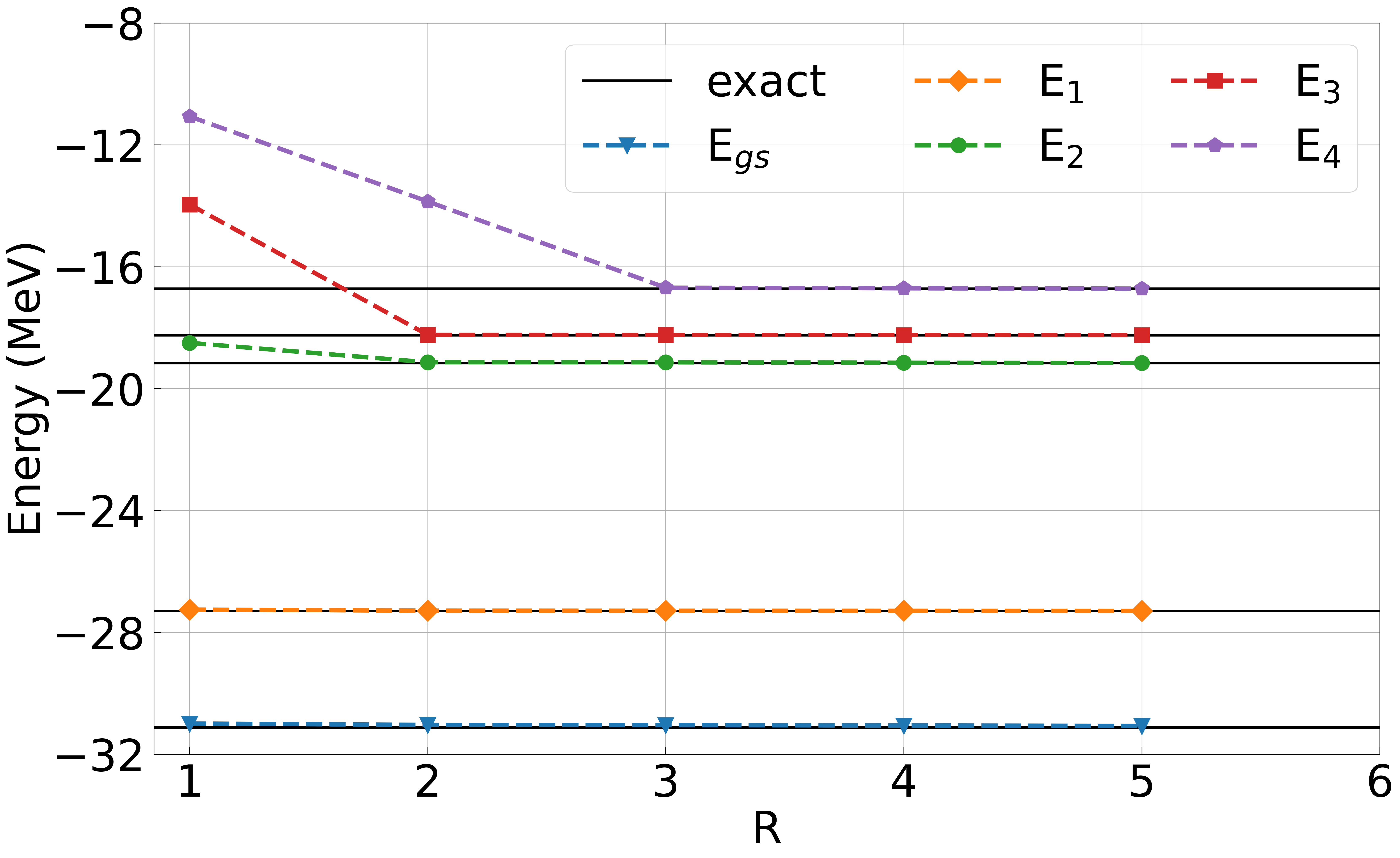}
    \subcaption[]{}
\end{subfigure}%
\begin{subfigure}{.5\textwidth}
    \centering
    \includegraphics[width=1\textwidth]{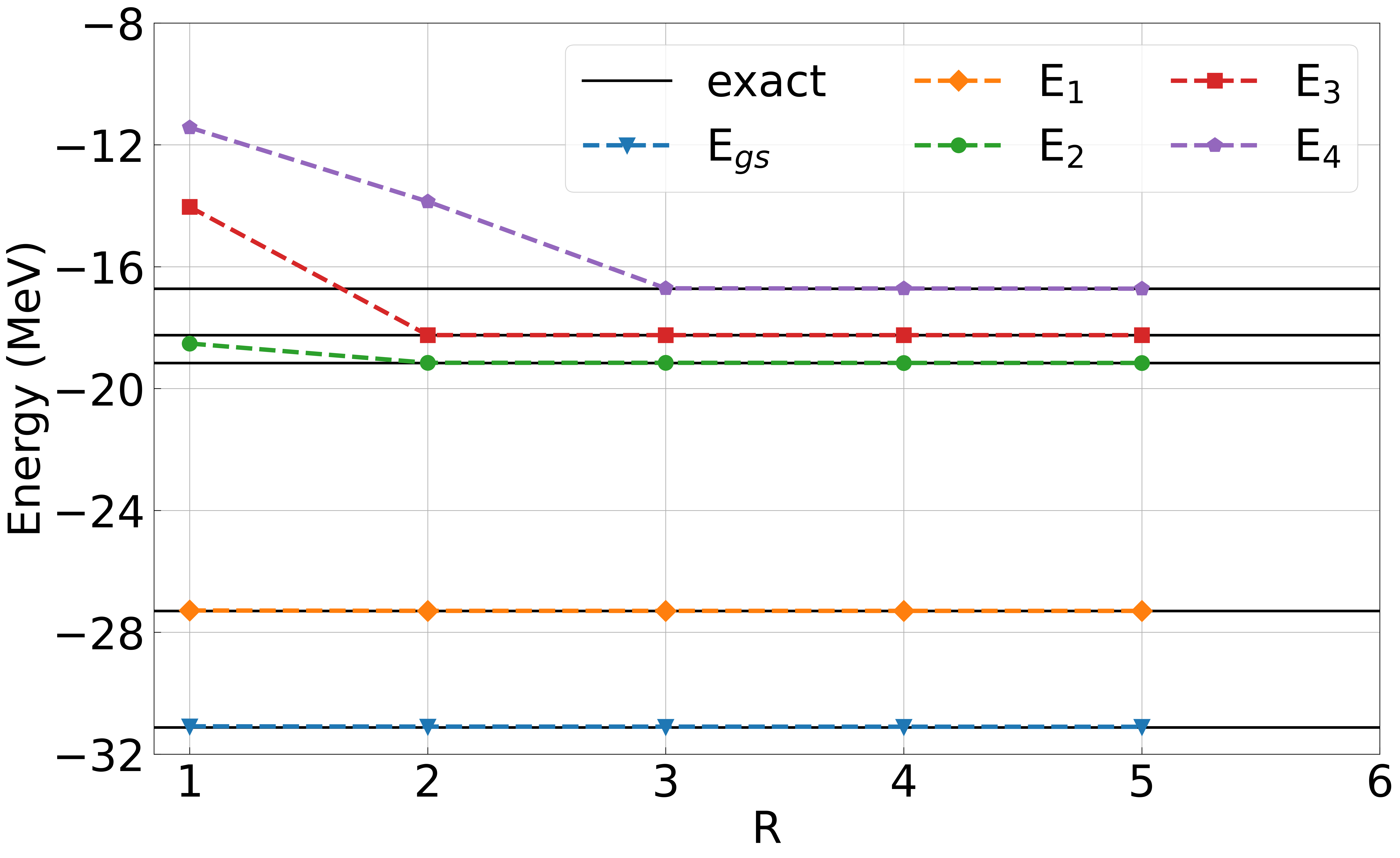}
    \subcaption[]{}
\end{subfigure}
\caption{Numerical simulations of the QLanczos algorithm with real-time evolution to solve for the lowest five energy states of the valence particles of $^8$Be (two protons and two neutrons in the full $0p$-shell) using different numbers of reference states, R. The Hamiltonian was put in the spherical basis, and the reference states were the lowest energy configurations. The simulation was run using (a) exact real-time evolution, (b) Trotterized real-time evolution with $N=1$, (b) Trotterized real-time evolution with $N=4$, and (d) Trotterized real-time evolution with $N=8$. The solid black lines are the exact energies. $E_{gs}$ is the ground state energy and $E_n$ labels the $n$th excited state. A fixed number of real-time evolution iterations was used ($S=8$) with a time step size of $\Delta t = 0.1 \textrm{MeV\textsuperscript{-1}}$.}
\label{fig:spectrum_be8_sph}
\end{figure}

\begin{figure}
\centering
\begin{subfigure}{.5\textwidth}
    \centering
    \includegraphics[width=1\textwidth]{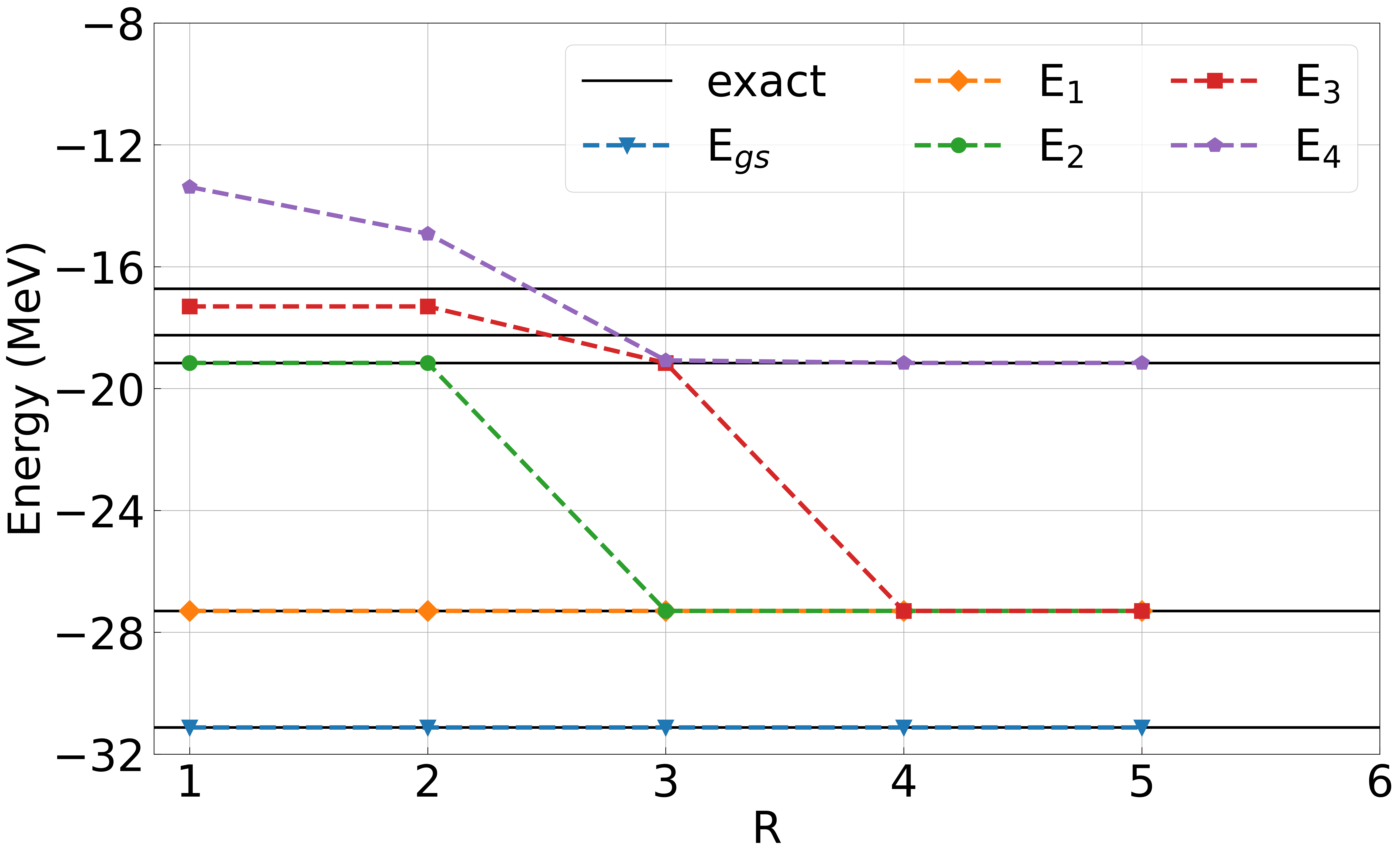}
    \subcaption[]{}
\end{subfigure}%
\begin{subfigure}{.5\textwidth}
    \centering
    \includegraphics[width=1\textwidth]{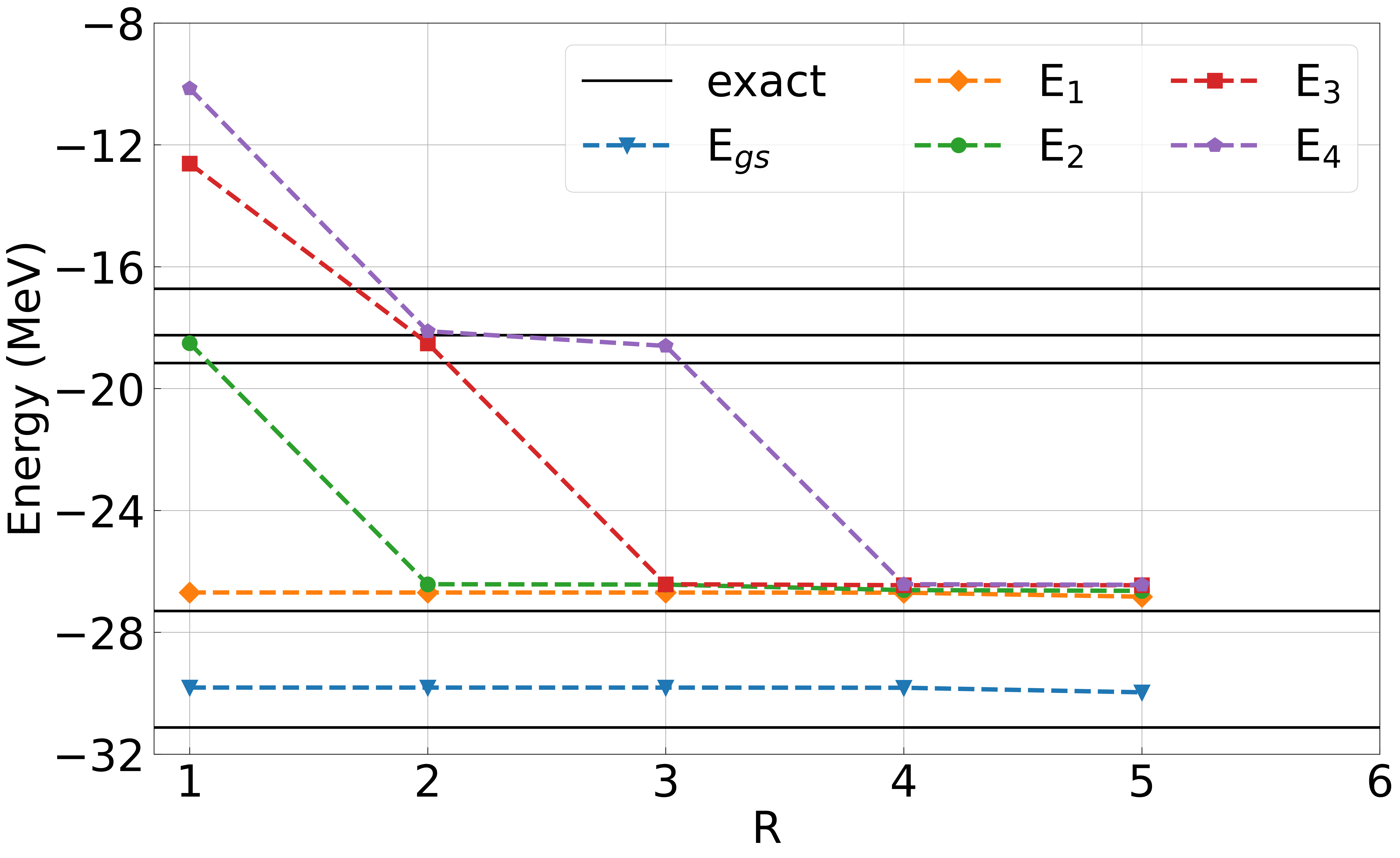}
    \subcaption[]{}
\end{subfigure}
\begin{subfigure}{.5\textwidth}
    \centering
    \includegraphics[width=1\textwidth]{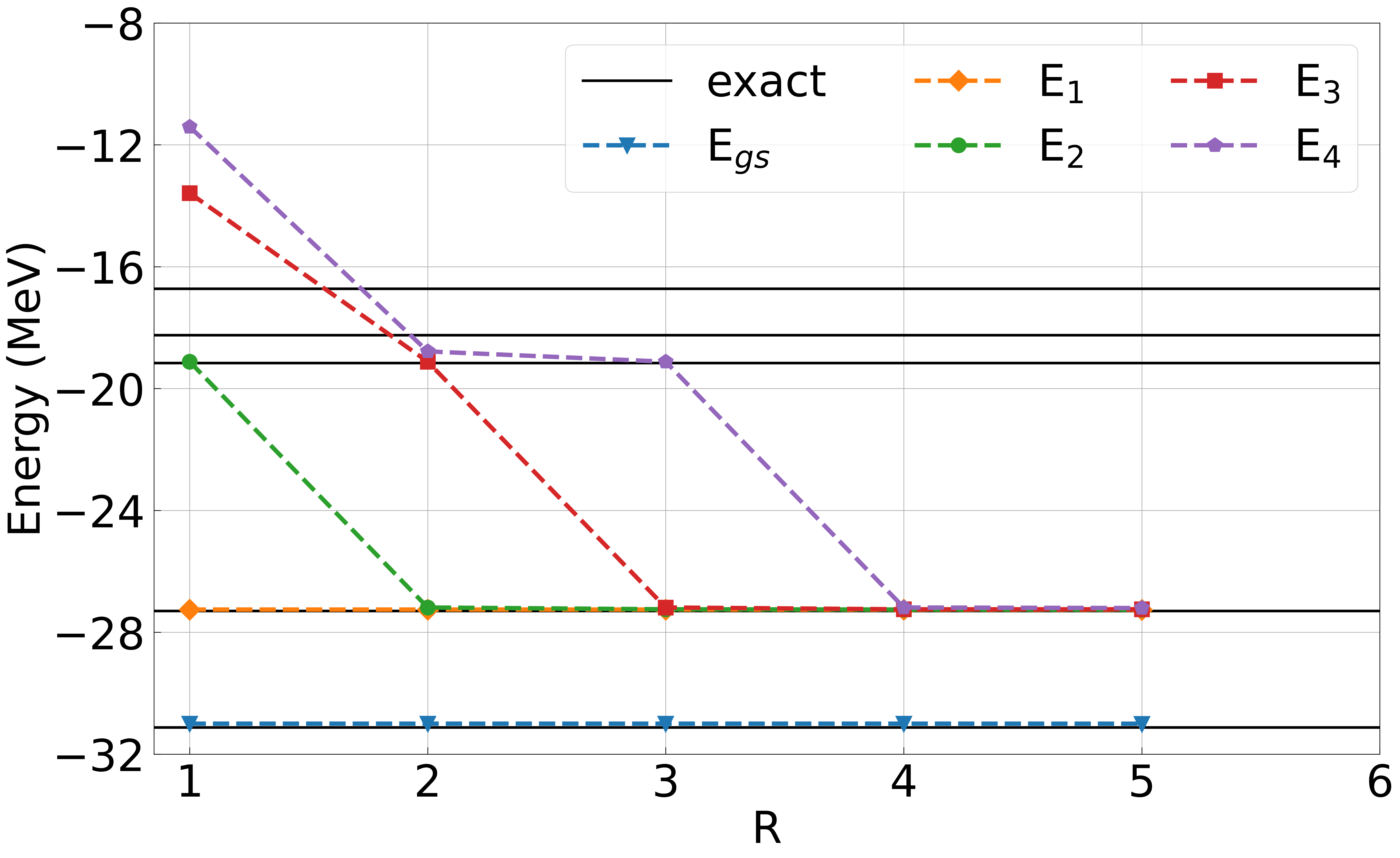}
    \subcaption[]{}
\end{subfigure}%
\begin{subfigure}{.5\textwidth}
    \centering
    \includegraphics[width=1\textwidth]{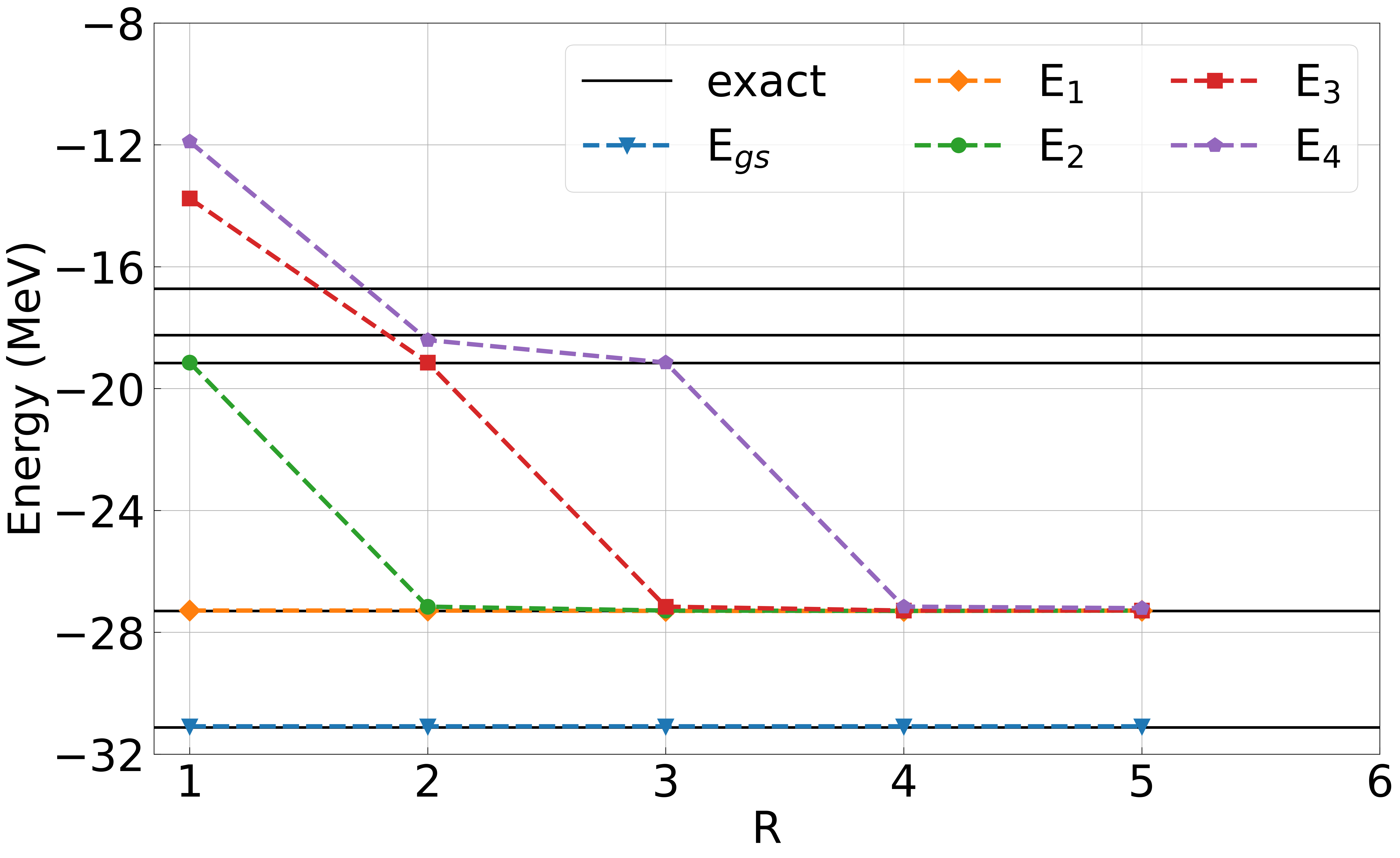}
    \subcaption[]{}
    
\end{subfigure}
\caption{Numerical simulations of the QLanczos algorithm with real-time evolution to solve for the lowest five energy states of the valence particles of $^8$Be (two protons and two neutrons in the full $0p$-shell) using different numbers of reference states, R. The Hamiltonian was put in the Hartree-Fock basis, and the reference states were the Hartree-Fock state and its excitations. The simulation was run using (a) exact real-time evolution, (b) Trotterized real-time evolution with $N=1$, (b) Trotterized real-time evolution with $N=4$, and (d) Trotterized real-time evolution with $N=8$. The solid black lines are the exact energies. $E_{gs}$ is the ground state energy and $E_n$ labels the $n$th excited state. A fixed number of real-time evolution iterations was used ($S=8$) with a time step size of $\Delta t = 0.1 \textrm{MeV\textsuperscript{-1}}$.}
\label{fig:spectrum_be8_hf}
\end{figure}

\noindent Stair et al. \cite{Stair2020} demonstrated that Trotter numbers of $N=4,8$ were sufficient to simulate exact real-time evolution. Additionally, they discuss the computational trade-off of using a smaller Trotter number, $N=1$, and more reference states. Based on these findings, I carried out these simulations for different Trotter numbers, $N=1,4,8$, and compared them to exact real-time evolution in the spherical basis (Figure \ref{fig:spectrum_be8_sph}) and the Hartree-Fock basis (Figure \ref{fig:spectrum_be8_hf}). The simulations ran for a fixed number of real-time iterations, $S=8$, with a time step size of $\Delta t = 0.1 \textrm{MeV\textsuperscript{-1}}$.

\pagebreak

\pagebreak

There are five degeneracies for the first excited state, $E_1$, nine for the second, $E_2$, five for the third excited state, $E_3$, and three for the fourth excited state, $E_4$. The degeneracies were avoided in the spherical basis because the chosen states preserved $M=0$. The Hartree-Fock basis produced degeneracies due to the mixing of states discussed in Section \ref{sec:HF}.

These results suggest that real-time evolution effectively generates Krylov basis states for the QLanczos algorithm. Using numerical simulations of the quantum operations, I found the low-lying energy states of $^8$Be in both the spherical basis and the Hartree-Fock basis. Using multiple references improved the accuracy of the energies. Simulations showed that a Trotter number of $N=1$ and a higher number of reference states lead to an accuracy comparable to a higher Trotter number ($N=4,8$) with fewer reference states. While the Hartree-Fock basis resulted in more accurate results, degeneracies were found once multiple references were introduced. In the next chapter, I will discuss what can be deduced from these results.

%%%%%%%%%%%%%%%%%%%%%%%%%%%%%%%%%%%%%%%%%%%%%%%%%%%%%%%%
%%%%%%%%%%%%%%%% Chapter:Conclusions %%%%%%%%%%%%%%%%%%
%%%%%%%%%%%%%%%%%%%%%%%%%%%%%%%%%%%%%%%%%%%%%%%%%%%%%%%%
 
\chapter{Conclusions}\label{chap:conclusions}

In this thesis, I investigated real-time evolution as an alternative method for generating Krylov basis states for the QLanczos algorithm to find low-lying energy states of nuclei. Additionally, I investigated using multiple reference states to find these low-lying eigenstates with fewer iterations. Finally, I compared using the spherical single-particle basis and the Hartree-Fock single-particle basis. Using Python on a classical computer, I ran numerical simulations of the quantum operations that would be used on a quantum computer. These simulations yielded promising results for this approach as a practical method for finding low-lying energy states of nuclei on a quantum computer.

Real-time evolution is easier to implement and more natural on a quantum computer and is a suitable alternative to imaginary-time evolution. Although real-time evolution does not have the advantage of projecting out excited states like imaginary-time evolution, the ground state energy can still be found within tens of iterations when using QLanczos with real-time evolution. Additionally, real-time evolution can be easily simulated on a quantum computer because it is unitary and can be directly translated onto a quantum computer using Trotterization.

The multiple reference generalization increased the accuracy of the computed energy eigenstates. Using multiple references with more modest Trotter numbers, such as $N=1$, allows for a less expensive computational cost on quantum computers but increases the cost of post-processing on a classical computer, which would be ideal for NISQ processors. The advantage of multiple references is the trade-off between more quantum processing or more classical processing, which makes it adaptable to near-term quantum computers and scalable for fault-tolerant quantum computers. Additionally, using multiple reference states increased the accuracy of the computed energy eigenstates when noise was added to the simulations.

Working in the spherical single-particle basis requires shallower circuits than the Hartree-Fock single-particle basis. Using the M-scheme, the spherical basis results in fewer Pauli strings and, therefore, fewer Pauli gates by a factor of 2.5 compared to the Hartree-Fock basis. On near-term hardware, this is a significant computational advantage. Additionally, the Hartree-Fock basis can produce degeneracies that can be avoided by using an M-scheme spherical basis. Although the lowest energy configuration in the spherical basis is higher than the Hartree-Fock state, the spherical basis only takes a few more real-time evolution iterations than the Hartree-Fock basis to converge to the ground state energy.

The following steps for this work would be to run the QLanczos with real-time evolution on a quantum computer, such as the ones provided by the IBM Quantum Experience \cite{Qiskit}, using the quantum circuits sketched in Chapter \ref{chap:quantum-circuits-qlanczos}. A smaller system such as the Ising model would be more suitable for the available NISQ processors.

%% file: thesis.bbl
\begin{thebibliography}{10}

\bibitem{Abrams1997}
{\sc D.~S. Abrams and S.~Lloyd}, {\em Simulation of many-body {F}ermi systems
  on a universal quantum computer}, Phys. Rev. Lett., 79 (1997),
  pp.~2586--2589.

\bibitem{Asupuru-Guzik2005}
{\sc A.~Aspuru-Guzik, A.~D. Dutoi, P.~J. Love, and M.~Head-Gordon}, {\em
  Simulated quantum computation of molecular energies}, Science, 309 (2005),
  pp.~1704--1707.

\bibitem{Blunt2015}
{\sc N.~S. Blunt, A.~Alavi, and G.~H. Booth}, {\em Krylov-projected quantum
  {M}onte {C}arlo method}, Phys. Rev. Lett., 115 (2015), p.~050603.

\bibitem{Bravyi2002}
{\sc S.~B. Bravyi and A.~Y. Kitaev}, {\em Fermionic quantum computation}, Ann.
  Phys., 298 (2002), pp.~210--226.

\bibitem{Brown2006}
{\sc B.~A. Brown and W.~A. Richter}, {\em New {``USD''} hamiltonians for the
  $\mathit{sd}$ shell}, Phys. Rev. C, 74 (2006), p.~034315.

\bibitem{Caffarel1991}
{\sc M.~Caffarel, F.~Gadea, and D.~M. Ceperley}, {\em Lancz{\'o}s-type
  algorithm for quantum {M}onte {C}arlo data}, EPL, 16 (1991), p.~249.

\bibitem{Cohen1965effective}
{\sc S.~Cohen and D.~Kurath}, {\em Effective interactions for the 1p shell},
  Nucl. Phys., 73 (1965), pp.~1--24.

\bibitem{Colless2018}
{\sc J.~I. Colless, V.~V. Ramasesh, D.~Dahlen, M.~S. Blok, M.~E.
  Kimchi-Schwartz, J.~R. McClean, J.~Carter, W.~A. de~Jong, and I.~Siddiqi},
  {\em Computation of molecular spectra on a quantum processor with an
  error-resilient algorithm}, Phys. Rev. X, 8 (2018), p.~011021.

\bibitem{Cortes2022QSD}
{\sc C.~L. Cortes and S.~K. Gray}, {\em Quantum {K}rylov subspace algorithms
  for ground- and excited-state energy estimation}, Phys. Rev. A, 105 (2022),
  p.~022417.

\bibitem{Dumitrescu2018}
{\sc E.~F. Dumitrescu, A.~J. McCaskey, G.~Hagen, G.~R. Jansen, T.~D. Morris,
  and T.~Papenbrock}, {\em Cloud quantum computing of an atom nucleus}, Phys.
  Rev. Lett., 120 (2018), p.~210501.

\bibitem{Epperly2022}
{\sc E.~N. Epperly, L.~Lin, and Y.~Nakatsukasa}, {\em A theory of quantum
  subspace diagonalization}, SIAM J. Matrix Anal. Appl., 43 (2022),
  pp.~1263--1290.

\bibitem{Farhi2022}
{\sc E.~Farhi, J.~Goldstone, S.~Gutmann, and L.~Zhou}, {\em The quantum
  approximate optimization algorithm and the {S}herrington-{K}irkpatrick model
  at infinite size}, Quantum, 6 (2022), p.~759.

\bibitem{Gocho2023}
{\sc S.~Gocho, H.~Nakamura, S.~Kanno, Q.~Gao, T.~Kobayashi, T.~Inagaki, and
  M.~Hatanaka}, {\em Excited state calculations using variational quantum
  eigensolver with spin-restricted ans{\"a}tze and automatically-adjusted
  constraints}, Npj Comput. Mater., 9 (2023), p.~13.

\bibitem{Higgot2019}
{\sc O.~Higgott, D.~Wang, and S.~Brierley}, {\em Variational quantum
  computation of excited states}, Quantum, 3 (2019), p.~156.

\bibitem{Huggins2020}
{\sc W.~Huggins, J.~Lee, U.~Baek, B.~O'Gorman, and K.~Whaley}, {\em A
  non-orthogonal variational quantum eigensolver}, New J. Phys., 22 (2020),
  p.~073009.

\bibitem{Qiskit}
{\sc IBM}, {\em Qiskit}.
\newblock https://qiskit.org/.

\bibitem{Johnson2013}
{\sc C.~W. Johnson, W.~E. Ormand, and P.~G. Krastev}, {\em Factorization in
  large-scale many-body calculations}, Comput. Phys. Commun., 184 (2013),
  pp.~2761--2774.

\bibitem{Johnson2018}
{\sc C.~W. Johnson, W.~E. Ormand, K.~S. McElvain, and H.~Shan}, {\em
  {BIGSTICK}: A flexible configuration-interaction shell-model code},
  arXiv:1801.08432,  (2018).

\bibitem{Jordan1928}
{\sc P.~{Jordan} and E.~{Wigner}}, {\em {\"U}ber das paulische
  {\"a}quivalenzverbot}, Z. Phys., 47 (1928), pp.~631--651.

\bibitem{Kandala2017}
{\sc A.~Kandala, A.~Mezzacapo, K.~Temme, M.~Takita, M.~Brink, J.~M. Chow, and
  J.~M. Gambetta}, {\em Hardware-efficient variational quantum eigensolver for
  small molecules and quantum magnets}, Nature, 549 (2017), pp.~242--246.

\bibitem{Kiss2022}
{\sc O.~Kiss, M.~Grossi, P.~Lougovski, F.~Sanchez, S.~Vallecorsa, and
  T.~Papenbrock}, {\em Quantum computing of the $^{6}\mathrm{Li}$ nucleus via
  ordered unitary coupled clusters}, Phys. Rev. C, 106 (2022), p.~034325.

\bibitem{Kitaev1995}
{\sc A.~Y. Kitaev}, {\em Quantum measurements and the {A}belian stabilizer
  problem}, Electron. Colloquium Comput. Complex., TR96-003 (1995).

\bibitem{Lanczos1950}
{\sc C.~Lanczos}, {\em An iteration method for the solution of the eigenvalue
  problem of linear differential and integral operators}, J. Res. Natl. Bur.
  Stand. B, 45 (1950), pp.~255--282.

\bibitem{McClean2016}
{\sc J.~McClean, J.~Romero, R.~Babbush, and A.~Aspuru-Guzik}, {\em The theory
  of variational hybrid quantum-classical algorithms}, New J. Phys., 18 (2016),
  p.~023023.

\bibitem{McClean2017}
{\sc J.~R. McClean, M.~E. Kimchi-Schwartz, J.~Carter, and W.~A. de~Jong}, {\em
  Hybrid quantum-classical hierarchy for mitigation of decoherence and
  determination of excited states}, Phys. Rev. A, 95 (2017), p.~042308.

\bibitem{Motta2020}
{\sc M.~Motta, C.~Sun, A.~T.~K. Tan, M.~J. O'Rourke, E.~Ye, A.~J. Minnich,
  F.~G. S.~L. Brand{\~a}o, and G.~K.-L. Chan}, {\em Determining eigenstates and
  thermal states on a quantum computer using quantum imaginary time evolution},
  Nat. Phys., 16 (2020), pp.~205--210.

\bibitem{Nakanishi2019}
{\sc K.~M. Nakanishi, K.~Mitarai, and K.~Fujii}, {\em Subspace-search
  variational quantum eigensolver for excited states}, Phys. Rev. Res., 1
  (2019), p.~033062.

\bibitem{Nielsen2010}
{\sc M.~A. Nielsen and I.~L. Chuang}, {\em Quantum Computation and Quantum
  Information: 10th Anniversary Edition}, Cambridge University Press,
  Cambridge, MA, 2010.

\bibitem{OMalley2016}
{\sc P.~J.~J. O'Malley, R.~Babbush, I.~D. Kivlichan, J.~Romero, J.~R. McClean,
  R.~Barends, J.~Kelly, P.~Roushan, A.~Tranter, N.~Ding, B.~Campbell, Y.~Chen,
  Z.~Chen, B.~Chiaro, A.~Dunsworth, A.~G. Fowler, E.~Jeffrey, E.~Lucero,
  A.~Megrant, J.~Y. Mutus, M.~Neeley, C.~Neill, C.~Quintana, D.~Sank,
  A.~Vainsencher, J.~Wenner, T.~C. White, P.~V. Coveney, P.~J. Love, H.~Neven,
  A.~Aspuru-Guzik, and J.~M. Martinis}, {\em Scalable quantum simulation of
  molecular energies}, Phys. Rev. X, 6 (2016), p.~031007.

\bibitem{Ortiz2001}
{\sc G.~Ortiz, J.~E. Gubernatis, E.~Knill, and R.~Laflamme}, {\em Quantum
  algorithms for fermionic simulations}, Phys. Rev. A, 64 (2001), p.~022319.

\bibitem{Parrish2019}
{\sc R.~M. Parrish and P.~L. McMahon}, {\em Quantum filter diagonalization:
  {Q}uantum eigendecomposition without full quantum phase estimation},
  arXiv:1909.08925,  (2019).

\bibitem{Peruzzo2014}
{\sc A.~Peruzzo, J.~McClean, P.~Shadbolt, M.-H. Yung, X.-Q. Zhou, P.~J. Love,
  A.~Aspuru-Guzik, and J.~L. O'Brien}, {\em A variational eigenvalue solver on
  a photonic quantum processor}, Nat. Commun., 5 (2014), p.~4213.

\bibitem{Preskill2018}
{\sc J.~Preskill}, {\em Quantum computing in the {NISQ} era and beyond},
  Quantum, 2 (2018), p.~79.

\bibitem{PérezObiol2023}
{\sc A.~Pérez-Obiol, A.~M. Romero, J.~Menéndez, A.~Rios, A.~García-Sáez,
  and B.~Juliá-Díaz}, {\em Nuclear shell-model simulation in digital quantum
  computers}, arXiv:2302.03641,  (2023).

\bibitem{Romero2022}
{\sc A.~M. Romero, J.~Engel, H.~L. Tang, and S.~E. Economou}, {\em Solving
  nuclear structure problems with the adaptive variational quantum algorithm},
  Phys. Rev. C, 105 (2022), p.~064317.

\bibitem{Seki2021}
{\sc K.~Seki and S.~Yunoki}, {\em Quantum power method by a superposition of
  time-evolved states}, PRX Quantum, 2 (2021), p.~010333.

\bibitem{Shen2023realtime}
{\sc Y.~Shen, K.~Klymko, J.~Sud, D.~B. Williams-Young, W.~A. de~Jong, and N.~M.
  Tubman}, {\em Real-time {K}rylov theory for quantum computing algorithms},
  arXiv:2208.01063,  (2023).

\bibitem{Stair2020}
{\sc N.~H. Stair, R.~Huang, and F.~A. Evangelista}, {\em A multireference
  quantum {K}rylov algorithm for strongly correlated electrons}, J. Chem.
  Theory Comput., 16 (2020), pp.~2236--2245.

\bibitem{Stetcu2002}
{\sc I.~Stetcu and C.~W. Johnson}, {\em Random phase approximation vs exact
  shell-model correlation energies}, Phys. Rev. C, 66 (2002), p.~034301.

\bibitem{Tranter2018}
{\sc A.~Tranter, P.~J. Love, F.~Mintert, and P.~V. Coveney}, {\em A comparison
  of the {B}ravyi--{K}itaev and {J}ordan--{W}igner transformations for the
  quantum simulation of quantum chemistry}, J. Chem. Theory Comput., 14 (2018),
  pp.~5617--5630.

\bibitem{Trotter1959}
{\sc H.~F. Trotter}, {\em On the product of semi-groups of operators}, Proc.
  Am. Math. Soc., 10 (1959), pp.~545--551.

\bibitem{Whitehead1977}
{\sc R.~R. Whitehead, A.~Watt, B.~J. Cole, and I.~Morrison}, {\em Computational
  Methods for Shell-Model Calculations}, Springer US, Boston, MA, 1977,
  pp.~123--176.

\bibitem{whitfield2011}
{\sc J.~D. Whitfield, J.~Biamonte, and A.~Aspuru-Guzik}, {\em Simulation of
  electronic structure hamiltonians using quantum computers}, Mol. Phys., 109
  (2011), pp.~735--750.

\bibitem{Yeter2020}
{\sc K.~Yeter-Aydeniz, R.~C. Pooser, and G.~Siopsis}, {\em Practical quantum
  computation of chemical and nuclear energy levels using quantum imaginary
  time evolution and lanczos algorithms}, Npj Quantum Inf., 6 (2020), p.~63.

\bibitem{Zhao2021}
{\sc L.~Zhao, Z.~Zhao, P.~Rebentrost, and J.~Fitzsimons}, {\em Compiling basic
  linear algebra subroutines for quantum computers}, Quantum Mach. Intell., 3
  (2021), p.~21.

\end{thebibliography}
